\documentclass[twocolumn]{svjour3} % smallcondensed, smallextended, twocolumn, draft, final

\usepackage{amsmath,amsfonts,amssymb,bm}
\usepackage{graphicx}
\usepackage{algorithm}
\usepackage{algpseudocode}
\usepackage{booktabs}
\usepackage{bm}
\usepackage{url}
\usepackage{ulem}

\usepackage{xcolor}
\colorlet{myGreen}{green}

\journalname{~}%Nonlinear Dynamics
\begin{document}

% --------------------------------------------------------------
\title{Uncertainty quantification in mechanistic epidemic models via cross-entropy approximate Bayesian computation}
%\subtitle{}
%\titlerunning{}

\author{
Americo Cunha~Jr \and 
David A. W. Barton\and
Thiago G. Ritto}
%\authorrunning{}

\institute{ A. Cunha~Jr (Corresponding author)\at
            Rio de Janeiro State University -- UERJ,
            Institute of Mathematics and Statistics,
		    Rio de Janeiro, Brazil\\
		    ORCID: 0000-0002-8342-0363\\
            \email{americo.cunha@uerj.br}\\
            \and 
            D. A. W. Barton \at 
            University of Bristol, 
            Faculty of Engineering, 
            Bristol, UK\\
            ORCID: 0000-0002-0595-4239\\
            \email{david.barton@bristol.ac.uk\\ \and
            T. G. Ritto \at
            Federal University of Rio de Janeiro -- UFRJ, 
            Department of Mechanical Engineering,
            Rio de Janeiro, Brazil\\
		    ORCID: 0000-0003-0649-6919\\
           \email{tritto@mecanica.ufrj.br}}
}

\date{Received: date / Accepted: date}

\maketitle
% --------------------------------------------------------------

% --------------------------------------------------------------------------
\begin{abstract}
This paper proposes a data-driven approximate Bayesian computation framework for parameter estimation and uncertainty quantification of epidemic models, which incorporates two novelties: (i) the identification of the initial conditions by using plausible dynamic states that are compatible with observational data; (ii) learning of an informative prior distribution for the model parameters via the cross-entropy method. The new methodology's effectiveness is illustrated with the aid of actual data from the COVID-19 epidemic in Rio de Janeiro city in Brazil, employing an ordinary differential equation-based model with a generalized SEIR mechanistic structure that includes time-dependent transmission rate, asymptomatics, and hospitalizations. A minimization problem with two cost terms (number of hospitalizations and deaths) is formulated, and twelve parameters are identified. The calibrated model provides a consistent description of the available data, able to extrapolate forecasts over a few weeks, making the proposed methodology very appealing for real-time epidemic modeling.

% (ii) the use of the CE method to build an informative prori distribution for the parameters of the epidemiological model, 

\end{abstract}
% --------------------------------------------------------------------------

% --------------------------------------------------------------------------
\section{Introduction}
\label{sec:introduction}
% --------------------------------------------------------------------------

Since the COVID-2019 outbreak became a public information in January 2020 \cite{WHO2020}, many researchers have contributed with a variety of epidemic models to deal with this epidemic spread. They seek a better understanding of the disease's propagation mechanism and make short-term forecasts to guide public and private agents in related decision-making. In this context, mechanistic compartmental models with classical structures such as the SIR (susceptible, infected and removed), SEIR (susceptible, exposed, infected and removed) \cite{Brauer:2017,Hethcote:2000,Vynnycky2010}, or their variants with additional compartments \cite{Costa2020p043306,Cotta2020p220,Lyra2020,Martcheva:2015,Muller:2015} has been widely explored in literature.

These models are exciting tools to aid an epidemiologist since they can explain the past and explore future scenarios for an epidemic outbreak from qualitative and quantitative points of view. Thus they generate insight and can support decision-making. Their balance between simplicity (fast to run simulations) and complexity (good to represent the phenomenology of the problem) may be an advantage for situations where analysis needs to be done in near-real-time (like in an evolving outbreak).

In this direction, Pacheco et al. \cite{Pacheco2021} analyzed an SEIR-type model and investigated different scenarios for Brazil, highlighting the importance of social isolation to avoid a collapse of the hospital infrastructure (in the early COVID-19 outbreak). In the meantime, Vyasarayani and Chatterjee \cite{Vyasarayani2020} studied an SEIR model with an additional compartment for quarantine, considering time delays for latency and an asymptomatic phase, while Yu et al. \cite{Yu2020} and Cai et al. \cite{Cai2022p071101} proposed fractional versions of the SEIR model that capture memory effects of epidemic dynamics.

At this point, it is essential to emphasize that to forecast real-world epidemic outbreaks, especially in real-time, the employed compartmental mechanistic models must be calibrated and validated with actual and reliable data. Often, this identification is a dynamic process, with the models being updated whenever new data becomes available. In addition, uncertainties play a significant role in epidemiological models \cite{Smith:2014,Taghizadeh2020p104011}. Values of the model parameters, the model structure, and the epidemic data are uncertain. Thus, beyond identifying values for parameters, it is crucial to perform an uncertainty quantification (UQ) study, which can take into account the variability of the parameters. Probability theory might be used in this endeavor \cite{cunhajr2017,Jaynes2003,Soize2017}, and the Bayesian learning strategy \cite{Kaipio2004,Sivia2006,Tenorio2017} is convenient because prior knowledge is updated consistently with data and UQ occurs automatically. Studies dealing with these aspects can be seen in the recent literature of computational epidemiology \cite{He2020,Jha2020,Kucharski2020,Kuhl2021,Libotte2022p1919,Lobato2021nody,Lyra2020,Zhang2021p1}.

For instance,  He et al. \cite{He2020} analyzed an SEIR model with hospitalization and quarantine, using the particle swarm optimization algorithm (a population-based stochastic optimization algorithm) to identify the model parameters from data and considering the stochastic nature of the infection by introducing a Gaussian white noise. Using Poisson and binominal processes to incorporate uncertainty in case observations within an SEIR model, Kucharski et al. \cite{Kucharski2020} describe the dynamics of newly symptomatic cases, reported onsets of new infections, reported confirmation of cases, and the infection prevalence on evacuation flights. A different stochastic system is used by Lobato et al. \cite{Lobato2021nody}, where a set of stochastic differential equations is employed to describe the random evolution of time-dependent parameters of a compartmental model.

On the other hand, Jha et al. \cite{Jha2020} employed a set of partial differential equations governing the spatial-temporal evolution of susceptible, exposed, infectious, recovered, and deceased individuals, considering a strategy for model calibration, validation, and UQ based on Bayesian learning. Within this UQ framework, they considered additive Gaussian noise to construct the likelihood function and assumed log-Normal priors. This UQ approach for computational epidemiology is very general and powerful, being considered the standard methodology to build data-driven mechanistic epidemiological models for use in real-time \cite{Kuhl2021,Roda2020p221}. Variations of this general methodology are also employed by Libotte et al. \cite{Libotte2022p1919}, Lyra et al. \cite{Lyra2020}, Zhang et al. \cite{Zhang2021p1} --- among many authors --- and the general setting is very well described in the excellent book by E. Kuhl \cite{Kuhl2021}.

Data-driven epidemic modeling via Bayesian learning has some natural advantages, which stand out: (i) combines the identification of the model parameters (model calibration) and quantifies the effects of underlying uncertainties (uncertainty quantification) into a single framework; (ii) allows new data (information) to be incorporated into the data-driven model in a very straightforward way, via Bayes theorem. However, some weaknesses of this framework often cannot be ignored, such as: (i) the use of sampling techniques like Markov Chain Monte Carlo, which often translates into a computationally intensive process; (ii) the great sensitivity of the inference results to the choice of the prior distribution, which encapsulates a priori knowledge about the model parameters; (iii) the typical difficulty of inferring the initial conditions of the dynamic model when information about these parameters is scarce.

The computational cost can be alleviated in several ways, for example, by exploring parallelization strategies, using surrogate models, or employing approaches that avoid evaluating the likelihood function (typically the most expensive step in the Bayesian inference process), etc. A technique that tackles this problem by replacing the evaluation of the likelihood function with the calculation of a computationally cheaper error metric is known as Approximate Bayesian Computation (ABC) \cite{Kypraios2017p42,Trevelyan2018p4,Minter2019p100368,Neal2019}. This is a likelihood-free learning strategy where the prior probability distribution of the parameters is updated, with the aid of available data, only comparing the discrepancy between predictions and observations. Nevertheless, this strategy maintains a prior sensitivity, making informative priors essential, and offers no advantage when inferring initial conditions with reduced information.

To construct an informative prior distribution for the dynamic model parameters, the present work exploits a non-convex optimization technique known as the cross-entropy (CE) method \cite{Cunha2021,Boer2005,kroese2011}. This iterative optimization technique starts with an initial probability distribution for the parameters and sequentially updates it, seeking to minimize a cost function that measures the discrepancy between model predictions and data observations, achieving the global optimum asymptotically. This metaheuristic is an exciting methodology to identify parameters in dynamical systems, primarily because of its simplicity and theoretical guarantees of convergence, with excellent results reported in recent literature \cite{Dantas2019,cunhajr_icedyn2019,cunhajr_cobem2019_2,Tosin2021,Wang2012}.

The identification of a reasonable initial condition from a small set of information can be made from the knowledge of dynamic states (obtained from the system of differential equations) that are compatible with observations of variables accessed via surveillance data. Starting the evolution of the dynamics from a state like this ensures that all temporal variables present, at the initial instant, plausible values, which increases the consistency of the simulation of the epidemic outbreak. The combination of this approach for initial conditions with a Bayesian inference process for the other parameters of the epidemiological model can be advantageous in several real problems of computational epidemiology, thus being the object of interest of this paper.

This paper proposes a novel methodology for calibrating and uncertainty quantifying a mechanistic epidemic model that combines the CE and ABC techniques with a clever strategy for inferring realistic initial conditions. First, the vector of initial conditions is estimated by a combination of dynamic states compatible with observations of the epidemic outbreak. The second step employs the CE method to construct (solving a non-convex optimization problem) an informative prior distribution that represents the parametric uncertainties. Then, it uses ABC to refine (update) the optimal parameter distributions and propagate the underlying uncertainties through the model. It can infer realistic initial conditions with a theoretical guarantee of building an informative a priori distribution. In addition to calibrating/updating the dynamic model, it also considers the effects of the parametric uncertainties underlying the problem. To the best of the authors' knowledge, this formulation of the Bayesian learning process for epidemiological models combining CE and ABC has not yet been explored, contributing towards improving the methodology's inference capacity and, thus, to developing a robust framework UQ on mechanistic epidemic models. The proposed methodology's effectiveness is illustrated with an SEIR-type epidemic model with seven compartments (susceptible, exposed, infectious, asymptomatic, hospitalized, recovered, and deceased) \cite{EPIDEMIC2022}, and actual data from the city of Rio de Janeiro.

The paper is organized as follows. Section \ref{sec_model} depicts the mechanistic epidemic model. The proposed methodology that combines CE with ABC is presented in section \ref{sec_ABCCE}. The results are shown in section \ref{sec_results}. The manuscript body is closed with the concluding remarks in section~\ref{sec_concl}.

%is robust due to two key ingredients: (i) the construction of an informative prior distribution for the model parameters with aid of the cross-entropy method (CE) ; (ii) the update of the dynamic model parameters estimation and the quantification of the underlying parametric uncertainties via approximate Bayesian computation (ABC) \cite{Toni2009}.

%In comparison with other UQ strategies for epidemics available in the literature \cite{He2020,Jha2020,Kucharski2020,Kuhl2021,Lobato2021nody,Lyra2020,Zhang2021p1}, the present methodology innovates by combining ABC and CE, resulting in a simple algorithm with a theoretical guarantee of obtaining a good model calibration in typical situations. 

% -- comment -- %
%Besides parameter estimation, ABC identifies correlation between the parameters and quantifies uncertainty in the response. Do other articles emphasise/explore this point? Should we put it in a digital twin framework? Is it too pretentious?
% --------------------------------------------------------------------------

% --------------------------------------------------------------------------
\section{SEIR(+AHD) epidemic model}
\label{sec_model}
% --------------------------------------------------------------------------

\subsection{Modeling of the contagion process}

\begin{figure*}
    \centering
    \includegraphics[scale=0.3]{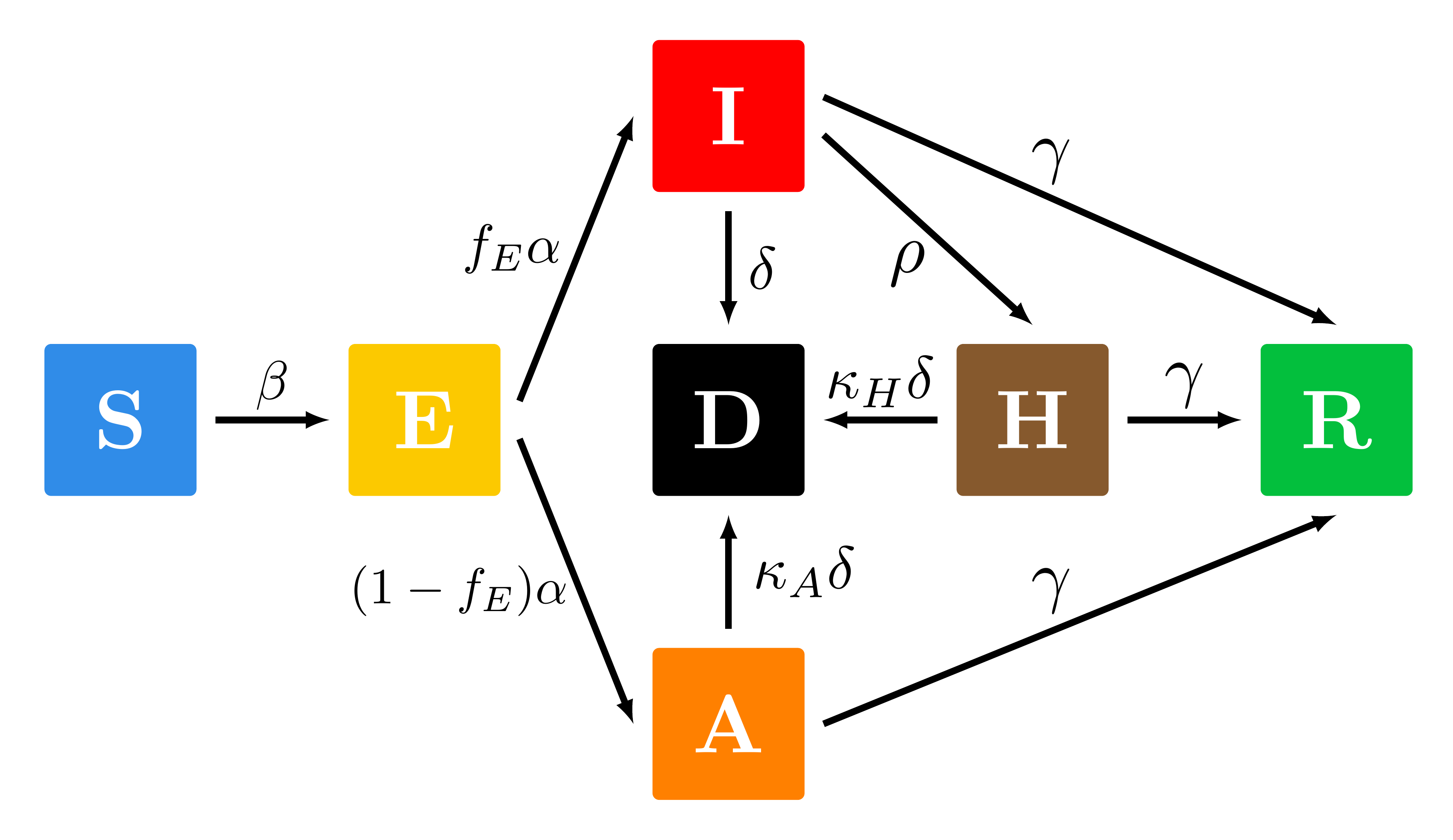}
    \caption{Schematic representation of the SEIR(+AHD) compartmental model considering latency period, asymptomatic individuals, hospitalizations, and deaths. This model is used here to describe the COVID-19 dynamics.}
    \label{fig_SEIARHD_model}
\end{figure*}

The compartmental model employed in this work to describe a COVID-19 outbreak in Rio de Janeiro city is schematically illustrated in Figure~\ref{fig_SEIARHD_model}, where the population is segmented into seven disjoint compartments: susceptible (S); exposed (E); infectious (I); asymptomatic (A); hospitalized (H); recovered (R); deceased (D). This model is dubbed here as the SEIR(+AHD) model.

In this dynamic contagion model, the infection spreads via direct contact between a susceptible and an infected (infectious, asymptomatic, or hospitalized) individual. For simplicity, it is assumed that infectious and asymptomatic individuals are equally likely to transmit the disease to a susceptible person, while this risk is reduced in hospitalized individuals. The latency period between a person becoming infected, starting to have symptoms, and transmitting the disease, is taken into account by the presence of an exposed compartment, which counts those individuals who, despite carrying the pathogen, still do not show symptoms nor can infect other people. Among the infected, some individuals are asymptomatic; only a fraction display symptoms after incubation; they are dubbed infectious. Asymptomatic individuals can recover or die (a rare event). On the other hand, infectious individuals, in addition to recovery and death, may result in hospitalization. Hospitalized people reduce their probability of dying from the disease, but they can still have this outcome or recover. The recovered compartment is just an accumulator receiving individuals from various groups but does not directly interfere with the dynamics. This model was proposed by Pavack et al. \cite{EPIDEMIC2022}, who were inspired by the age-structured model presented in \cite{Lyra2020}, and its variant which considers ICU admissions presented in \cite{Oliveira2021p333}.

%This assumption of the model is hard to explain. If an asymptomatic individual can die he will probably go to the hospital first. Concerning the model, maybe it is not too bad because $kappa_A$ is very small.

The population in each of the compartments at time $t$ is measured by the following state variables: susceptible $S(t)$; exposed $E(t)$; infectious $I(t)$; asymptomatic $A(t)$; hospitalized $H(t)$; recovered $R(t)$; and deceased $D(t)$. Variable $N=N(t)$ represents the alive population size at time $t$. This contagion model has the following parameters: initial alive population $N_0$ (number of individuals); transmission rate $\beta$ (days$^{-1}$); latent rate $\alpha$ (days$^{-1}$); fraction of symptomatic $f_E$ (non-dimensional); recovery rate $\gamma$ (days$^{-1}$); hospitalization rate $\rho$ (days$^{-1}$); death rate  $\delta$ (days$^{-1}$); asymptomatic mortality-factor  $\kappa_A$ (non-dimensional); hospitalization mortality-factor $\kappa_H$ (non-dimensional); and hospitalization infectivity-factor $\epsilon_H$ (non-dimensional).

The deterministic non-autonomous dynamical system associated to this compartmental model (see \cite{EPIDEMIC2022} for details) is written as
\begin{equation}
    \begin{array}{lcl}
         \vspace{2mm}
         \dot{S} & = & - \beta(t) \, S \, (I+A+ \epsilon_H \, H)/N \, ,\\ 
         \vspace{2mm}
         \dot{E} & = & ~~ \beta(t) \, S \, (I+A+\epsilon_H \, H)/N - \alpha \, E \, ,\\
         \vspace{2mm}
         \dot{I} & = &  f_E \, \alpha \, E - (\gamma+\rho+\delta) \, I \, ,  \\ \vspace{2mm}
         \dot{R} & = &  \gamma \, (I+A+H) \, , \\ 
         \vspace{2mm}
         \dot{A} & = &  (1-f_E) \, \alpha \, E - (\kappa_A \, \delta + \gamma) \, A \, , \\
         \vspace{2mm}
         \dot{H} & = & \rho \, I - (\gamma+\kappa_H \, \delta) \, H \, , \\ \vspace{2mm}
         \dot{D} & = &  \delta \, (I+\kappa_A \, A+\kappa_H \, H) \, ,\\
         \vspace{2mm}
         \dot{N} & = &  - \dot{D} \, ,
    \end{array}
    \label{eq_dyn}
\end{equation}
where the corresponding initial conditions are given by $\bm{u}(0) = \left(S_0,E_0,I_0,A_0,H_0,R_0,D_0,N_0 \right)$. Obviously, if convenient, the size of this system can be reduced by one unit if the last equation is replaced by the algebraic constraint $N = N_0 + D_0 - D$, which represents the total population evolution over time.

\subsection{Time dependence of the transmission rate}

As the disease spreads, the parameter $\beta$ might change, and this temporal dependence can be taken into account through the following expression (taken from \cite{Vasconcelos2021}):
\begin{equation}
\beta(t)=\beta_0 + \frac{(\beta_{\infty}-\beta_0)}{2}\left(1+\tanh\left(\eta \, \frac{(t-t_{\beta})}{2}\right)\right) \, ,
\label{eq_beta}
\end{equation}
where $\beta_0$ is the initial value of $\beta$, $\beta_{\infty}$ the final value, the adaptation time $\eta$ defines how fast $\beta$ reaches $\beta_{\infty}$, and $t_{\beta}$ is the transition time (when $t=t_{\beta}$ then $\beta=(\beta_0+\beta_{\infty})/2$). This model allows $\beta$ to smoothly vary between two distinct levels of disease transmission (from lower to higher, or vice versa), a situation typically encountered in the \mbox{COVID-19} contagion dynamics \cite{Gianfelice2022p031101,Vasconcelos2021}.

\subsection{Associated dynamic system}
\label{sec:dyn_system}

The dynamic state of the epidemic system (\ref{eq_dyn}) at time $t$ can be represented, in a compact way, by the time-dependent vector
\begin{equation}
\bm{u}(t) =  \left(S, E, I, A, H, R, D, N\right) \, ,
\label{eq_dyn_state}
\end{equation}
\noindent
while the model parameters may be lumped into the parameter vector
\begin{equation}
\bm{x} =  \left(\beta_0,\alpha,f_E,\gamma,\rho,\delta,\kappa_A,\kappa_H,\epsilon_H,\beta_{\infty},\eta, t_{\beta} \right) \, ,
\end{equation}
so that the dynamic model can be written as
\begin{equation}
    \dot{\bm{u}}(t) = F(t, \, \bm{u}(t), \, \bm{x}) \, ,
\end{equation}
\noindent
where the map $(t, \, \bm{u}(t), \, \bm{x}) \mapsto F(t, \, \bm{u}(t), \, \bm{x}) \in  \mathbb{R}^{8}$ represents the nonlinear evolution law defined by the right hand side of the dynamical system in (\ref{eq_dyn}).

\subsection{Applicability and limitations}
\label{sec:appl_limit}

The system of differential equations defined in (\ref{eq_dyn}) gives rise to a predictive computational model to describe the dynamics of COVID-19 contagion in a context where tracking the number of hospitalizations is essential. 

Such a model can help study possible epidemiological scenarios and may lead to qualitative and quantitative insights into the epidemic dynamics. Such information can help guide decision-makers in managing their local health system. For instance, the model can check whether there is a risk of overloading the hospitals in a particular city. Also, they could estimate when they will suffer the most significant demand. From a more qualitative perspective, the model can assess the impact on hospitalizations of different strategies to mitigate (or even suppress) the epidemic.

But like any computational model, it is subject to limitations, and its use outside the proper context can lead to entirely erroneous predictions \cite{Tolles2020p2515}. Since it is a deterministic compartmental model, obtained as a mean-field approximation in the thermodynamic limit, it is only applicable in regions where the population density can be modeled as a continuous function, a situation typically valid in urban centers of large cities. As it is a model derived from the SEIR family, it assumes a population with a homogeneous contact structure, which does not correspond to the reality of practically anywhere. Therefore, one must care about the potential effects of population heterogeneity.

Other unmodeled effects that may be significant are related to social behavior change due to the course of the epidemic (e.g. risk perception, change in the pattern of social iterations, mask use, etc.) \cite{Weitz2020p32764}, reinfections \cite{Rahman2022p1438}, etc. These can be included in the model, but this is not the goal of the present paper.
% --------------------------------------------------------------

% --------------------------------------------------------------------------
\section{Uncertainty quantification framework}
\label{sec_ABCCE}
% --------------------------------------------------------------------------

\subsection{Quantities of interest}

Among all the quantitative information that can be estimated with the epidemic model in (\ref{eq_dyn}), this paper is particularly interested in two time-dependent quantities, the number of hospitalizations, and the total number of deaths, i.e., the quantities of interest (QoIs) here are the time series $H(t)$ and $D(t)$.

None of these time series, individually or together, correspond to the response of the dynamic model itself. The model response is given by the parametric curve $t \mapsto \bm{u}(t)$, so that the above time series correspond to a derived quantity $t \mapsto (H(t), D(t))$, extracted from $\bm{u}(t)$ through a projection.

On the theoretical plane, $t \mapsto (H(t), D(t))$ is defined over a continuous-time domain and, consequently, is an infinite-dimensional object. However, for computational purposes, it is necessary to discretize both time-series so that, in practice, the dynamic model returns finite-dimensional representations of them. Once the computational representation of each time series materializes itself in the form of an n-dimensional numerical sequence, one might think that the model's discrete response is given by the quantities of interest vector
\begin{equation}
    \bm{y} = [H(t_1), \, \cdots, \, H(t_n), \, D(t_1), \, \cdots, \, D(t_n)] \, ,
\end{equation}
where $t_1, \, \cdots, \, t_n$ are the time-instants underlying the temporal discretization. If other observables become the quantities of interest, the vector $\bm{y}$ can be modified straightforwardly. Just as if observables defined in different temporal grids are needed.

\subsection{Abstraction of the epidemic model}

In an abstract perspective, the computational model can be represented by an equation of form
\begin{equation}
    \bm{y} = \mathcal{M}(\bm{x}) \, ,
\end{equation}
\noindent
which indicates that the vector $\bm{y}$ is obtained from the vector of parameters $\bm{x}$ through a mapping $\mathcal{M}$, which represents the discretized version of the dynamic model that is coded in the computer. Therefore, whenever convenient, the notation $\bm{y}(\bm{x})$ is adopted.

Furthermore, if is necessary to distinguish the components of $\bm{y}$ that are related to $H(t)$ from those associated with $D(t)$, the following partition is adopted
\begin{equation}
    \bm{y}(\bm{x}) = [\bm{y}^H (\bm{x}) \, \, \bm{y}^D(\bm{x})] \, .
    \label{eq_model_resp_2}
\end{equation}

In a case where there are $K$ quantities of interest, the response vector is written
\begin{equation}
    \bm{y}(\bm{x}) = [\bm{y}^1 (\bm{x}) \, \, \cdots \, \, \bm{y}^K(\bm{x})] \, .
    \label{eq_model_resp_K}
\end{equation}

This abstract representation helps to simplify the formulation of the uncertainty quantification framework presented in sequence.

\subsection{Data from the epidemic surveillance system}

Epidemiological surveillance data, represented in this paper by the vector quantity $\bm{y}_{\, data}$, can be used to monitor and understand (in real-time or a posterior) the course of an epidemic through direct observation, or in conjunction with computational models that can be calibrated and validated against them.

\begin{figure}
    \centering
    \includegraphics[scale=0.5]{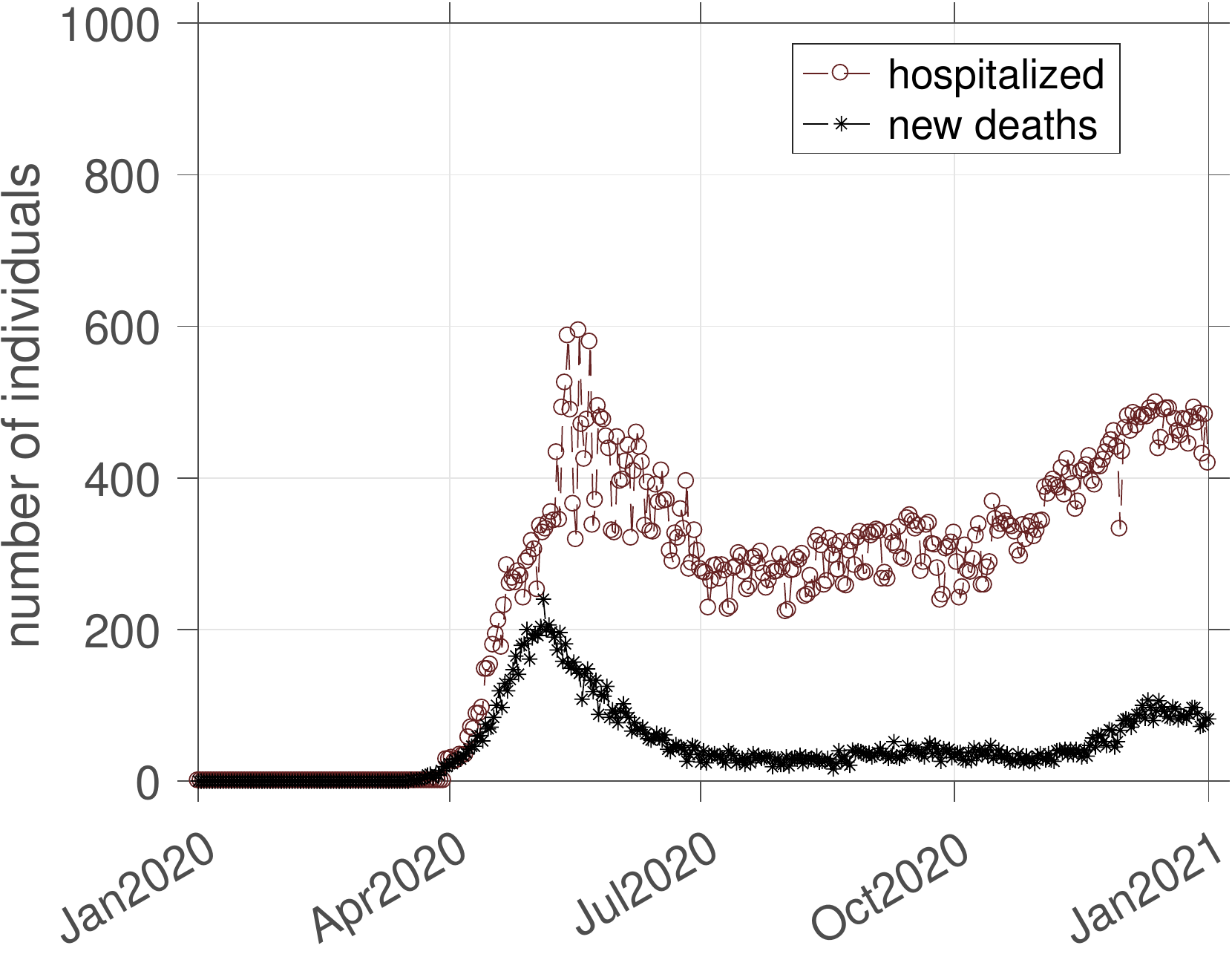}
    \includegraphics[scale=0.5]{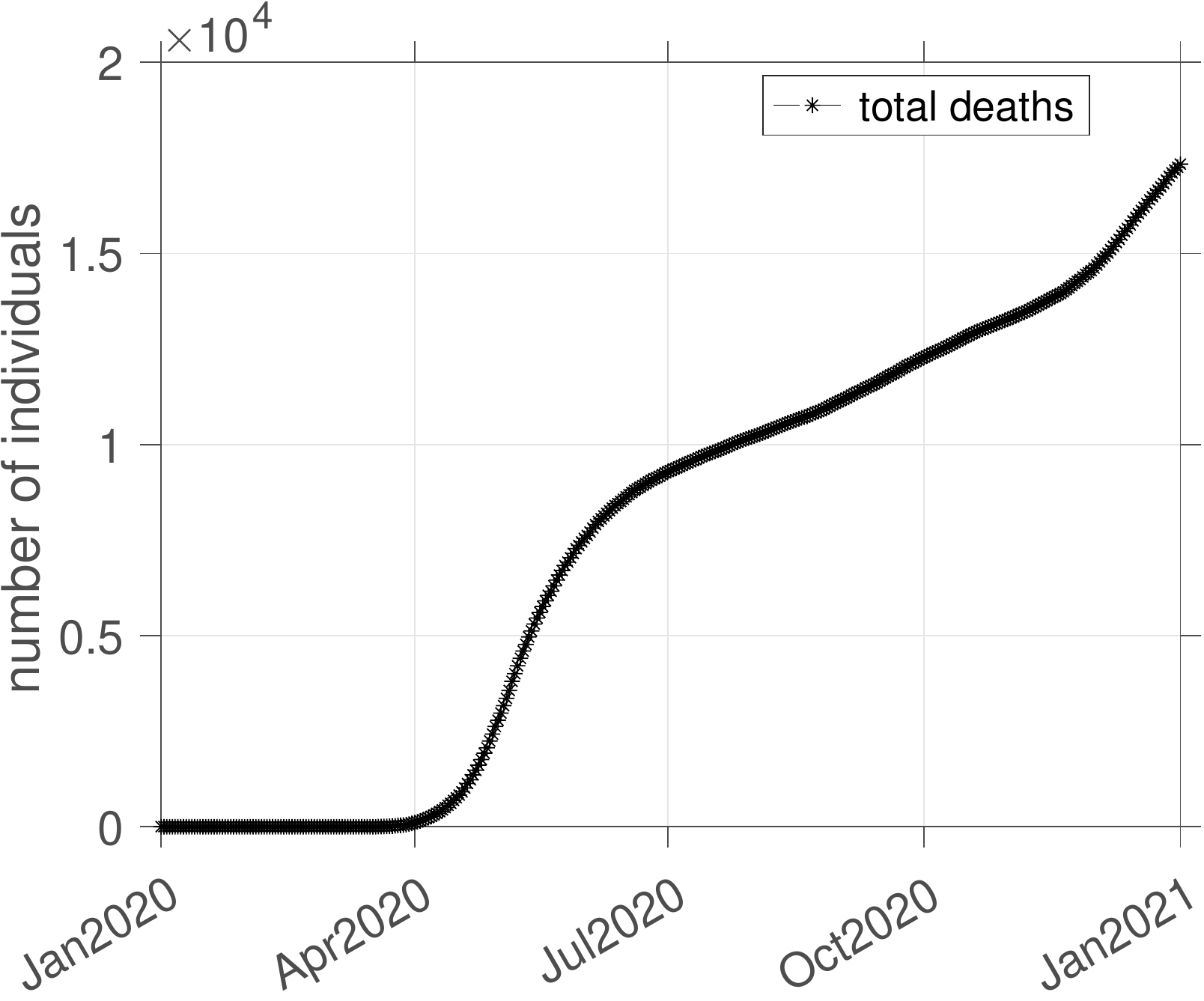}
    \caption{Surveillance data of COVID-19 outbreaks in Rio de Janeiro city between Jan 01, 2020, until Dec 31, 2020 \cite{PainelCovidRJ}. The number of hospitalized individuals and new deaths appears at the top and the total number of deaths at the bottom.}
    \label{fig_data}
\end{figure}

In this work, the data used are related to the records of hospitalizations and deaths due to COVID-19 in the city of Rio de Janeiro city, from Jan 01, 2020, until Dec 31, 2020. The choice of data from this period, rather than more recent observations, is motivated by the fact that this was one of the critical moments of the COVID-19 pandemic in the city of Rio de Janeiro, with high pressure in both health and funeral systems. These data, shown in Figure~\ref{fig_data}, are cataloged and made available by local health authorities \cite{PainelCovidRJ}, being anonymous for ethical reasons and patient privacy.

For the sake of compatibility with the structure of $\bm{y}(\bm{x})$, the data vector $\bm{y}_{\,data}$ which lumps hospitalizations and total deaths (or simply deaths) time series is partitioned as follows
\begin{equation}
    \bm{y}_{\,data} = [\bm{y}^H_{\,data} \, \, \, \bm{y}^D_{\,data}] \, .
\end{equation}

A combination between these data and the epidemic model predictions is done in the uncertainty quantification framework presented below.

\subsection{Quantification of the discrepancy between the mathematical model and available data}

The comparison between data and predictions can be done by means of the following discrepancy (error estimation) function
\begin{equation}
\begin{array}{l}\vspace{.4cm}
\mathcal{J}(\bm{x})=
\omega \, 
\displaystyle\frac{||\bm{y}_{\,data}^{H} -\bm{y}^{H}(\bm{x})||^2}{||\bm{y}_{\,data}^H||^2}\quad +\\ \quad\quad\quad\quad\quad\quad\quad\quad\quad
(1-\omega) \, 
\displaystyle\frac{||\bm{y}_{\,data}^D -\bm{y}^D(\bm{x})||^2}{||\bm{y}_{\,data}^D||^2}\,,
\label{eq_misfit_2}
\end{array}
\end{equation}
where $\omega \in [0,1]$ is a weight parameter which controls how the hospitalization/deaths data contributes to this discrepancy function. For $\omega=0$, only death data are taken into account. Conversely, for $\omega=1$ only hospitalization data matter. Between these extremes the error metric considers a balance between the two data sets. If $\omega=0.5$ they have the same weight. It is worth mentioning that we could identify $\omega$ together with the other model parameters, including it in vector $\bm{x}$, possibility not explored in this paper.

In the case where $K$ quantities of interest are available in the form of data, partitioned as
\begin{equation}
    \bm{y}_{\,data} = [\bm{y}^{1}_{\,data} \, \, \cdots \, \, \bm{y}^{K}_{\,data}] \, ,
\end{equation}
so that the model response reads as in Eq.(\ref{eq_model_resp_K}), and the discrepancy function is written as
\begin{equation}
\begin{array}{l}\vspace{.4cm}
\mathcal{J}(\bm{x}) = 
\displaystyle\sum_{k=1}^{K} \omega_k \, 
\displaystyle\frac{||\bm{y}_{\,data}^{k} -\bm{y}^{k}(\bm{x})||^2}{||\bm{y}_{\,data}^k||^2} \, ,
\end{array}
\label{eq_misfit_K}
\end{equation}
with the weights defining a convex combination, i.e.,
\begin{equation}
    \omega_1 + \cdots + \omega_K = 1 \, .
\end{equation}

\subsection{Baseline calibration of model parameters via the Cross-Entropy (CE) method for optimization}

The process of calibrating the computational model against the available data requires that the discrepancy function defined by Eq.(\ref{eq_misfit_2}) (or by Eq.(\ref{eq_misfit_K})) be minimized by an optimal choice of parameters, a task that is mathematically formulated as the following optimization problem
\begin{equation}
\bm{x}^{\, \star} = \arg \min_{\displaystyle \bm{x} \in \mathcal{X}} \, \mathcal{J}(\bm{x}) \, ,
\label{eq_opt_CE}
\end{equation}
where the set of admissible parameters is defined by
\begin{equation}
\mathcal{X} = \left\lbrace ~ \bm{x} ~ |~  \bm{x}_{min} \preceq \bm{x} \preceq \bm{x}_{max} ~ \right\rbrace \, ,
\label{eq_feasible_set}
\end{equation}
where $\bm{x}_{max}$ and $\bm{x}_{min}$ represent, respectively, upper and lower bound vectors for the model parameters, and the generalized inequality $\preceq$ is understood to hold for each component of the vectors.

In general, the optimization problem defined by (\ref{eq_opt_CE}) is nonconvex, so the use of gradient-based techniques may not be effective in capturing the parameter configuration that best fits the model to the data. Due to the nonconvexity, the solution obtained may be a local optimum (perhaps quite distinct from the global optimum). To avoid this type of situation, the present work tackles this optimization problem with the aid of the cross-entropy method \cite{Boer2005,Rubinstein1999p127,Rubinstein2004}, a simplistic gradient-free iterative procedure for global optimization that has guarantees of convergence in certain typical situations \cite{Rubinstein2016,Rubinstein2004}. This method has been successfully used in recent literature for the identification of parameters in nonlinear computational models \cite{Dantas2019,cunhajr_icedyn2019,cunhajr_cobem2019_2,Tosin2021,Wang2012}.

The fundamental idea of the cross-entropy method is to transform the optimization problem, defined by (\ref{eq_opt_CE}), into a rare event estimation problem. In this way, a sequence of approximations to the global optimum is constructed with the aid of an importance sampling process, where the probability of sampling the rare event (global optimum) grows over time \cite{Rubinstein2016,Rubinstein2004}. 

The CE algorithm consists of two phases: 
\begin{itemize}
    \item \emph{Sampling} -- where the objective function domain is sampled according to a certain distribution to explore the feasible region;
    \item \emph{Learning} -- where the distribution parameters are updated, with the aid of a set of elite samples. The goal is to shrink the distribution simultaneously as it is translated towards the global optimum.
\end{itemize}

To mathematically analyze the behavior of this algorithm, let $\Gamma^{\, \star} = J(\bm{x}^{\, \star})$ be the global minimum sought, and $\bm{X}$ a randomized version of $\bm{x}$, with probability distribution characterized by the probability density function (PDF) $p(\cdot, \bm{v})$, i.e., $\bm{X} \sim p(\cdot, \bm{v})$, for a parameter vector $\bm{v} = (\bm{\mu}, \bm{\sigma})$, with mean vector $\bm{\mu}$ and standard deviation vector $\bm{\sigma}$. 

Since $\Gamma^{\, \star}$ is the global optimal value of $\mathcal{J}$, there are few points $\bm{x}$ in the domain $\mathcal{X}$ that produce a value $\Gamma = \mathcal{J}(\bm{x})$ very close to $\Gamma^{\, \star}$. In this way,
\begin{equation}
\mathcal{P} \left\lbrace \mathcal{J}(\bm{x}) \leq \Gamma \right\rbrace \approx 0 \, ~\mbox{for}~ \Gamma \approx \Gamma^{\, \star} \, ,
\label{eq_prob_ce}
\end{equation}
which states that $\mathcal{J}(\bm{x}) \leq \Gamma$ is a rare event for $\Gamma \approx \Gamma^{\, \star}$.

The solution to this rare event estimation problem involves sampling the domain $\mathcal{X}$ with $N_{ce}$ samples -- drawn according the distribution $p(\cdot, \bm{v})$, evaluate the objective function at the samples $\bm{X}_k$, and then construct of a sequence of estimators $(\hat{\Gamma}_{\ell}, \hat{\bm{v}}_{\ell})$ such that
\begin{equation}
\hat{\Gamma}_{\ell} \xrightarrow{~a.s.~} \Gamma^{\, \star}
~~ \mbox{and} ~~ 
p \, (\bm{x},\hat{\bm{v}}_{\ell}) \xrightarrow{~a.s.~} \delta \left(\bm{x} - \bm{x}^{\, \star} \right),
\end{equation}
where the parameter vector $\bm{v} = (\bm{\mu}, \bm{\sigma})$ is updated by solving the following nonlinear program
\begin{equation}
	\hat{\bm{v}}_{\ell} =
	\arg\,\max_{\displaystyle \bm{v}} \, \displaystyle \sum_{\displaystyle k \in \mathcal{E}_{\ell}}
	\bm{1}\left\{ \mathcal{J}(\bm{X}_k) \leq \hat{\Gamma}_{\ell} \right\} \ln p \, (\bm{X}_k; \, \bm{v}) \, ,
	\label{eq_v_MLE}
\end{equation}
being $\bm{1} \left\{ \cdot \right\}$ the indicator function, and $\mathcal{E}_{\ell}$ an elite sample set, defined by a fixed percentage of the samples $\bm{X}_k$ that produced the values closest to the global optimum. Among the values associated with the elite set, the largest one defines the estimator $\hat{\Gamma}_{\ell}$. 

%The original problem (15) is to minimize. Why maximize here? In the sequence we want to minimize the cross entropy...

The above sequence is optimal in the sense that the importance sampling process tries to minimize the Kullback-Leibler divergence (also known as the cross-entropy function) between the sampling distribution $p \, (\cdot; \, \bm{v})$ and a Dirac delta function centered on the global optimum \cite{Rubinstein2016,Rubinstein2004}.

For sake of numerical implementation, $p \, (\cdot, \, \bm{v})$ is assumed as a truncated Gaussian distribution with bounds defined in a conservative way (assuming broad intervals for the random variables supports). Distributions from the exponential family, like truncated Gaussian, allow the nonlinear program from Eq.(\ref{eq_v_MLE}) to be solved analytically \cite{Rubinstein2016,Rubinstein2004}, so that the low-order statistics in the parameters vector $\bm{v} = (\bm{\mu}, \bm{\sigma})$ are updated by simply calculating the sample mean and sample standard deviation from the elite sample set $\mathcal{E}_{\ell}$.

From a theoretical point of view, the process described above is guaranteed to converge to the global optimum \cite{Rubinstein2016,Rubinstein2004}. However, in practical terms, the distribution may degenerate before it gets close enough to this optimum point \cite{kroese2011,Rubinstein2004}. To avoid this situation, the following smoothing scheme is employed
\begin{equation}
\hat{\bm \mu}_{\ell} := a \, \hat{\bm \mu}_{\ell} + (1-a) \, \hat{\bm \mu}_{\ell-1} \, ,
\end{equation}
\begin{equation}
\hat{\bm \sigma}_{\ell} := b_{\ell} \, \hat{\bm \sigma}_t + (1-b_{\ell}) \, \hat{\bm \sigma}_{\ell-1} \, ,
\end{equation}
\begin{equation}
b_{\ell}  =  b - b \, \left(1 - \frac{1}{\ell} \right)^q \, ,
\end{equation}
for a set of smooth parameters such that $0 < a \leq 1$, $0.8 \leq b \leq 0.99$ and $5 \leq q \leq 10$ \cite{kroese2011,Rubinstein2004}, with the estimations at iterations $\ell$ and $\ell-1$ obtained by solving the Eq.(\ref{eq_v_MLE}) analytically.

The convergence of the sampling process is controlled by the test
\begin{equation}
 ||\bm{\sigma}_{\ell} - \bm{\sigma}_{\ell-1}||_{w} \leq 1 \, ,
 \label{eq_w_norm}
\end{equation}
where the weighted root-mean-square norm of the difference vector $\bm{x}-\bm{y} \in \mathbb{R}^N$ is defined as
\begin{equation}
 ||\bm{x} - \bm{y}||_{w} = \sqrt{ \frac{1}{N} \, \sum_{j=1}^{N} \, \left (  w_j \, (x_j - y_j) \right )^{\, 2} } \, ,
\end{equation}
for the error weights
\begin{equation}
w_j = \frac{1}{\texttt{atol}_j + 0.5 \, |x_j + y_j| \, \texttt{rtol}} \, ,
\label{eq_weights}
\end{equation}
with $\texttt{atol}_j$ and $\texttt{rtol}$ denoting absolute and relative tolerances, respectively. Due to the normalization provided by the weights of Eq.(\ref{eq_weights}), a weighted norm of the order of 1 in (\ref{eq_w_norm}) can be considered small. This type of convergence test, frequently used in the best differential equation solvers \cite{hindmarsh2005p410,Shampine1997p1},  provides robust error control.

\begin{figure*}
    \centering
    \includegraphics[scale=0.24]{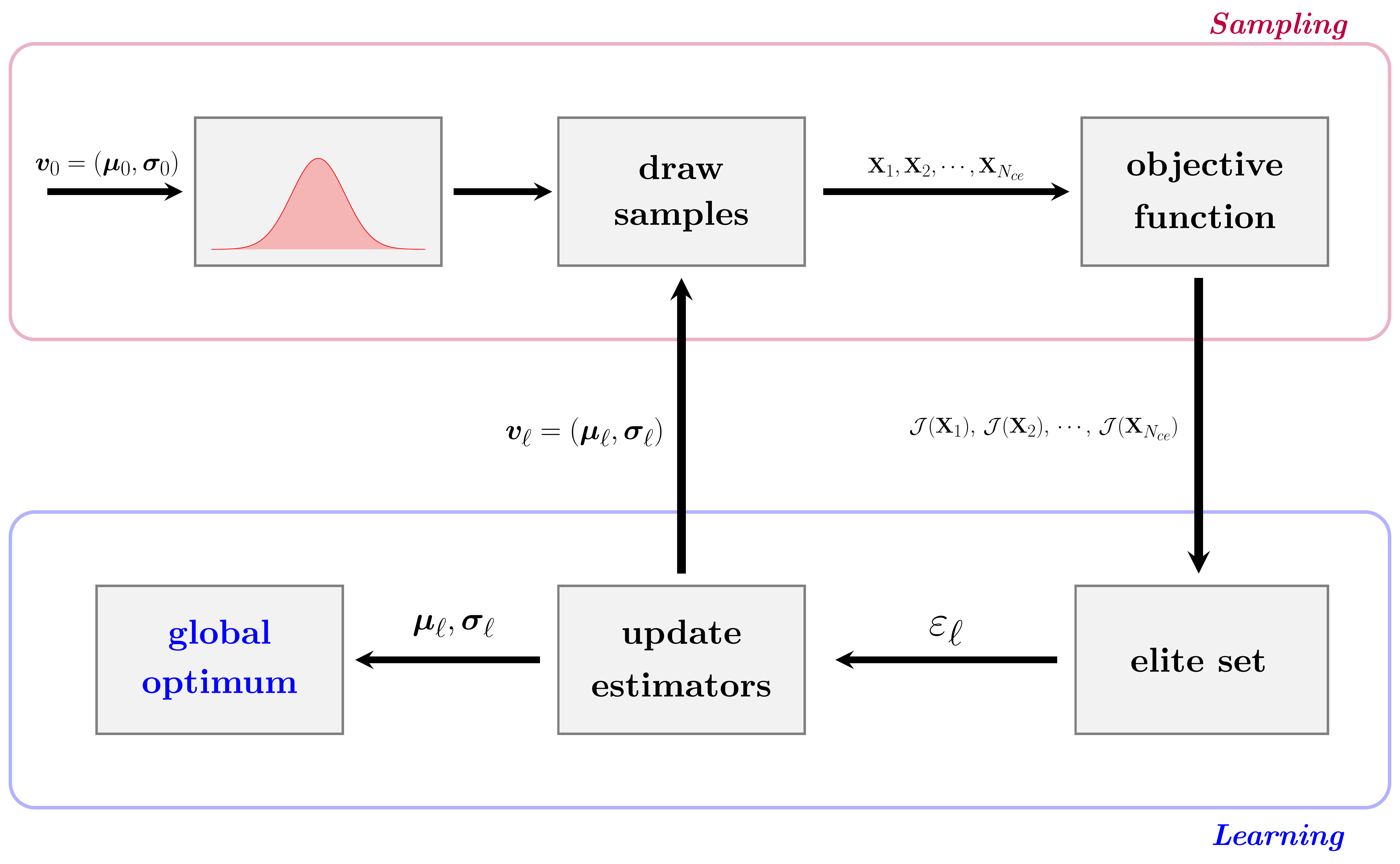}
    \caption{Schematic representation of the two-phases CE algorithm: (i) sampling -- where the domain is sampled according to a given distribution to explore the feasible region, and (ii) learning -- where the distribution parameters are updated with the aid of an elite set, to improve the optimum estimation.}
    \label{fig_CE_method}
\end{figure*}

An overview of the cross-entropy method can be seen in Figure~\ref{fig_CE_method}. More details about the CE implementation can be seen in Algorithm~\ref{algorithm1}, section~\ref{secAlgorithm}, and in the references \cite{Cunha2021,Rubinstein2016,Rubinstein2004}.

\begin{figure*}
    \centering
    \includegraphics[scale=0.21]{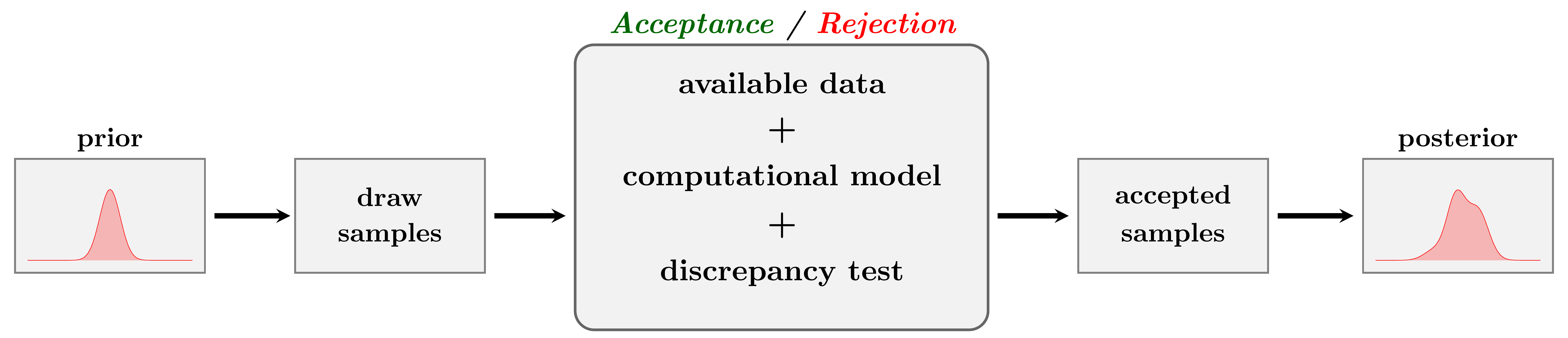}
    \caption{Schematic representation of the ABC algorithm. An a priori distribution is used to generate samples that are selected, if a discrepancy function is small, in an acceptance/rejection process, to generate a posterior distribution.}
    \label{fig_ABC_method}
\end{figure*}

\subsection{Model update and uncertainty quantification through Approximate Bayesian Computation (ABC)}

The update of the model calibration process involves a Bayesian inference scheme \cite{Kaipio2004,Roda2020p221,tarantola2005}, where the data set $\bm{y}_{\,data}$ and a prior distribution for the parameters $\pi (\, \bm{X} \,)$ are combined with the aid of a likelihood function $\pi (\, \bm{y}_{\,data} ~|~ \bm{X} \,)$ to estimate a posterior parameter distribution $\pi (\, \bm{X} ~|~ \bm{y}_{\,data} \,)$ through Bayes' theorem
\begin{equation}
    \pi (\, \bm{X} ~|~ \bm{y}_{\,data} \,) \propto 
    \pi (\, \bm{y}_{\,data} ~|~ \bm{X} \,) \,\, \pi (\, \bm{X} \,) \, ,
    \label{eq_bayes}
\end{equation}
which combines prior information and available data in an optimal way \cite{tarantola2005,Tenorio2017}.

For inference purposes in this setting, the approximate Bayesian computation (ABC) scheme proposed by Toni et al. \cite{Toni2009} is employed. A likelihood function form is not assumed, so the usual hypothesis of additive independent Gaussian noise is unnecessary. Alternatively, the model prediction and the epidemic data are directly compared with the aid of a discrepancy function $\mathcal{J}(\bm{x})$ --- such as those defined by Eq.(\ref{eq_misfit_2}) or Eq.(\ref{eq_misfit_K}) --- to measure the representation quality of the drawn model.  Monte Carlo simulation \cite{cunhajr2014p1355,kroese2011}, employing an acceptance-rejection sampling strategy, is used in the inference process, in a way that a sample $\bm{X}_k$ drawn from the prior distribution $\pi(\bm{X})$ is accepted only if $|\mathcal{J}(\bm{X}_k)| < \texttt{tol}$, where $\texttt{tol}$ is a (problem-dependent) tolerance prescribed by the user. Once the discrepancy function of Eq.(\ref{eq_misfit_2}) is defined as a kind of relative error, it is not necessary to employ two tolerances to control the convergence of the ABC process, as done in the case of CE. But for other definitions of $\mathcal{J}(\bm{x})$ this kind of convergence criterion may be helpful.

The good practice of this technique dictates that all known information about the model parameters should be encapsulated into a prior distribution $\pi(\bm{X})$, to obtain an informative inference process. 

Typically, the iterative process of the CE method provides a lot of information about the parameters, so it is beneficial to take advantage of this knowledge to build the prior. Therefore, the methodology proposed in this paper adopts as prior distribution, for the ABC inference step, the truncated Gaussian distribution with support bounds $\bm{x}_{min}$, and $\bm{x}_{max}$, central tendency $\bm{\mu}$, and dispersion information $\bm{\sigma}$ that comes from the last iteration of CE algorithm, i.e.,
\begin{equation}
    \pi (\, \bm{X} \,) \sim \mathcal{TN} \left( \bm{\mu}, diag(\bm{\sigma}),\bm{x}_{min},\bm{x}_{max} \right) \, .
\end{equation}

It is important to note that although this prior distribution is defined in the same (broad) region where the initial truncated Gaussian of the CE method was defined, it is much more informative. Despite the support limits being kept invariant, the central tendency encapsulated in the mean $\bm{\mu}$, and the dispersion defined by the standard deviation $\bm{\sigma}$ are updated by the CE iteration several times, obtaining a substantial gain of information in this process.

Conceptually, the posterior distribution for the computational model parameters $\pi (\, \bm{X} ~|~ \bm{y}_{\,data} \,)$ is obtained from the samples accepted in the acceptance/rejection process, through some technique of statistical inference. Armed with this probabilistic distribution, in theory any statistical information about the model parameters can be obtained, as well as the intrinsic uncertainty of the parameters can be propagated to the response of the dynamic system. In practice, this distribution is not always inferred, and it is very common that only partial statistical characterizations (e.g. low-order moments) are calculated, or that the accepted samples are used directly in the uncertainty propagation process that follows the definition of the distribution of the model input.

It should be noted at this point that the calibration process described above concerns only the coefficients of the system of differential equations that define the epidemiological model. The initial conditions of the dynamics were not considered. They are inferred by a heuristic process, guided by intuition about the behavior of nonlinear systems, which is described in section~\ref{initial_condit_inference}.

An overview of the ABC can be seen in Figure~\ref{fig_ABC_method}. Further details are available in Algorithm~\ref{algorithm1} from section~\ref{secAlgorithm}, and in the references \cite{Kypraios2017p42,Trevelyan2018p4,Minter2019p100368,Neal2019}.

\subsection{The novel metaheuristic CE-ABC framework for model calibration and uncertainty quantification}
\label{secAlgorithm}

\begin{figure*}
    \centering
    \includegraphics[scale=0.3]{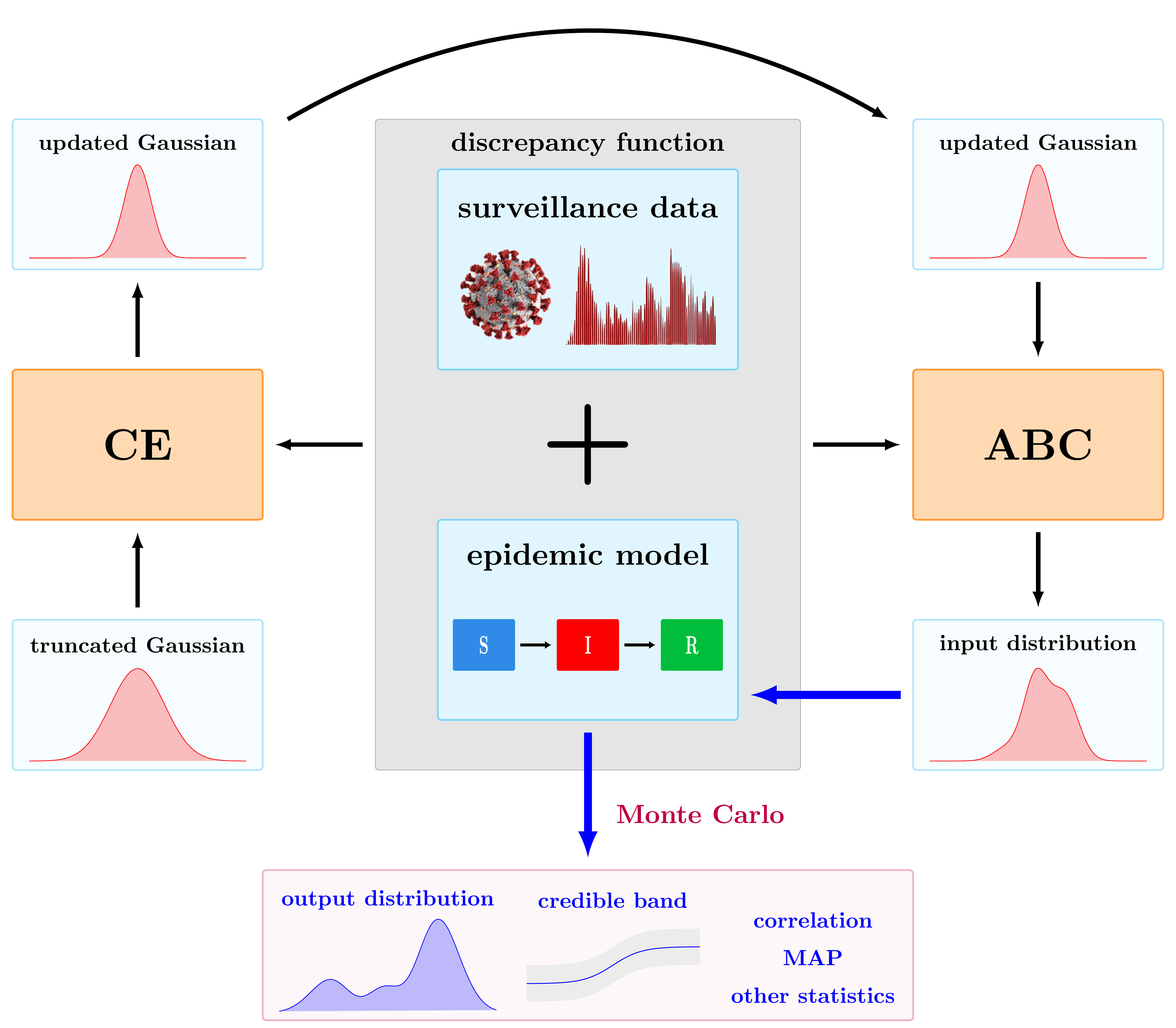}
    \caption{Schematic representation of the CE-ABC framework for estimating parameters and uncertainty quantification in mechanistic epidemic models. First, starting from a truncated Gaussian distribution, an estimate for the model parameters is obtained with the cross-entropy (CE) method. Then, the estimation of the parameters is refined through an inference process employing approximate Bayesian computation (ABC), which also propagates the uncertainties through an acceptance-rejection Monte Carlo simulation to obtain, in the end, a statistical characterization of the model output uncertainty.}
    \label{fig_CE-ABC_diagram}
\end{figure*}

The combination of CE and ABC gives rise to a novel algorithm for epidemic model calibration and UQ. A schematic representation of this new UQ framework, called here CE-ABC, can be seen in Figure~\ref{fig_CE-ABC_diagram}, where the available data (from epidemic surveillance system) and the computational model --- defined by Eq.(\ref{eq_dyn}) --- are combined to evaluate the discrepancy function $\mathcal{J}(\bm{X})$ --- defined by Eq.(\ref{eq_misfit_2}).

First, a truncated Gaussian distribution, defined with aid of conservative (board) bounds and informative values for central tendency, is used by CE method to sample the domain and obtain a first informative estimation for the model parameters values. After the convergence of this iterative process, the updated truncated Gaussian is used to define a prior distribution to be used in the ABC algorithm. Then the ABC combine this informative prior distribution with data, using a discrepancy function, to obtain a posterior distribution of the model parameters. The accepted samples from the Monte Carlo sampling, which define the posterior, are also used to draw credible envelopes. Other statistical information (e.g. low-order moments, MAP, etc) may be obtained in the same way. The computational recipe for the CE-ABC procedure is shown in Algorithm~\ref{algorithm1}.

\begin{algorithm*}
\caption{CE-ABC is a metaheuristic that combines CE and ABC for parameters estimation and uncertainty quantification in mechanistic epidemic models. It receives as input the computational model $\mathcal{M}$, the discrepancy function $\mathcal{J}$, the sampling distributions bounds $\bm{x}_{min}$ and $\bm{x}_{max}$, the number of CE samples $N_{\text{ce}}$, the elite sample set size $N_{\mathcal{E}_{\ell}}$, the number of ABC samples $N_{\text{abc}}$, an absolute and a relative tolerance for CE $\texttt{atol}$ and $\texttt{rtol}$, a tolerance for ABC $\texttt{tol}$,  and an upper bound for CE iterations $\texttt{maxiter}$. The algorithm returns the best parameter estimate obtained by both CE and ABC, and the samples accepted during the ABC iteration process.}
\label{algorithm1}
\begin{algorithmic}[1]
    
\Procedure{CE-ABC}{$\mathcal{M}$, $\mathcal{J}$, $\, \bm{x}_{min}, \, \bm{x}_{max}, \, N_{\text{ce}}, \, N_{\mathcal{E}_{\ell}}, \, N_{\text{abc}}, \, \texttt{atol}, \, \texttt{rtol}, \, \texttt{tol}$, \, $\texttt{maxiter}$}
    \Require $\bm{x}_{min} \preceq \bm{x}_{max}$
    \Require $N_{\text{ce}} > 0$
    \Require $N_{\text{abc}} > 0$
    \Require $N_{\mathcal{E}_{\ell}} > 0$ and $N_{\mathcal{E}_{\ell}} < N_{\text{ce}}$
    \Require $\texttt{atol} \geq 0$
    \Require $\texttt{rtol} > 0$
    \Require $\texttt{tol} > 0$
    \Require $\texttt{maxiter} > 0$

    \textcolor{brown}{//- - - - - CE step - - - - -//}
    \State $\ell := 0$
    \State  $\bm{\mu} := (\bm{x}_{max} + \bm{x}_{min})/2$
    \State  $\bm{\sigma} := (\bm{x}_{max} - \bm{x}_{min})/\sqrt{12}$
    \State Draw $\bm{X} \sim \mathcal{TN}(\bm{\mu},
                                    diag(\bm{\sigma}), 
                                    \bm{x}_{min}, 
                                    \bm{x}_{max})$
    \textcolor{brown}{// total of $N_{\text{ce}}$ samples}
    \While{$||\bm{\sigma}_{\ell}-\bm{\sigma}_{\ell-1}||_{w} > 1$ and $\ell < \texttt{maxiter}$}
        \State $\ell := \ell + 1$
        \State Evaluate $\bm{Y}_k = \mathcal{M}(\bm{X}_k)$ for $k=1:N_{ce}$
    	\State Evaluate $\mathcal{J}(\bm{X}_k)$  for $k=1:N_{ce}$
    	\State Define elite sample set $\mathcal{E}_{\ell}$
        \State Update $\bm{\mu}$ and $\bm{\sigma}$ using $N_{\mathcal{E}_{\ell}}$ samples from $\mathcal{E}_{\ell}$
    \EndWhile
    
    \textcolor{brown}{//- - - - - ABC step - - - - -//}
    \State $\mathcal{J}_{min} = \infty$
    \State $\bm{X}_{best} = \texttt{NaN}$
    \State $\bm{Y}_{best} = \texttt{NaN}$
    \State  Define prior $\pi(\bm{X}) = \mathcal{TN}(\bm{\mu},
                                    diag(\bm{\sigma}), 
                                    \bm{x}_{min}, 
                                    \bm{x}_{max})$
    \State Draw $\bm{X} \sim \pi(\bm{X})$ \textcolor{brown}{// total of $N_{\text{abc}}$ samples}
    \For{$k=1:N_{\text{abc}}$}
        \State Evaluate $\bm{Y}_k = \mathcal{M}(\bm{X}_k)$
    	\State Evaluate $\mathcal{J}(\bm{X}_k)$
    	\If{$\mathcal{J}(\bm{X}_k) < \texttt{tol}$}
    	
            \State Accept $\bm{X}_k$
            \State Save $\bm{X}_k$ and $\bm{Y}_k$
            \If{$\mathcal{J}(\bm{X}_k) < \mathcal{J}_{min}$}
                \State $\bm{X}_{best} := \bm{X}_k$
                \State $\bm{Y}_{best} := \bm{Y}_k$
                \State $\mathcal{J}_{min} := \mathcal{J}(\bm{X}_k)$
            \EndIf
        \Else
            \State Reject $\bm{X}_k$
        \EndIf
    \EndFor
    \State Return $(\bm{X},\bm{Y})_{opt}^{ce}$ \, ,  $(\bm{X},\bm{Y})_{best}^{abc}$ \, , and $(\bm{X},\bm{Y})_{saved}^{abc}$
\EndProcedure
\end{algorithmic}
\end{algorithm*}

\subsection{Remarks about the CE-ABC algorithm}

The novel CE-ABC framework presented here is based on two general statistical methods which have already been applied to several complex problems \cite{Minter2019p100368,Rubinstein2004,Sunnaker2013p1,Toni2009}. The resulting algorithm inherits two interesting properties from these methods: (i) from CE, the guarantee of convergence to the global optimum in typical situations; and (ii) from ABC, the likelihood shape independence and relative computational efficiency, when compared to approaches that require a direct assessment of the likelihood function. Such a mixture of good properties generates a robust framework for stochastic simulations involving epidemic models, which are typically difficult to calibrate and have a limited predictability horizon, requiring quantification of uncertainties for any minimally reliable forecast.

Although the good theoretical properties of the CE-ABC framework are observed in the numerical studies developed by the authors with the epidemic model employed in this paper, its use in conjunction with other types of computational models (e.g. computational mechanics) requires a more comprehensive theoretical analysis. Such a formal analysis for a broad class of models is beyond the scope of this work and the present journal, but it would be a fascinating work on applied mathematics, which the authors leave as a suggestion for future work.

Due to the generality of CE and ABC methods, but in a context where a rigorous mathematical analysis to ensure algorithm functionality for a broad class of computational models is missing, the authors consider that the proposed CE-ABC framework is a metaheuristic\footnote{A technique for efficiently solving a computational problem (approximately) that is generally suboptimal in some sense for practical use.} for model calibration and UQ.

Despite the CE-ABC algorithm's excellent convergence properties, it is impossible to make an accurate inference if ``bad" values (physically/biologically inconsistent or very discrepant with reality) are assigned to the model's nominal parameters, initial conditions, and bounds. Defining the bounds and nominal values for the parameters and initial conditions is an important task that must be done carefully. It is necessary to have biological intuition (in parallel to the importance of physical intuition when in the context of computational physics). The analyst's experience with the problem of interest is essential; it is a kind of expert knowledge that must be embedded into the priors distributions. Besides, exploratory tests with the computational model and information from previous works may be precious to discover a suitable interval of values. 

The results obtained with the CE-ABC framework also strongly depend on the tolerances $\texttt{atol}$, $\texttt{rtol}$, and $\texttt{tol}$, chosen by the user. There are no canonical values for these parameters that are valid for all types of inference; good values are problem-dependent. In this way, the analyst's experience and intuition are crucial in defining these values and a little numerical experimentation with the computational model. In the numerical experiments repeated in the session~\ref{sec_results} these tolerances are defined as being $\texttt{atol} = 0.001$, $\texttt{rtol} = 0.05$, and $\texttt{tol} = 0.1$.

Once again, it is worth emphasizing the observations made in section~\ref{sec:appl_limit} about the limitations and applicability of the model. No matter how robust the calibration and uncertainty propagation algorithm is, how good is the choice of model parameters and bounds. If the model does not describe the reality in a minimally reliable way, terribly wrong (or in the limit nonsense) predictions will emerge from the simulations. Choosing a suitable model is a primary exercise and of great importance in this type of analysis.

% --------------------------------------------------------------------------

% --------------------------------------------------------------------------
\section{Results and discussion}
\label{sec_results}
% --------------------------------------------------------------------------

This section presents several numerical experiments conducted with the SEIR(+AHD) epidemic model and the proposed CE-ABC algorithm. The plausible nominal values used in the integration of the dynamics of a virgin population for COVID-19 infections are presented in Table~\ref{tab_nominal_parameters}, which also shows numerical bounds (upper and lower) that are used to delineate the feasible domain limits in the CE method. The plausible values from Table~\ref{tab_nominal_parameters} correspond to a COVID-19 outbreak in a virgin population to the disease, such as those observed worldwide in 2020. They were determined by information from the literature \cite{Byambasuren2020p223,Cheng2021,Grasselli2020p1574,MIDAS2020,Nogrady2020,PainelCovidRJ,Statistica2021,Vasconcelos2021,Wang2020,Wu2020,Zhou2020p1054} or numerical experimentation.

Numerical experiments with these parameters are not focused on being very reliable reproductions of the COVID-19 outbreaks in 2020. They only aim to have the main characteristics of the epidemic dynamics so that they offer a good test for the methodology proposed in this paper. 

The objective is to show that the CE-ABC framework is a powerful tool for data-driven epidemic modeling and can be used, together with a suitable epidemic model, in near real-time to predict the course of an epidemic outbreak of an emerging disease (such as COVID-19) in a time horizon compatible with the limits of predictability of the underlying dynamics.

\begin{table}
\caption{Plausible nominal values and bounds for the parameters of the SEIR(+AHD) epidemic model.}
\centering
\begin{tabular}{llllll}
\toprule
 & Unit & Nominal & Min    & Max   & Refs \\
\midrule
$\beta$ or $\beta_0$ & 1/day & 1/7      & 1/14    & 1/2    & \cite{MIDAS2020,Wu2020}\\
$\alpha$             & 1/day & 1/5      & 1/10    & 1/2    & \cite{Cheng2021,MIDAS2020}\\
$f_E$                & ---   & 0.8      & 0.7     & 0.9    & \cite{Byambasuren2020p223,MIDAS2020,Nogrady2020}\\
$\gamma$             & 1/day & 1/14     & 1/21    & 1/7    & \cite{MIDAS2020,Zhou2020p1054}\\
$\rho$               & 1/day & 1/700    & 1/2100  & 1/100  & \cite{MIDAS2020,Wang2020,Zhou2020p1054}\\
$\delta$             & 1/day & 1/14000  & 1/21000 & 1/100  & \cite{Statistica2021}\\
$\kappa_A$           & ---    & 0.0010  & 0.0005  & 0.0050 & --- \\
$\kappa_H$           & ---    & 0.05    & 0.01    & 0.10   & \cite{Grasselli2020p1574,MIDAS2020,Wu2020}\\
$\epsilon_H$         & ---    & 0.2     & 0.1     & 0.5    & --- \\
$\beta_{\infty}$     & 1/day & 1/7      & 1/14    & 1/2    & \cite{Vasconcelos2021} \\
$\eta$               & 1/day & 5        & 0       & 10     & \cite{Vasconcelos2021} \\
$\tau_{\beta}$       & day   & 60       & 0       & 120    & \cite{PainelCovidRJ} \\
\bottomrule
\end{tabular}
\label{tab_nominal_parameters}
\end{table}

% old table values
%$\beta$ or $\beta_0$ & 1/day & 1/7     & 1/10   & 1/2   & \cite{MIDAS2020,Wu2020}\\
%$\alpha$             & 1/day & 1/5     & 1/10   & 1/2   & \cite{Cheng2021,MIDAS2020}\\
%$f_E$                & ---    & 0.8     & 0.7    & 0.9   & \cite{Byambasuren2020p223,MIDAS2020,Nogrady2020}\\
%$\gamma$             & 1/day & 1/14    & 1/21   & 1/7   & \cite{MIDAS2020,Zhou2020p1054}\\
%$\rho$               & 1/day & 1/1000  & 1/3500 & 1/800 & \cite{MIDAS2020,Wang2020,Zhou2020p1054}\\
%$\delta$             & 1/day & 1/5000  & 1/8000 & 1/100 & \cite{Statistica2021}\\
%$\kappa_A$           & ---    & 0.001   & 0.0005 & 0.005 & --- \\
%$\kappa_H$           & ---    & 0.05    & 0.01   & 0.1   & \cite{Grasselli2020p1574,MIDAS2020,Wu2020}\\
%$\epsilon_H$         & ---    & 0.2     & 0.1    & 0.5   & --- \\
%$\beta_{\infty}$     & 1/day & 1/4     & 1/10   & 1/2   & \cite{Vasconcelos2021} \\
%$\eta$               & 1/day & 1       & 0      & 2     & \cite{Vasconcelos2021} \\
%$\tau_{\beta}$       & day   & 60      & 0      & 120   & \cite{PainelCovidRJ} \\

\subsection{Dynamic evolution of a fully susceptible population subjected to an initial infection}
\label{sec:VirginPop}

The first analysis presented here concerns the situation of a population virgin to COVID-19 infections, where the disease is introduced into the community by a single individual externally exposed to the viral agent that causes the disease.

This is a hypothetical case (possibly unrealistic ), as it does not consider any measures to mitigate or suppress the outbreak during its occurrence. However, its study may be essential to delineate a possible baseline behavior related to a potential epidemic of COVID-19, providing projections of a worst-case scenario and some intuition about the free evolution of the disease.

In this scenario, a constant value for $\beta = \beta_0$ is considered, as well as an initial population $N_0=5.5 \times 10^{6}$ (compatible with the city of Rio de Janeiro, Brazil), a single exposed individual $E_0 = 1$, and all other initial conditions are set to zero, except the suscetibles, which is set as the difference between $N_0$ and all other variables. The time-step for simulation is equal to 1 day. The model parameters values are defined in the third column of Table~\ref{tab_nominal_parameters}. 

\begin{figure}
    \centering
    \includegraphics[scale=0.5]{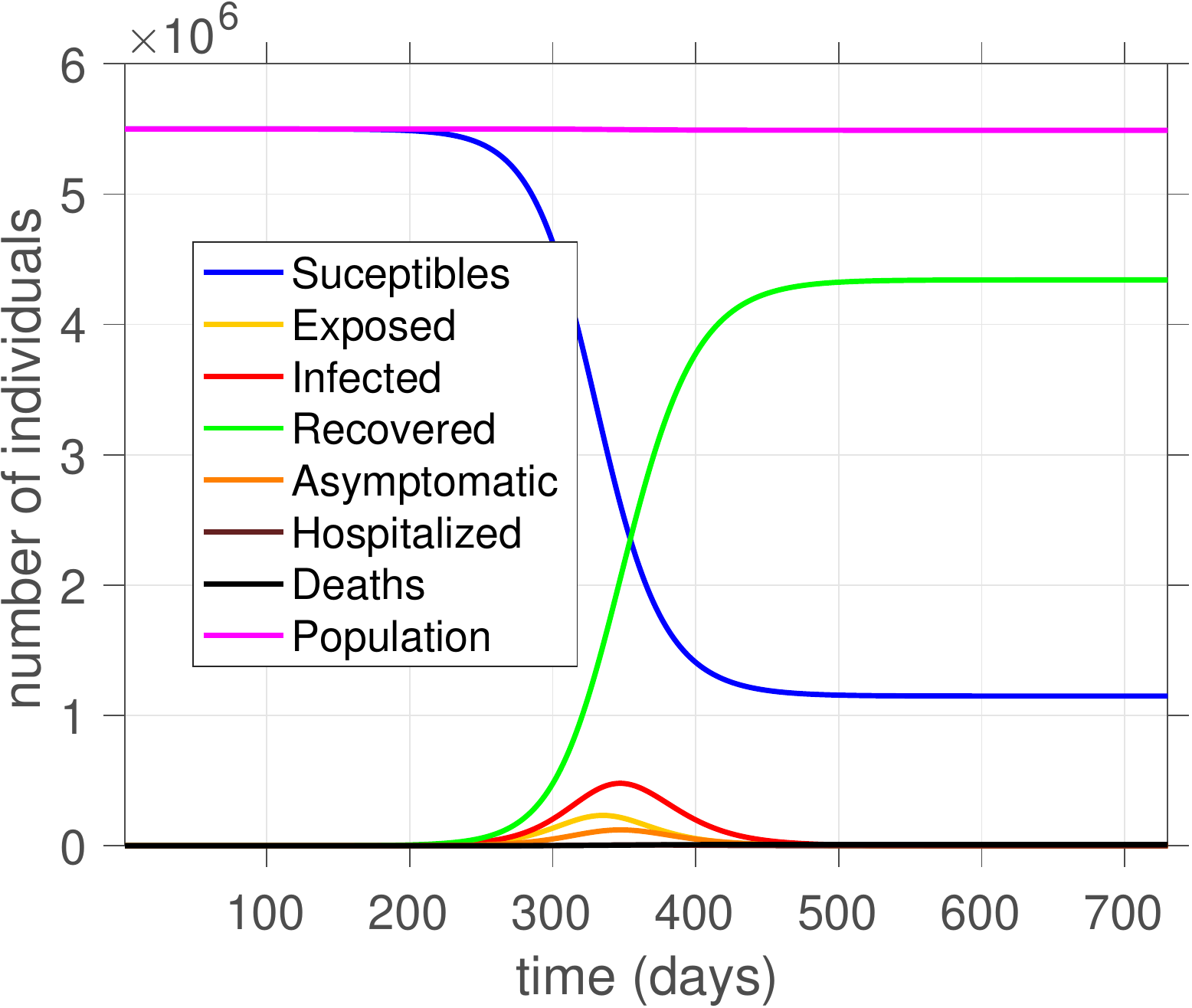}
    \includegraphics[scale=0.5]{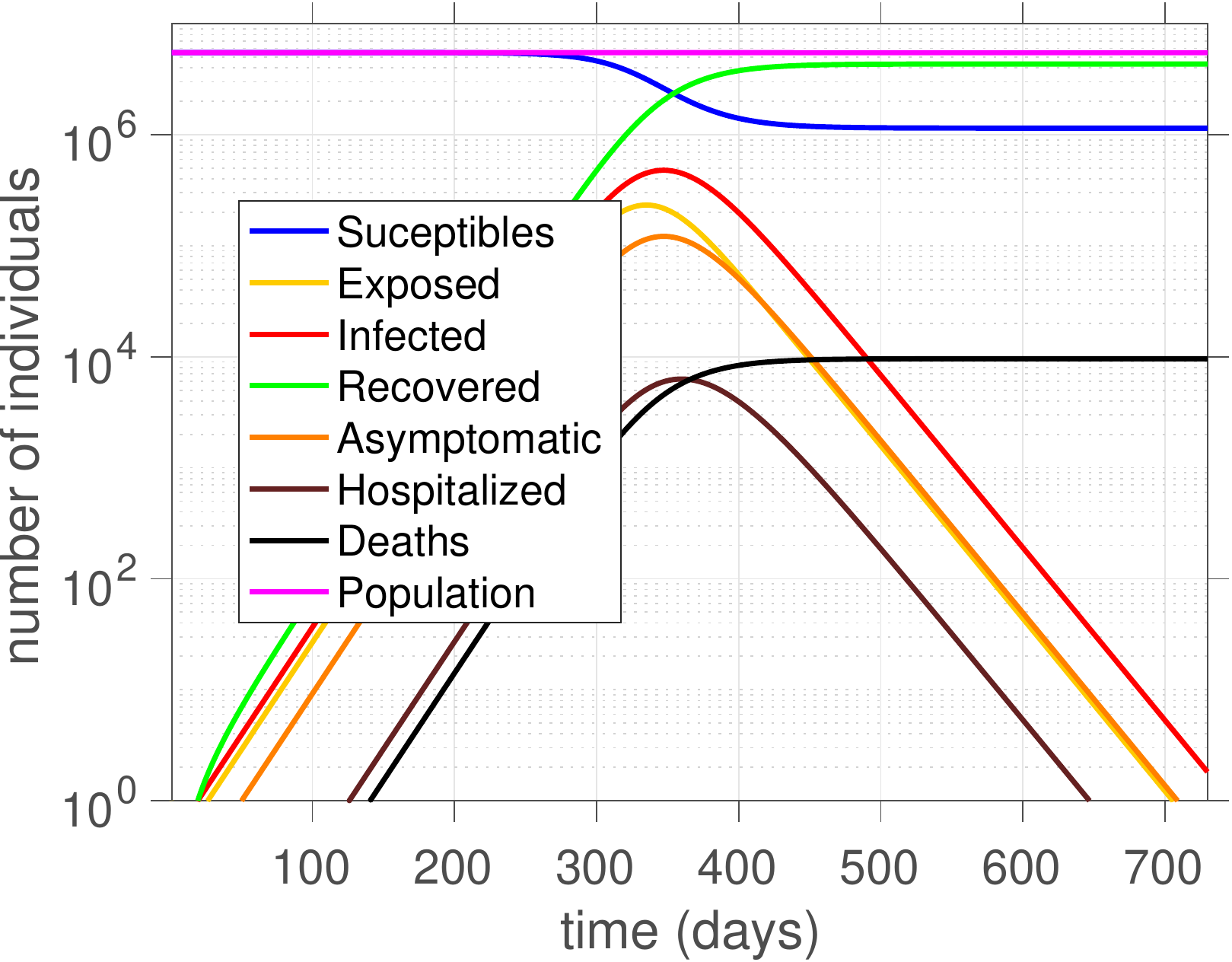}
    \caption{Dynamic response of the SEIR(+AHD) epidemic model in a scenario of a totally susceptible population, with a single individual exposed. Time series in linear scale are shown at the top and in logarithmic scale at the bottom, which allows displaying better the curves in which maximum values are small compared to the initial population value.}
    \label{fig_ivp_all}
\end{figure}

The dynamic evolution of the SEIR(+AHD) model, for a temporal interval of 2 years, can be seen in Figure~\ref{fig_ivp_all}, which shows the corresponding time series in linear (top) and logarithmic (bottom) scales.  In the bottom part, it is possible o better observe the trends of time series in which the maximum values are small compared to the initial size of the population.

Despite community transmission starting in the first moments of the dynamics, due to the transmission structure of this type of model, an outbreak only takes on notable proportions after 200 days of disease circulation in the population. In other words, it may take more than six months after the start of transmission of the disease within this population for the outbreak to be noted by the major public.

However, after the outbreak became noticeable, a wave of contagion by COVID-19 quickly emerged, characterized by a rapid increase in exposed compartments, concomitant with a decrease at the same rate in the susceptible population. Most recover directly, while a small portion dies without medical care. The other part of those infected are hospitalized, most of them recover, and a small amount dies.

The peak of infections occurs around 350 days after the insertion of the first exposed case in the population, almost a year after the disease arrives in the community. There were about 700 thousand people with active disease (exposed, infectious and asymptomatic) in the community during the peak, almost 13\% of the initial population. The susceptible corresponded to close to 20\% of the people after two years, while the recovered account for almost 80\% of the people.

\begin{figure}
    \centering
    \includegraphics[scale=0.5]{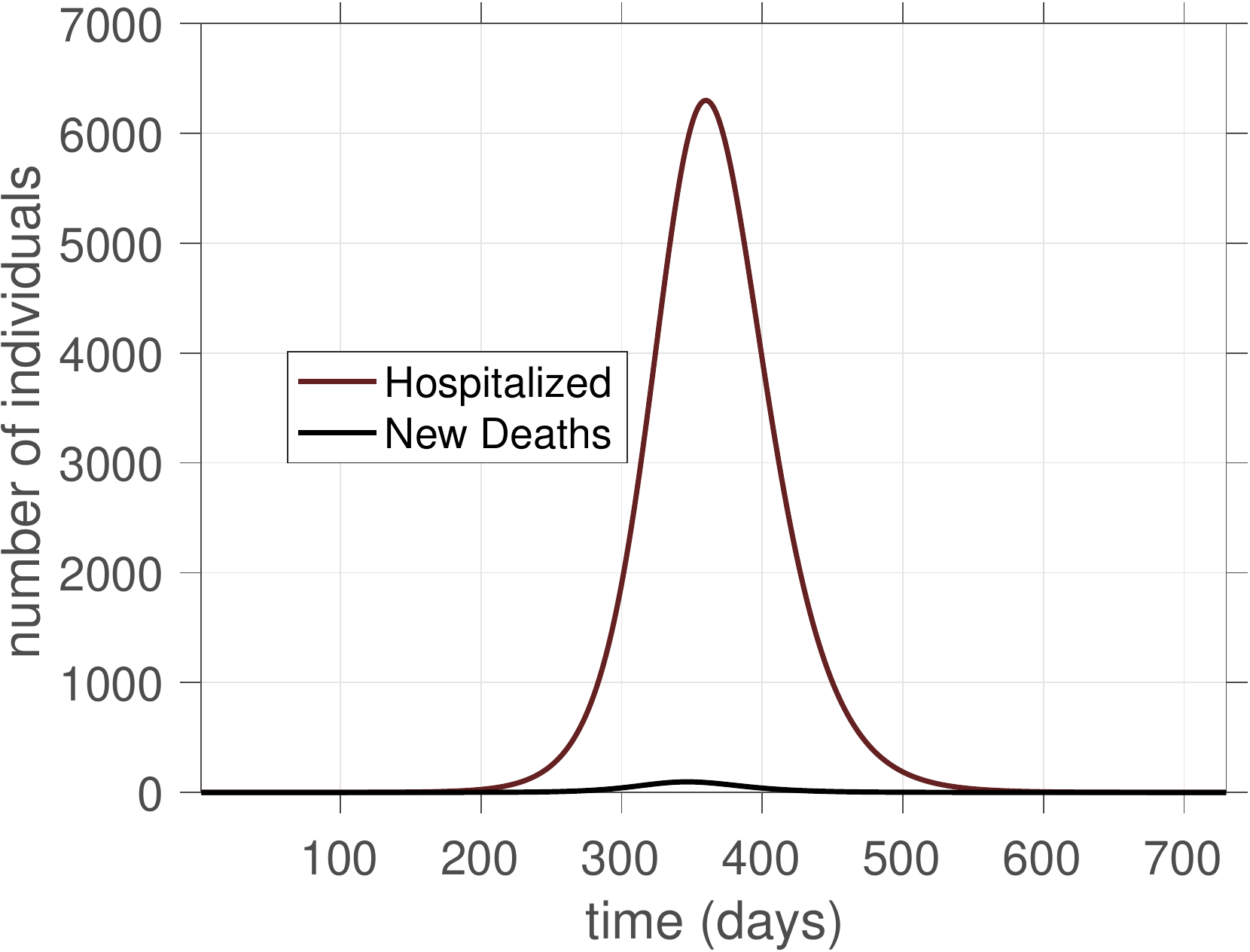}
    \includegraphics[scale=0.5]{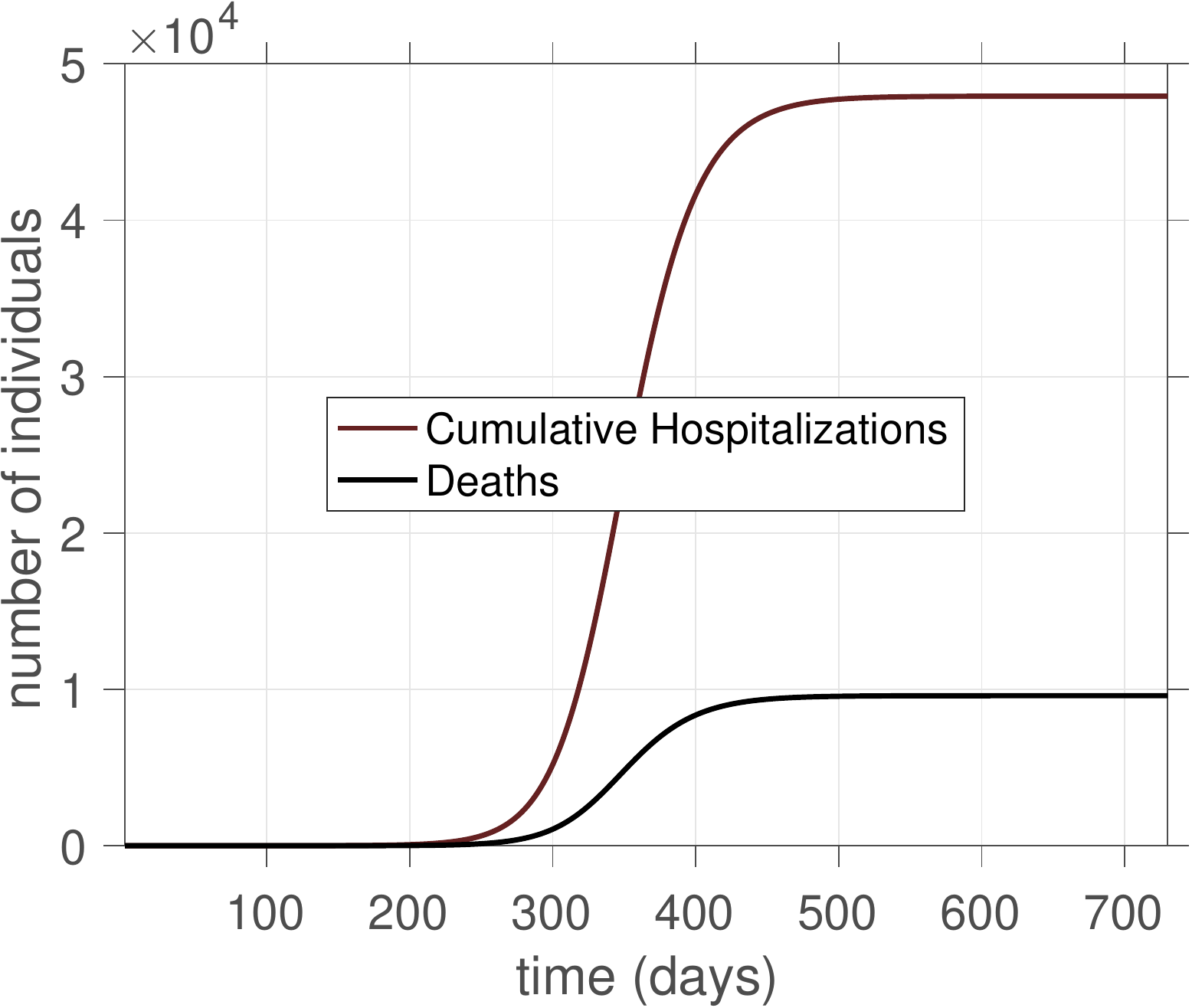}
    \caption{Dynamic response of the two QoIs for the SEIR(+AHD) epidemic model in a scenario of a totally susceptible population, with a single individual exposed. Time series for the number of hospitalized (upper brown curve) and new deaths (lower black curve) are shown at the top, while the corresponding cumulative numbers can be seen at the bottom.}
    \label{fig_ivp_QoI}
\end{figure}

The reader can better appreciate the evolution of the number of hospitalizations and new deaths per day in Figure~\ref{fig_ivp_QoI} (top), as well as their respective cumulative values throughout the epidemic outbreak (bottom). The peaks of hospitalizations and deaths occur a few days after the peak of infections, involving more than 6000 people under medical care and around 100 deaths. At the end of the two-year window, almost 48,000 people were hospitalized at some point, and another 10,000 people died from complications inherent to the disease.

Throughout the outbreak, the variation in population size is slight compared to its initial size (10 thousand is a small number compared to 5.5 million, around 0.2\%) but highly significant in demographic terms. This hypothetical outbreak is responsible for losing approximately 10 thousand lives in about 300 days. This value would correspond to something around 15\% of all deaths that occurred in the city of Rio de Janeiro in 2019\footnote{Demographic data for the city of Rio de Janeiro can are available at \url{https://transparencia.registrocivil.org.br}.}. But note that, in this case, such an unusual amount of deaths is due to a single disease.

\subsection{Determination of a dynamically consistent initial state for the epidemic model calibration}
\label{initial_condit_inference}

Typically, in the process of calibrating a dynamic model with the aid of data, the bottleneck is identifying the initial conditions since, often, the initial state of the system of interest is partially (or even totally) unknown. This is the case when dealing with compartmentalized epidemic systems in the form of an SEIR model or its variants. Observations on the infected compartment are usually available (subject to delay and underreporting), but data from recovered rarely (and even when they are, they are often unreliable). However, for the practical impossibility of measuring them, the susceptible and exposed practically are never known directly. Other possible compartments can also be challenging to measure in practice \cite{Gamerman2022,Kuhl2021}.

In this scenario, the determination of initial conditions (or part of its components) is usually done via direct inference from the data \cite{cunhajr2018p249,cunhajr2019p91,Kucharski2016,cunhajr2020p051103} or by indirect means, with plausible assumptions or educated estimates about the actual values \cite{Costa2020p043306,Pacheco2021}. But while the latter approach is highly subject to epistemic errors, the former may suffer from identifiability issues. Thus, novel methodologies to identify initial conditions of epidemiological systems are welcome.

By the existence and uniqueness theorem for ODEs \cite{arnold1992,hirsch2012,perko2006,strogatz2014,Verhulst2012b}, the dynamical system defined by (\ref{eq_dyn}) and a given initial condition has only one dynamic state for each instant of time. Once the value of one of the components is fixed (e.g. hospitalizations) for a particular moment, only one combination of values in the other compartments produces a dynamic state compatible with the fixed value. For this reason, it is practically impossible to infer a consistent initial condition from assumptions or ansatz to values (especially in an actual setting where surveillance data are imperfect representations of the dynamics of interest, and there are compartments for which data are not available).

To avoid the above difficulties, a three-step procedure to determine a suitable set of initial conditions that is compatible with the observed data is proposed:
\begin{enumerate}
    \item Given a reference value for hospitalizations $H_{ref}$, the dynamics of a population virgin to the disease (such as presented in section~\ref{sec:VirginPop}) is used to determine the time instant for which $H(t)$ is closest to $H_{ref}$. The corresponding dynamic state is recorded;
    \item Analogously, given a reference value for deaths $D_{ref}$, the dynamic state corresponding to the shortest distance between $D(t)$ and $D_{ref}$ is obtained and recorded;
    \item Finally, a dynamic state corresponding to a weighted average between the two states determined above is calculated and assumed to be a dynamically consistent initial condition. 
\end{enumerate}

When using the dynamics of a population totally susceptible to the disease to identify a dynamic state close to the reference values for $H$ (or for $D$), the procedure guarantees that this state is ``dynamically consistent'', as it is a solution to the initial value problem associated with the epidemic model. Although, in general, such a state does not exactly satisfy the reference value, by the continuous dependence of the solutions on the initial conditions, one can guarantee that such a dynamic state is ``sufficiently close'' to the state associated with the exact value of the reference. By making a convex combination of initial conditions obtained this way, we still have a dynamic state close to all the reference values. In this way, the procedure described above can generate an initial condition that is ``dynamically consistent'' with the available data. The procedure is naturally generalized if there are reference values for other compartments.

\begin{figure}
    \centering
    \includegraphics[scale=0.5]{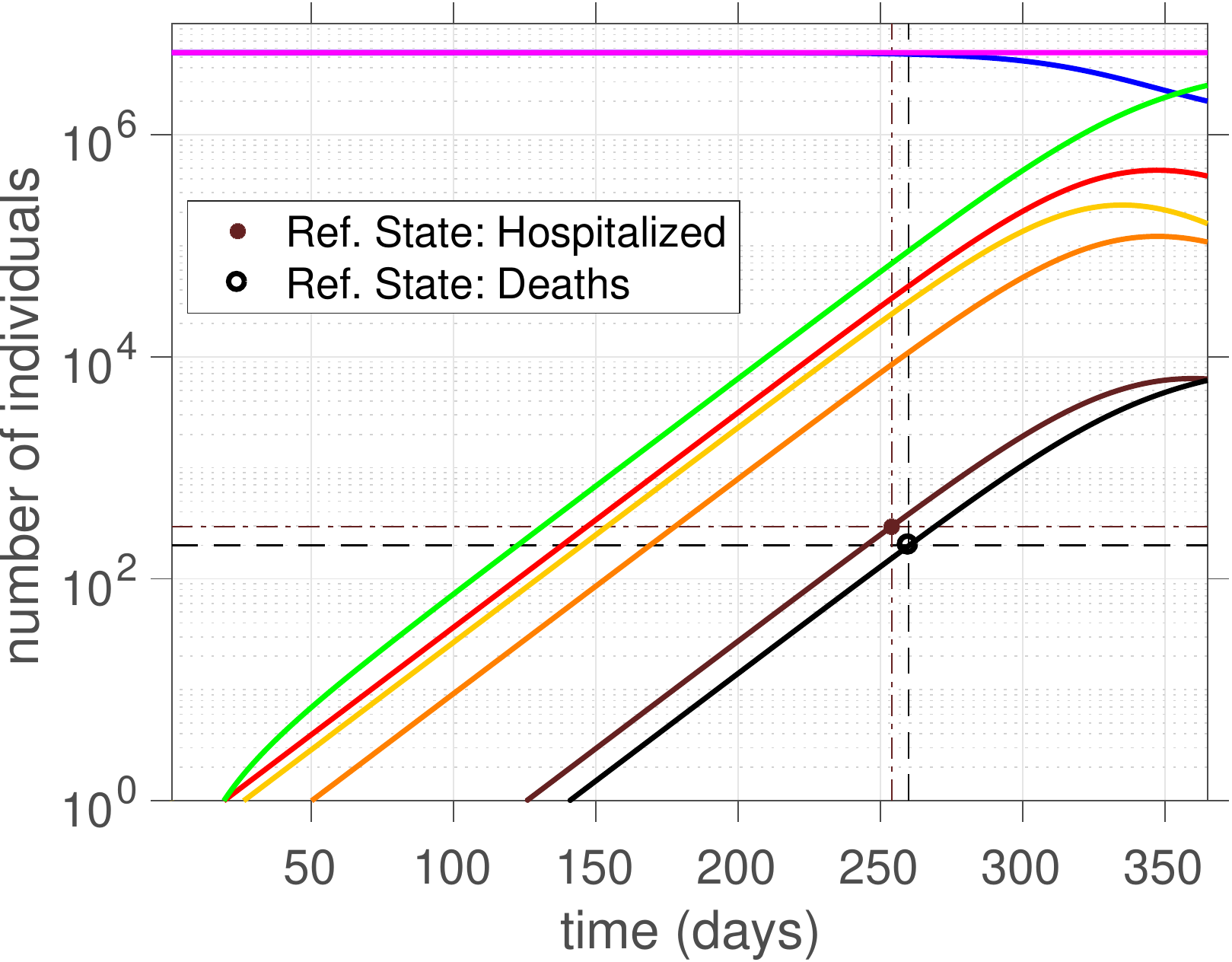}
    \caption{Dynamic response of the SEIR(+AHD) epidemic model in a scenario of a totally susceptible population, with a single individual exposed. Two states (for different times) are highlighted, corresponding to prescribed values of $H$ and $D$. The initial condition of subsequent simulations is defined by a convex combination of these two dynamic states. The colors of the curves follow the same scheme as in Figure~\ref{fig_ivp_all}.}   
    %Two distinct dynamic states of reference, the first obtained from hospitalization data and the second from death data. A convex combination of these two dynamic states defines the initial condition used in the simulation of the epidemic dynamics. The figure also shows the time series of the SEIR(+AHD) model, on a semi-logarithmic scale, in a scenario of a totally susceptible population with a single exposed individual.}
    \label{fig_IC_state}
\end{figure}

To illustrate of the methodology, the reader can observe Figure~\ref{fig_IC_state}, which shows two distinct dynamic states obtained from the data on hospitalizations and deaths together with the time series on a semi-logarithmic scale, corresponding to the response of a virgin population to the disease with a single exposed individual. The convex combination of these two dynamic states, considering the same weights used in Eq.(\ref{eq_misfit_2}), is used as an initial condition in the numerical experiments presented in the following sections.

\subsection{Calibration and validation of the SEIR(+AHD) epidemic model and its descriptive capacity}
\label{calib_valid_model}

In this section the proposed CE-ABC framework is used to calibrate and quantify the parametric uncertainties inherent to the SEIR(+AHD) dynamic model. To this end, the following hyperparameters are adopted in the CE-ABC algorithm: discrepancy function weight $\omega = 0.75$ for hospitalizations (and thus, $1 - \omega = 0.25$ for deaths); CE samples $N_{ce} = 100$; CE elite sample set size $N_{\mathcal{E}_{\ell}} = 10\%$ of $N_{ce}$; CE absolute tolerance $\texttt{atol} = 0.001$; CE relative tolerance $\texttt{rtol} = 0.05$; CE mean value smoothing parameter $a = 0.7$; CE variance dynamic smoothing parameters $b = 0.8$ and $q = 5$; CE maximum number of iterations $\texttt{maxiter} = 150$; ABC samples $N_{abc} = 2000$; ABC tolerance $\texttt{tol} = 0.1$. The bounds for the model parameters are adopted according to the values shown in Table~\ref{tab_nominal_parameters}.

The data considered here for the training of the dynamic model include the records of hospitalizations and total deaths in the city of Rio de Janeiro between May 1 and 31, 2020. The statistical validation process of the calibrated model uses the data corresponding to the following month, between June 1 and 30, 2020. Data for April 2020 are ignored because they are unreliable since the city's epidemiological surveillance system was still adapting to the new reality at the beginning of the pandemic.

\begin{figure}
    \centering
    \includegraphics[scale=0.45]{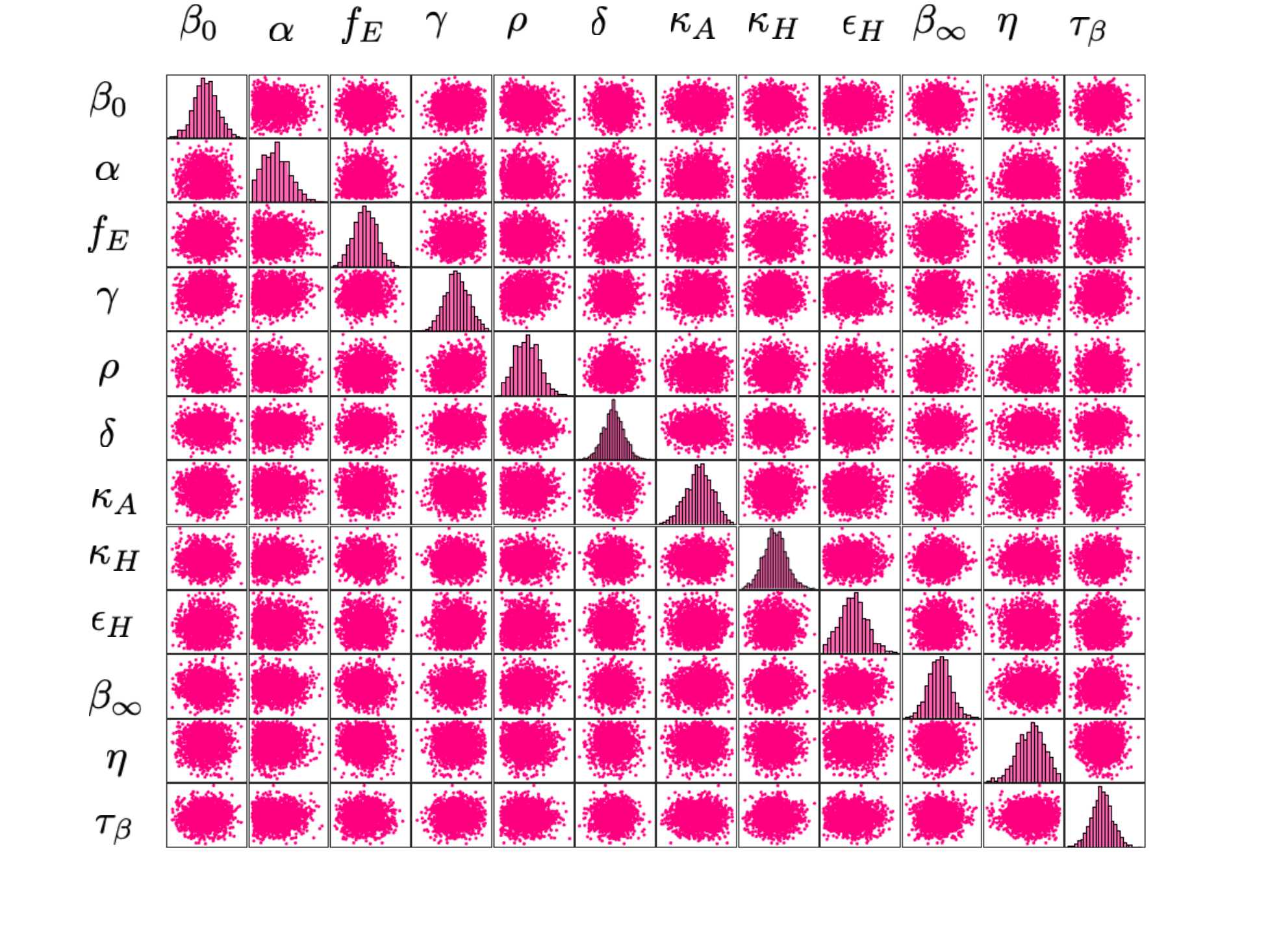}
    \caption{Histograms and scatter plots of the SEIR(+AHD) epidemic model, estimated with the samples accepted by the ABC simulations. The order of the parameters is the same as shown in Table~\ref{tab_CE-ABC_parameters}.}
    \label{fig_param_samples}
\end{figure}

\begin{figure*}
    \centering
    \includegraphics[scale=0.45]{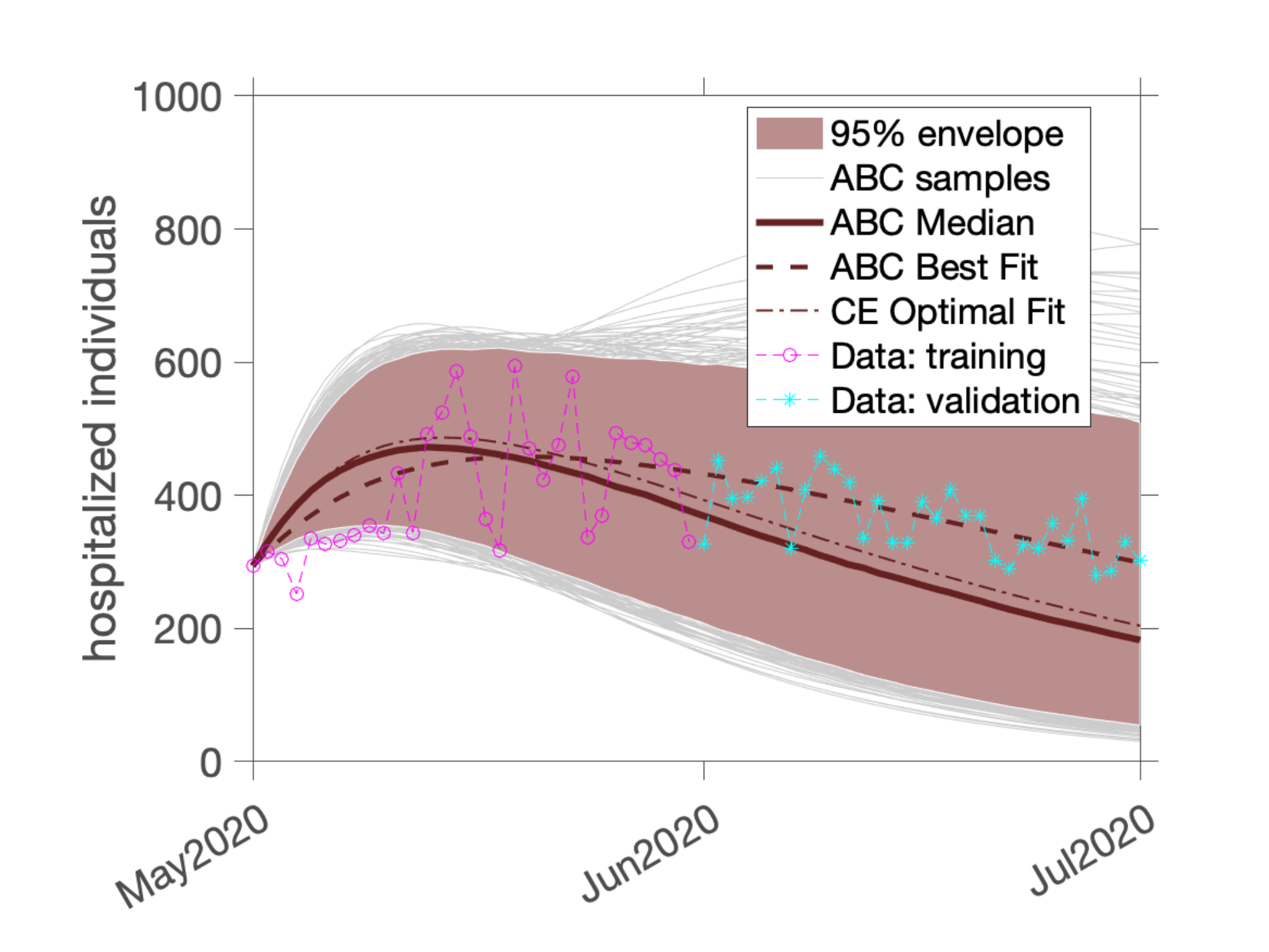}~
    \includegraphics[scale=0.45]{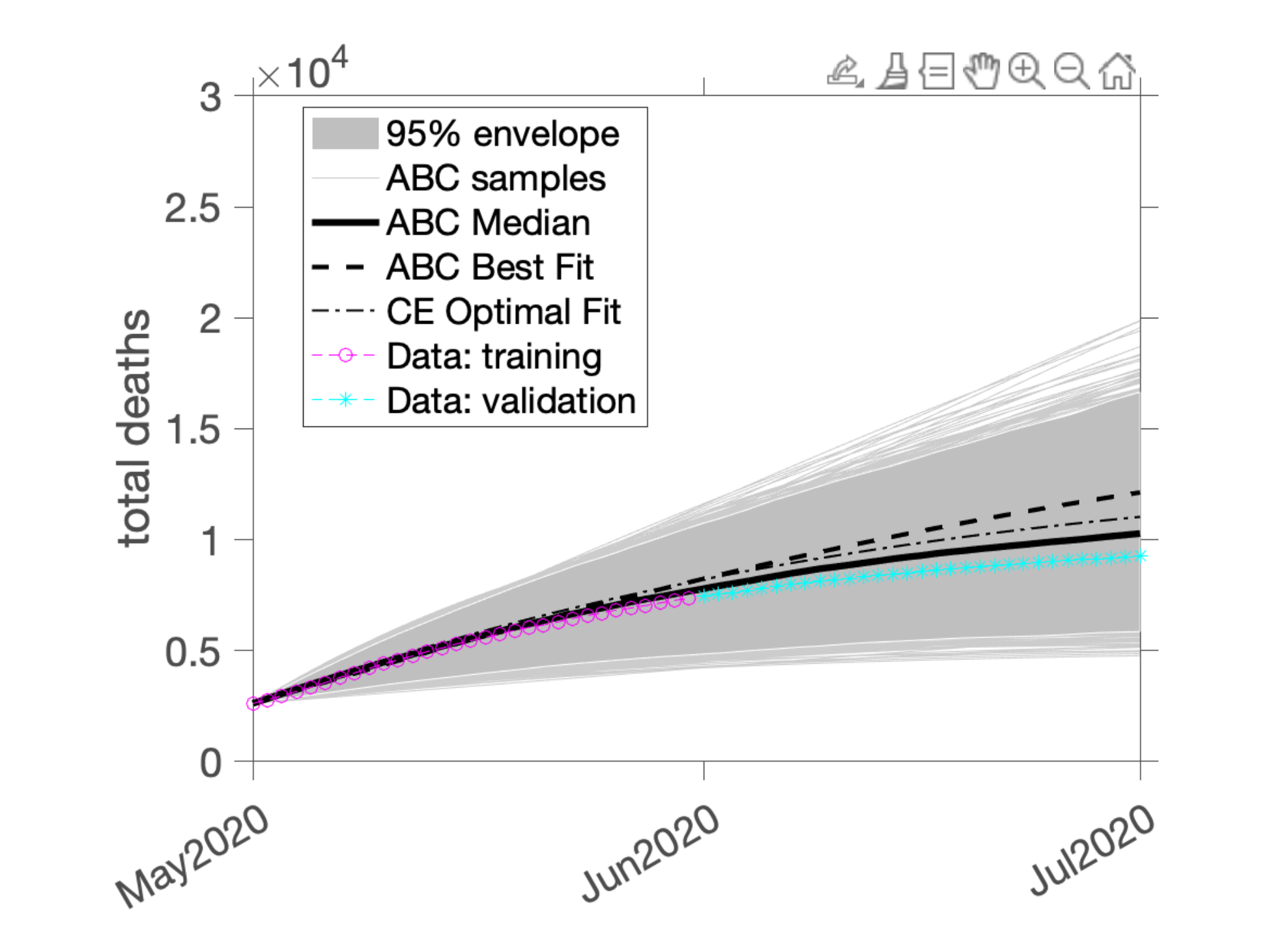}
    \caption{Time series generated by the CE-ABC algorithm for the number of hospitalized individuals (left) and total deaths (right) obtained with the SEIR(+AHD) model, which is calibrated with Rio de Janeiro epidemic data from May 2020 and validated for a temporal window covering the month of June 2020. Here the discrepancy function weight is $\omega = 0.75$, and ABC acceptance rate is 87\%.}
    \label{fig_HD_calib_valid}
\end{figure*}

The results regarding the calibration, quantification of uncertainties, and validation of the SEIR(+AHD) model with the aid of the CE-ABC algorithm can be seen in Table~\ref{tab_CE-ABC_parameters} and Figures~\ref{fig_param_samples} and \ref{fig_HD_calib_valid}.

Table~\ref{tab_CE-ABC_parameters} shows the values calculated by the CE-ABC algorithm for the parameters of the SEIR(+AHD) model, showing the estimates obtained by the CE optimizer in the third column; the respective standard deviation values in the fourth column; the best sample of the ABC simulation in the fifth column; and the standard deviation values of the posterior distributions obtained by ABC in the sixth column. The two sets of parameters identified present very close values, and the ABC result is a kind of refinement of the estimate obtained by the CE.

Regarding the posterior joint distribution of the model parameters, the reader finds this information in Figure~\ref{fig_param_samples}, which presents the histograms and scatter plots for each of the model parameters (estimated with the samples accepted by the ABC simulation). Scatter plots give information about the correlation between the parameters. In this figure, the order of the parameters is the same as shown in Table~\ref{tab_CE-ABC_parameters}.

\begin{table}
\caption{Parameters identified for SEIR(+AHD) epidemic model via CE-ABC framework, and the respective standard deviation values.}
\centering
\begin{tabular}{llcccc}
\toprule
 &      & CE  & CE      & ABC  & ABC\\
 & Unit & Optimal & std dev & Best & std dev\\
\midrule
$\beta_0$ & 1/day  & 0.12   & 0.02 & 0.13     & 0.02\\
$\alpha$             & 1/day  & 0.20   & 0.07 & 0.27     & 0.06\\
$f_E$                & ---    & 0.81   & 0.03 & 0.84     & 0.03\\
$\gamma$             & 1/day  & 0.13   & 0.01 & 0.12     & 0.01\\
$\rho$               & 1/day  & 0.0006 & 0.0001 & 0.0005 & 0.0001\\
$\delta$             & 1/day  & 0.0021 & 0.0004 & 0.0015 & 0.0004\\
$\kappa_A$           & ---    & 0.0026 & 0.0008 & 0.0027 & 0.0008\\
$\kappa_H$           & ---    & 0.0563 & 0.0130 & 0.0575 & 0.0128\\
$\epsilon_H$         & ---    & 0.25   & 0.07   & 0.33   & 0.07\\
$\beta_{\infty}$     & 1/day  & 0.31   & 0.06   & 0.43   & 0.06\\
$\eta$               & 1/day  & 5.8    & 1.9    & 6.2    & 1.8\\
$\tau_{\beta}$       & day    & 146    & 7      & 153    & 7\\
\bottomrule
\end{tabular}
\label{tab_CE-ABC_parameters}
\end{table}

In Figure~\ref{fig_HD_calib_valid} the reader can see the time series of hospitalizations (left) and total deaths (right) for the city of Rio de Janeiro in a time window that covers the months of May and June 2020. Time trajectories, accepted by the ABC simulations (87\% of the 2000 total)\footnote{This high acceptance rate, which may seem very high at first glance, is due to the informative prior obtained by CE.}, are displayed as thin solid lines in light gray; the trajectory that corresponds to the optimal set of parameters obtained by the CE optimizer is displayed as a dash-dotted line; the best sample trajectory obtained by the ABC simulation (the one with the smallest error) is indicated as a dashed line; while the median calculated with the samples accepted in the ABC simulation is indicated as a thick solid line. In addition, a 95\% credibility envelope is displayed in the form of a filled region above the ABC samples. Training data are displayed as magenta circles, while validation data are shown as cyan asterisks.

The comparison between the CE-ABC time series and the training data shows that, in this scenario, the dynamic model can reproduce well the epidemic outbreak experienced by the city of Rio de Janeiro in May 2020. For both time series, the curve corresponding to the optimal set of parameters identified by the CE optimizer, the best scenario simulated by ABC, and the median calculated with the cases accepted in the ABC simulation provide good descriptions of the epidemic data trend. Furthermore, the training data fit is robust once it includes the 95\% credibility interval obtained by ABC accepted samples, covering most of data fluctuations. 

In terms of validation, by comparing the predictions (extrapolations) made by the dynamic model and future data (not used in the calibration), one can note that the dynamic model captures the trend of the outbreak. It takes over the data due to a 95\% credibility band around the calculated evolution curves. Strictly speaking, the forecasts are reasonably accurate for the first seven days of extrapolation, starting to shift from the simulated curves from this point onwards. In what follows, hospitalizations are slightly underestimated by the ABC median by approximately 25\%, while the median overestimates total deaths up to a limit of around 10\%.

In light of the minimal horizon of predictability that epidemic systems present\footnote{In an epidemic where people are aware of what is happening, there is a feedback between the rate of infection and people's social behavior. Being aware of the severity of the outbreak beforehand can help reduce its intensity or vice versa. The great difficulty in modeling such feedback is one of the factors (perhaps the main one) that limits the predictability horizon of epidemic models.}, these predictions can be considered very good, as they provide accurate values in the short term (one week) and bring some reasonable information in the medium term (one month). Although 10-25\% uncertainty in forecasting the number of hospital beds/expected deaths is not highly accurate for an immediate sizing of hospitals or funeral units, it is still informative in indicating to decision makers the correct order of magnitude for these outcomes. For instance, knowing a month in advance, in the course of an epidemic, that a few hundred (not thousands, or vice-versa) of hospital beds/burials will be required per day can be crucial information to prevent a hospital or cemetery from collapsing. Added to this is that the model can be recalibrated weekly (or daily), updating the short/medium forecasts whenever new data become available.

\begin{figure}
    \centering
    \includegraphics[scale=0.45]{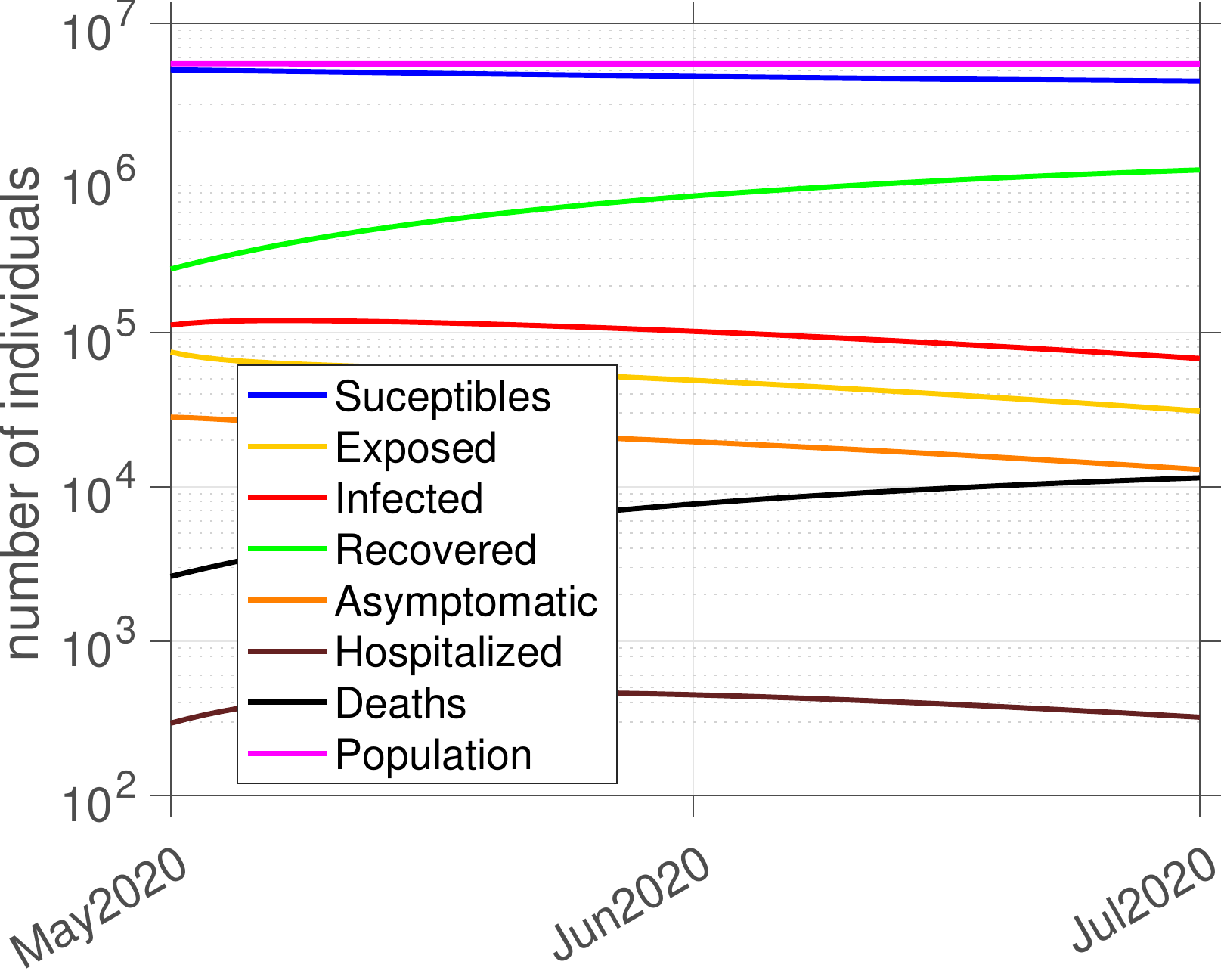}
    \caption{Dynamic response of the SEIR(+AHD) epidemic model, in a time window that includes the months of May and June 2020, considering the best estimate of the ABC simulation for the model parameters. Here the discrepancy function weight is $\omega = 0.75$.}
    \label{fig_SEIRA}
\end{figure}

It is also worth mentioning that, in addition to making predictions about the QoIs for which epidemic data are available, a well-calibrated mechanistic model can provide information on latent quantities (for which data are not available), such as the number of susceptible, exposed, asymptomatic, etc. In this sense, to illustrate this possibility, the reader is invited to observe Figure~\ref{fig_SEIRA}, which presents the evolution of the time series associated with eight dynamics state coordinates of the epidemic model in a time window that includes the months of May and June 2020, considering the best estimate of the ABC simulation for the model parameters.

From a qualitative point of view, this simulation allows the analyst to infer that, in this two-month interval, there is a slight but notable decrease in the number of susceptible people in the general population due to the increase in COVID-19 infections, followed by an increase in total recoveries. The number of symptomatic infected is always more significant than the number of exposed, greater than the asymptomatic infected. Quantitatively, it can be seen that these last three groups have sizes of the same order of magnitude, which is hundreds of times greater than the number of hospitalized patients. At the end of this two-month interval, the total number of accumulated deaths reaches a value comparable to the active infected.

Of course, the accuracy of such information largely depends on the extent to which the structure of the epidemic model provides a reliable representation of epidemic dynamics. If it is a good representation, the simulations should offer great insight; if it is a feeble representation, the simulations do not tell anything useful at the limit. In intermediate cases, where the model is more or less accurate, useful information can be obtained, but not all the information from the simulations is reliable. Separating what is helpful from what is not requires deep knowledge of some basic principles of epidemiology.

\subsection{Influence of CE-ABC hyperparameters on the description of the epidemic dynamical system}

The CE-ABC framework combines two advanced stochastic simulation techniques, thus inheriting all the control parameters (a.k.a. hyperparameters) underlying the two methods. Consequently, the model calibration process and the propagation of parametric uncertainties depend, nonlinearly, on these hyperparameters. Thus, a study of how such quantities affect the modeling is desirable and recommendable. The present section of the manuscript seeks to shed light on this.

\begin{figure*}
    \centering
    \includegraphics[scale=0.3]{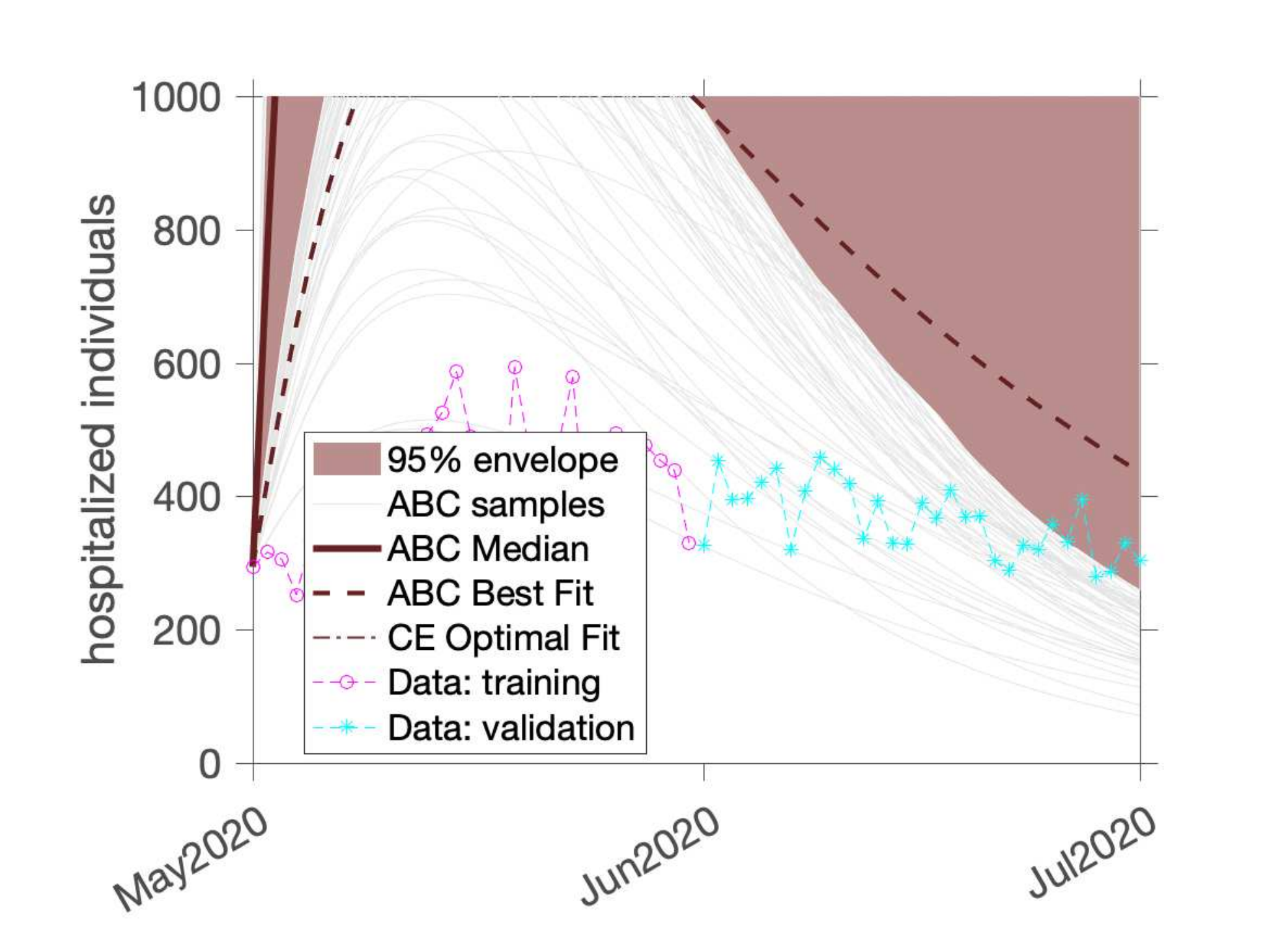}
    \includegraphics[scale=0.3]{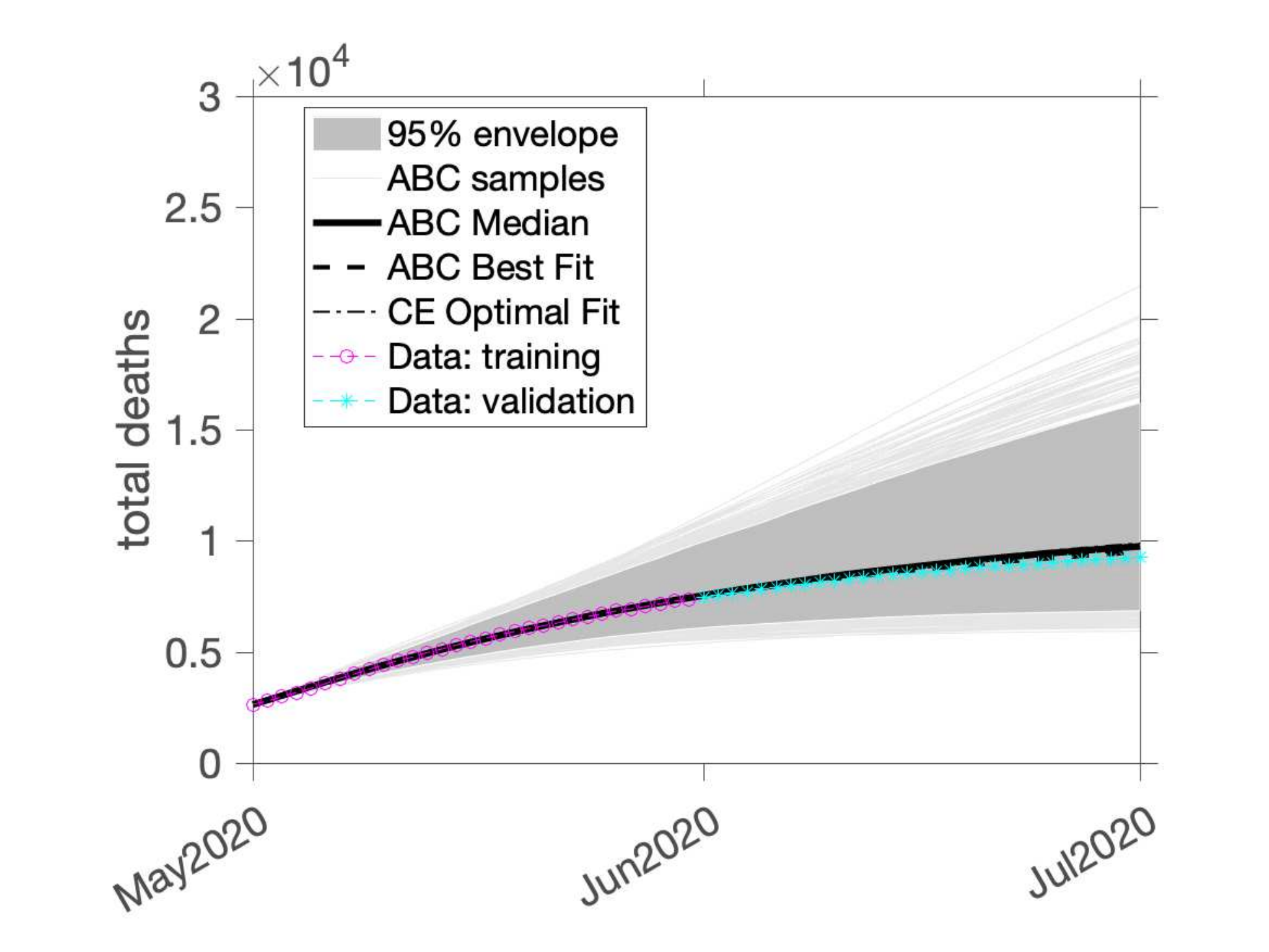}\\
    \includegraphics[scale=0.3]{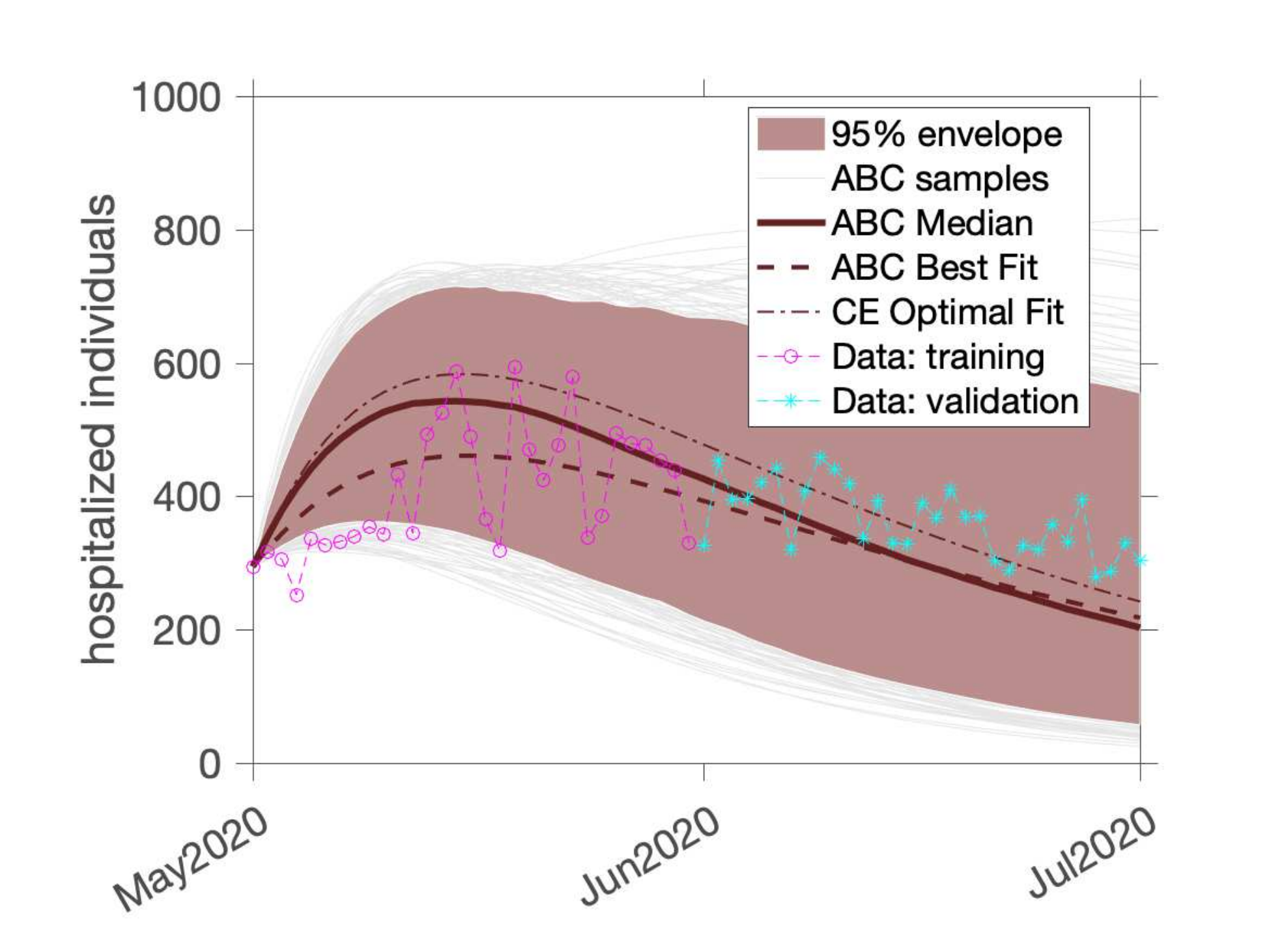}
    \includegraphics[scale=0.3]{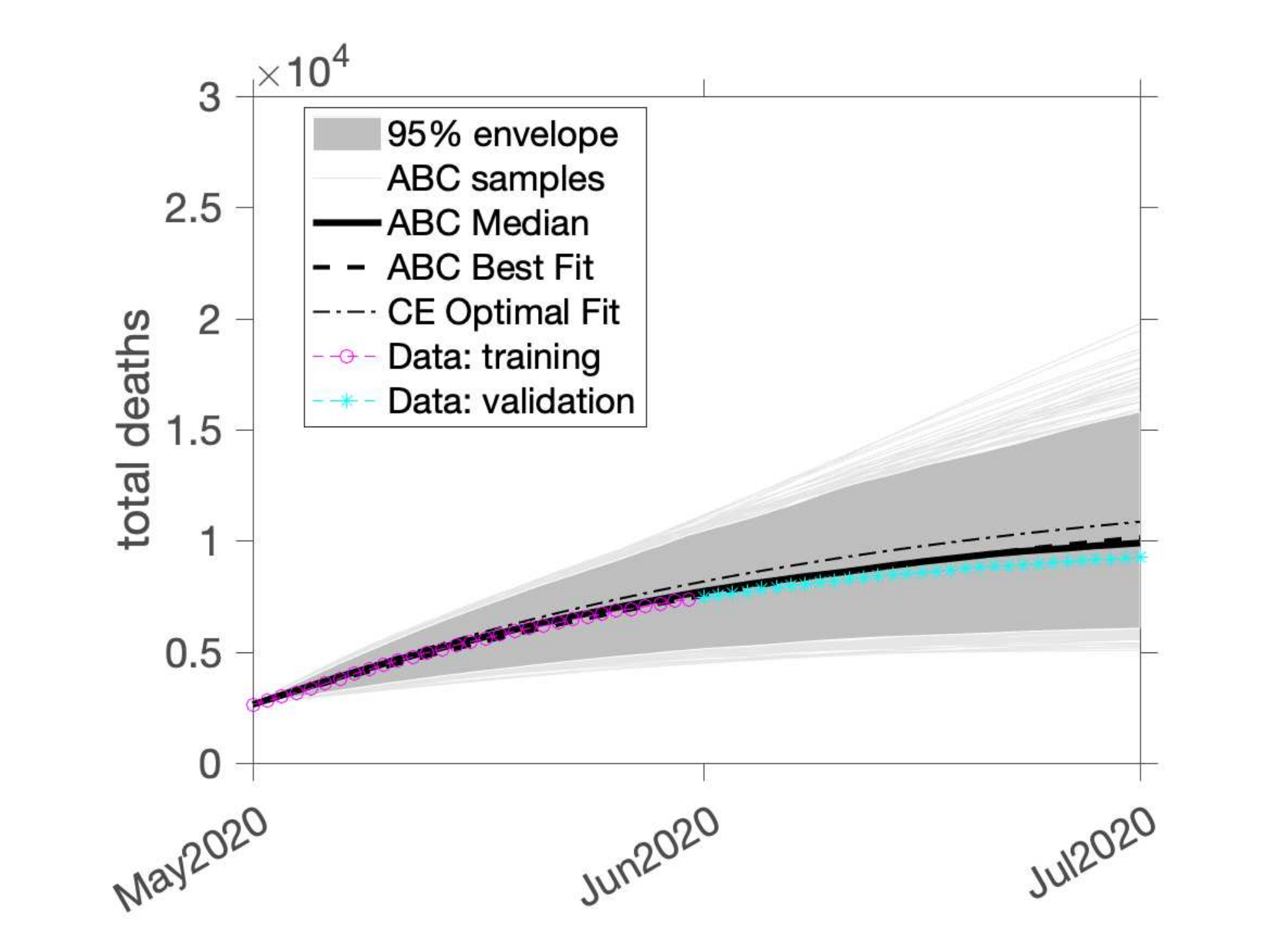}\\
    \includegraphics[scale=0.3]{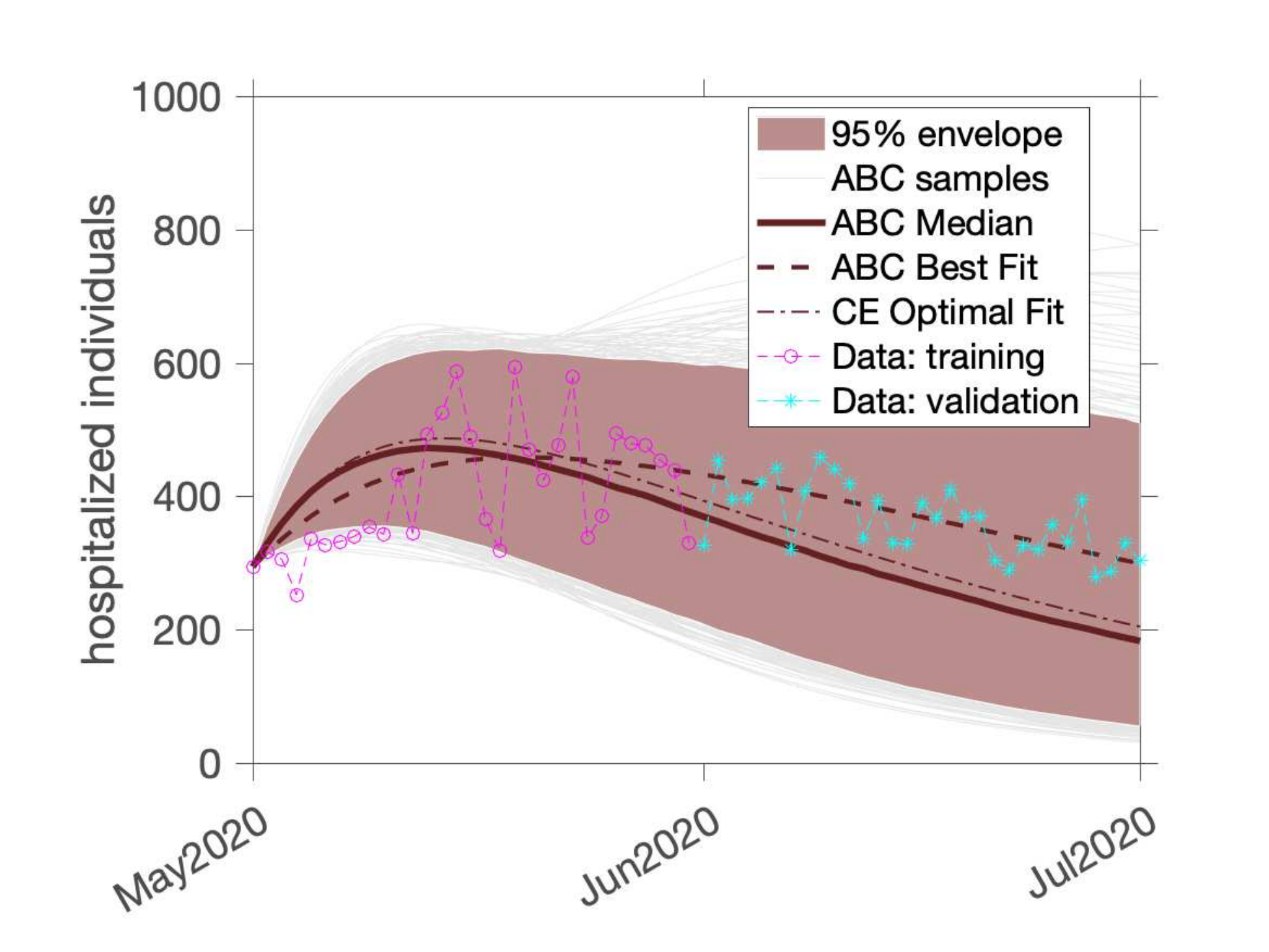}
    \includegraphics[scale=0.3]{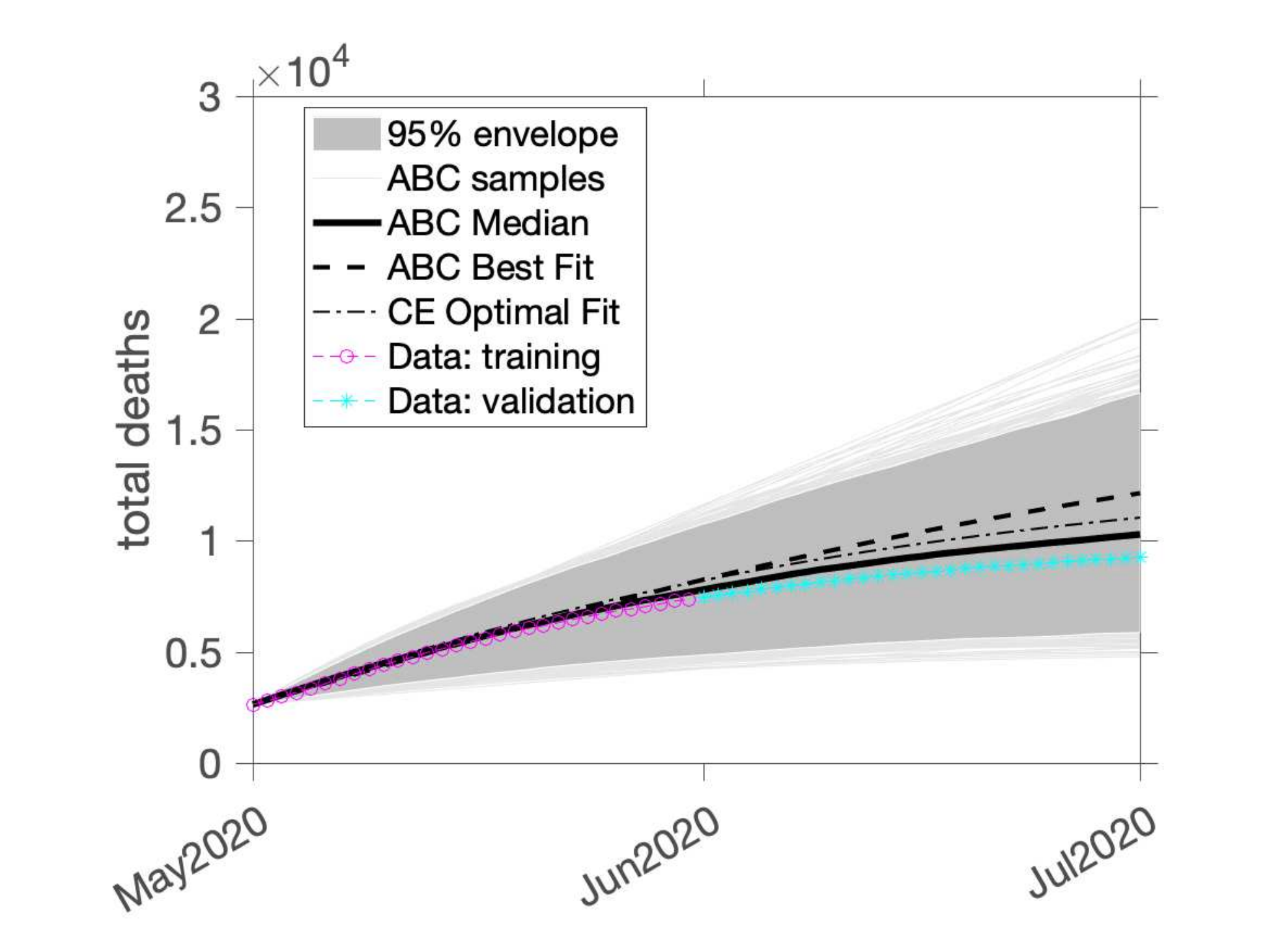}\\
    \includegraphics[scale=0.3]{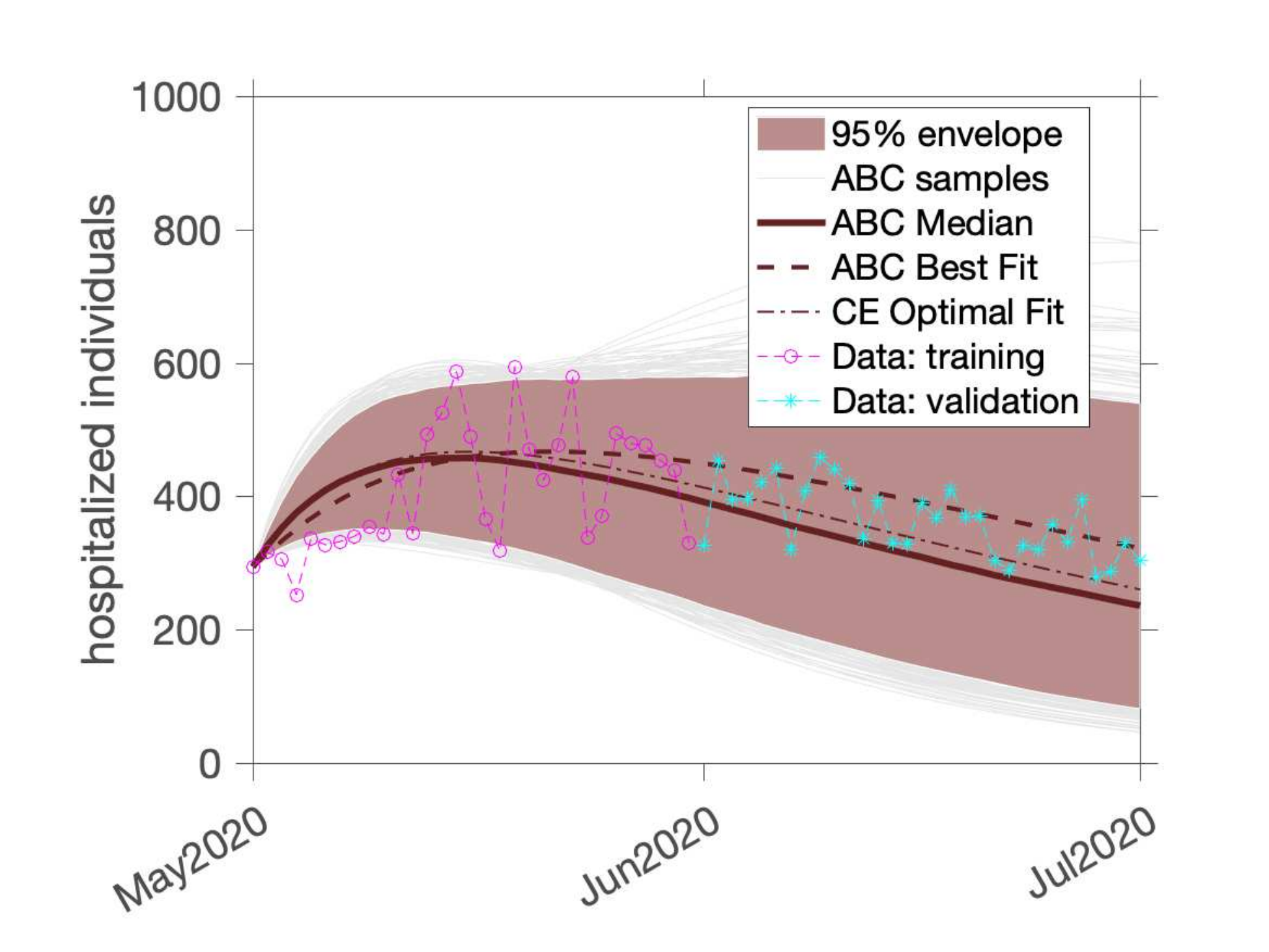}
    \includegraphics[scale=0.3]{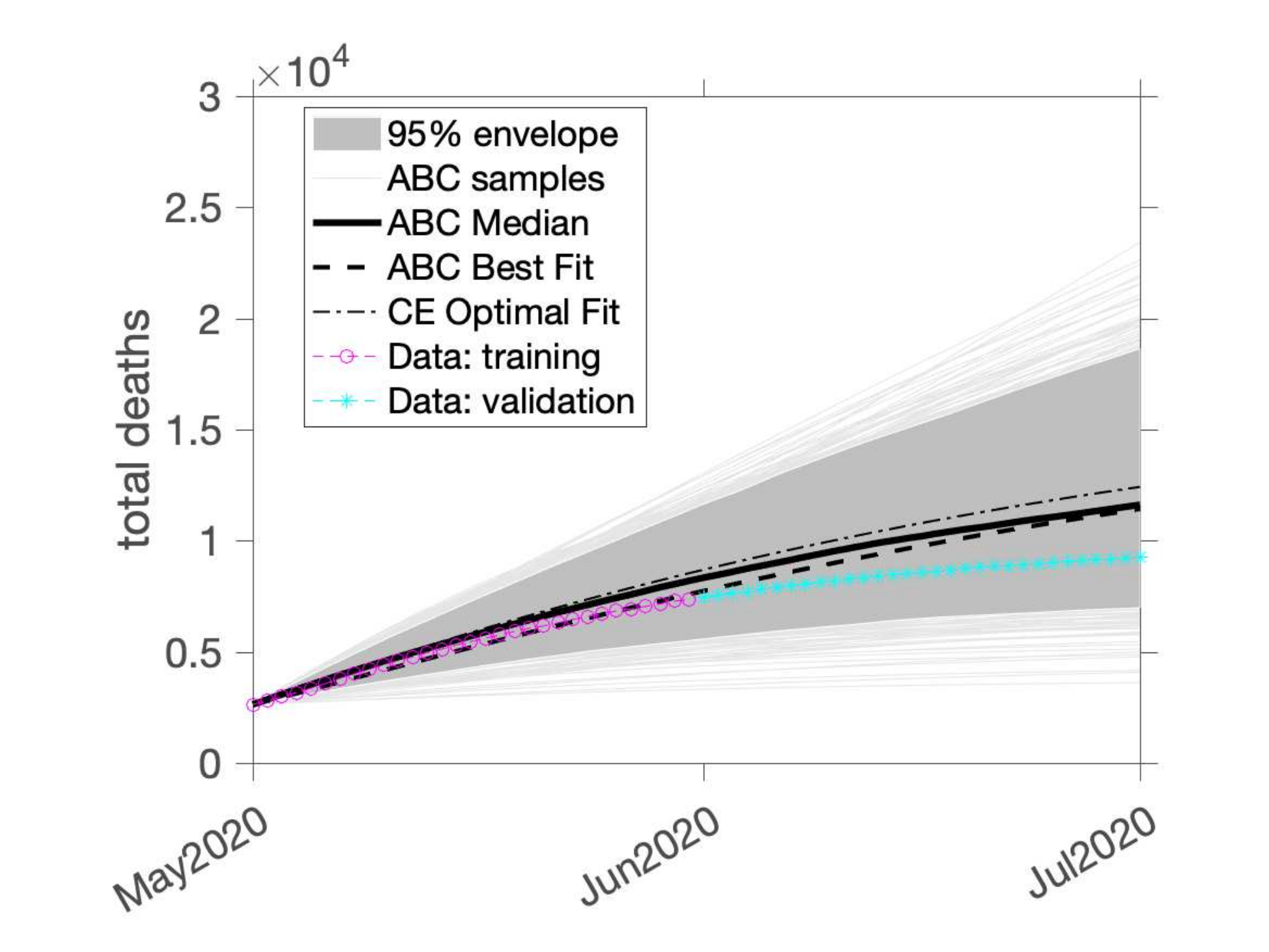}\\
    \includegraphics[scale=0.3]{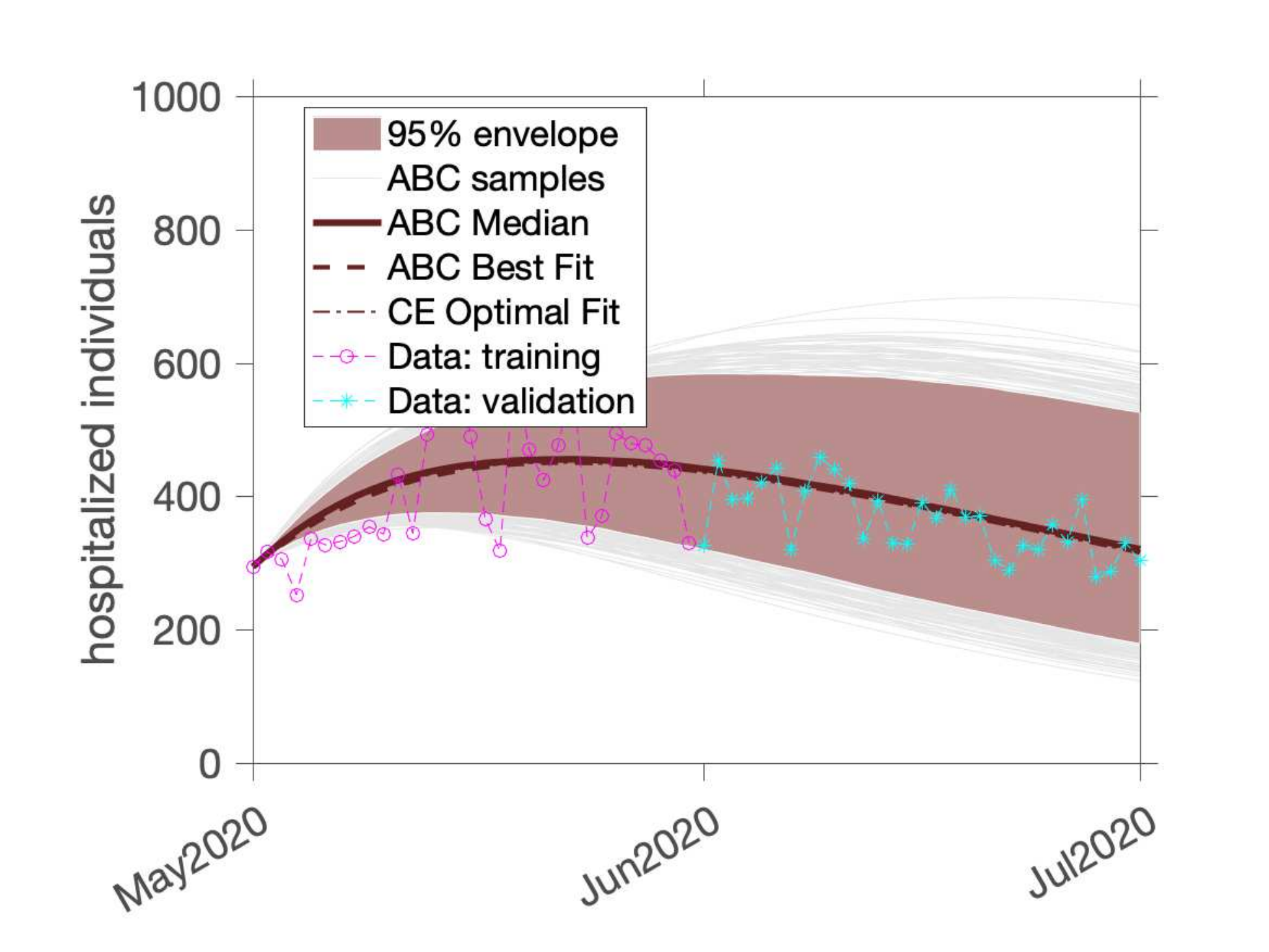}
    \includegraphics[scale=0.3]{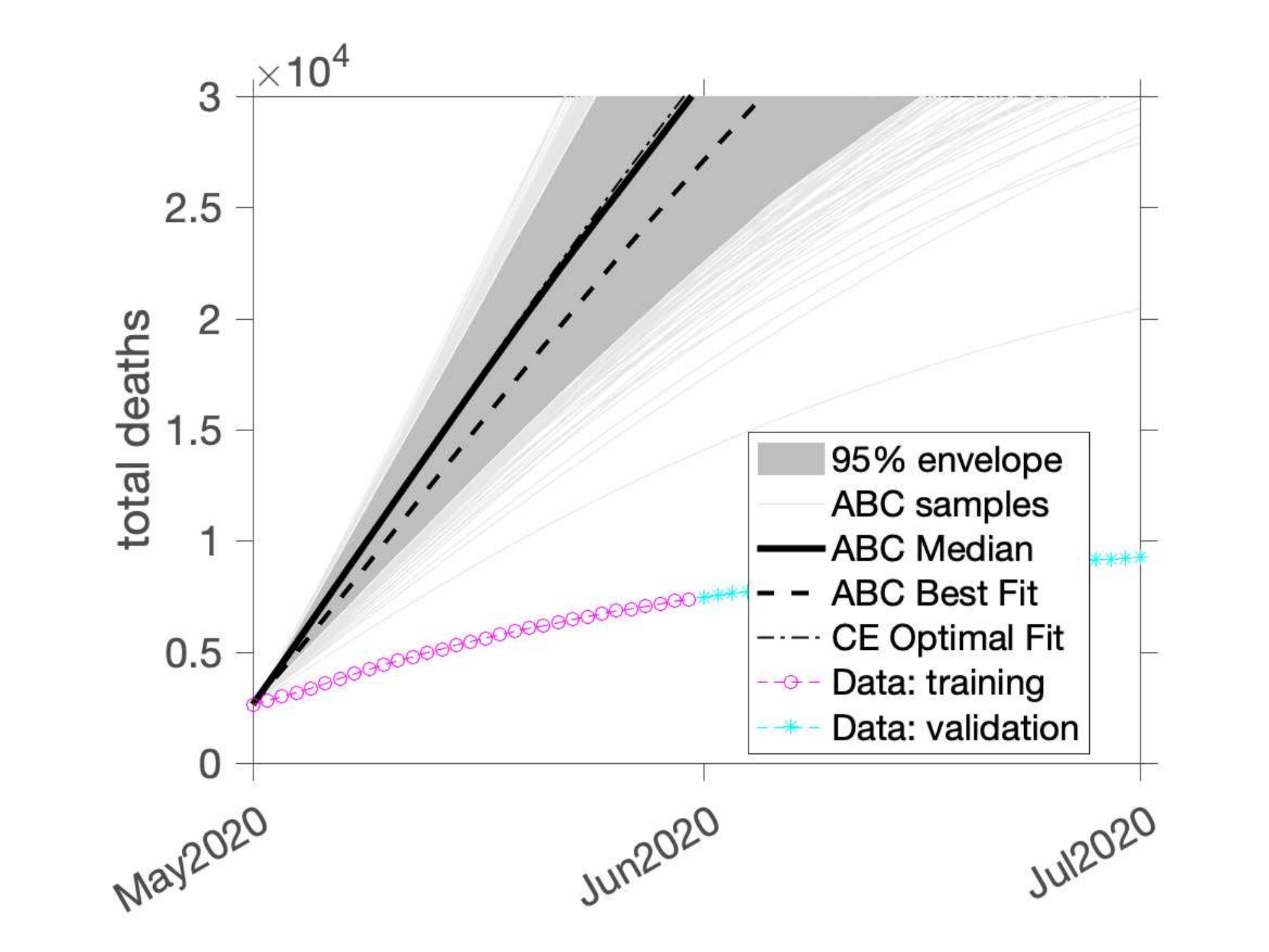}
    \caption{Quantities of interest calculated by the SEIR(+AHD) epidemic model with different values of the weight parameter: $\omega=0$ (first line); $\omega=0.25$ (second line); $\omega=0.5$ (third line); $\omega=0.75$ (fourth line); $\omega=1$ (fifth line).}
    \label{fig_HD_w}
\end{figure*}

Initially, we investigated the effect of the discrepancy function weight parameter $\omega$ on the results. For that, Figure~\ref{fig_HD_w} presents the two QoIs calculated by the SEIR(+AHD) epidemic model, considering five different values for the weight: $\omega \in \{0, \, 0.25, \, 0.5, \, 0.75, 1 \}.$

By visual inspection, it is possible to see that the best fits are obtained when $\omega = 0.25$, $\omega = 0.5$ or $\omega = 0.75$. The first and the last case favor one QoI, but without totally disregarding the effect of the other, while the intermediate case balances both. In this setting, deciding which QoI should be given more weight is a matter of convenience. If the information on hospitalizations is more important, higher $\omega$ values should be considered. The opposite is true if the primary interest is to follow the evolution of deaths.

%Surprisingly, the case $\omega = 0.5$, which seeks to balance both QoIs, does not have the best fit for either data. This fact may indicate an asymmetry in the quality of the two data series or a potential model deficiency in representing the dynamics of interest.

The $\omega = 0$ scenario considers a limiting case, where the discrepancy function defined by Eq.(\ref{eq_misfit_2}) considers only the total number of deaths. In contrast, considering only hospitalizations, the other extreme situation is counted when $\omega = 1$. These two limit cases have a terrible fit in the disregarded QoI, and should only be considered in situations in which only one of the QoI is of reliable.

Once the value of $\omega$ is fixed, the results are also affected by the choices of $\bm{x}_{min}$, $\bm{x}_{max}$, $N_{\text{ce}}$, $N_{\mathcal{E}_{\ell}}$, $N_{\text{abc}}$, $\texttt{atol}$, $\texttt{rtol}$,  $\texttt{tol}$, and $\texttt{maxiter}$.

\begin{figure*}
    \centering
    \includegraphics[scale=0.45]{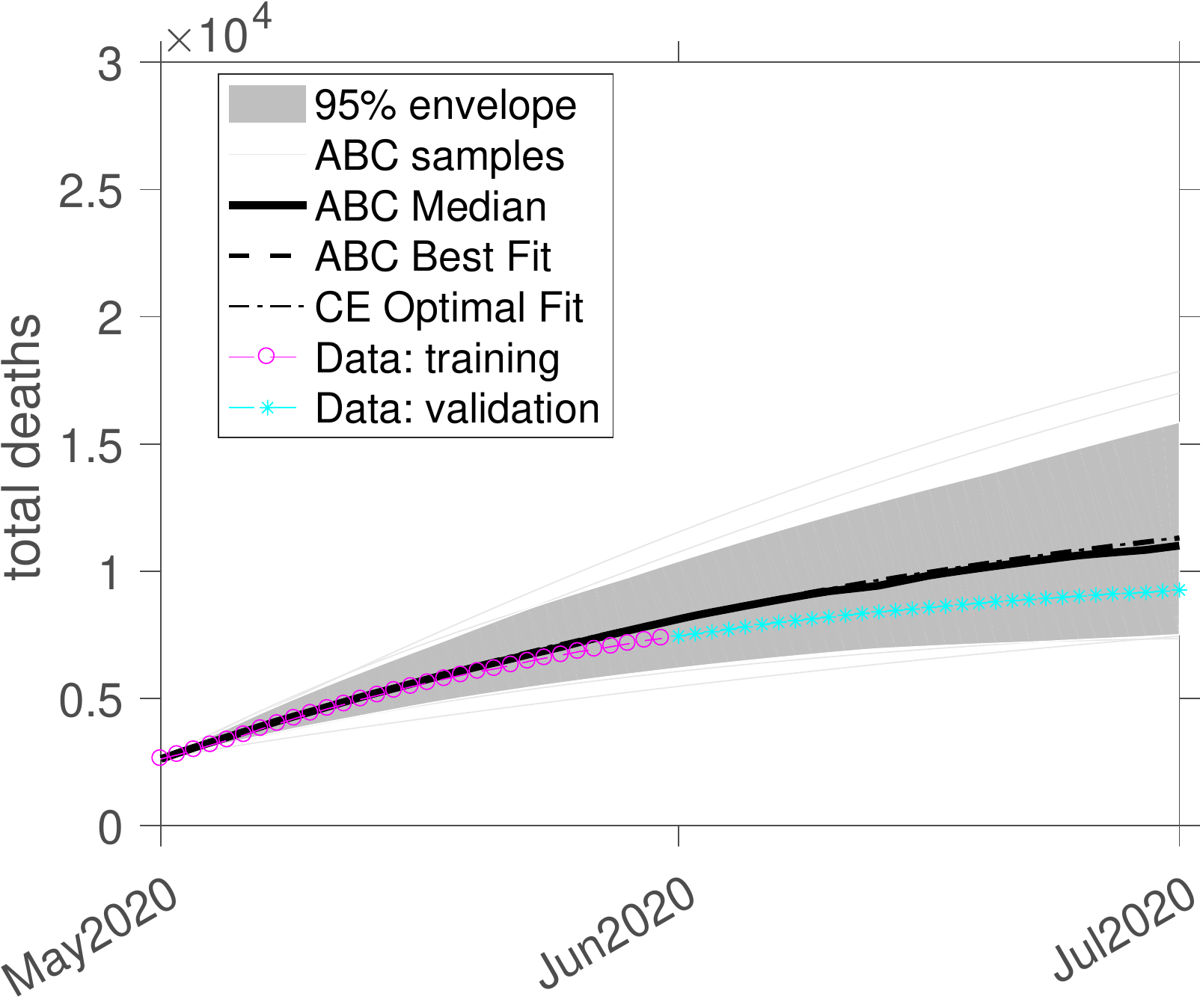}~~~~~~
    \includegraphics[scale=0.45]{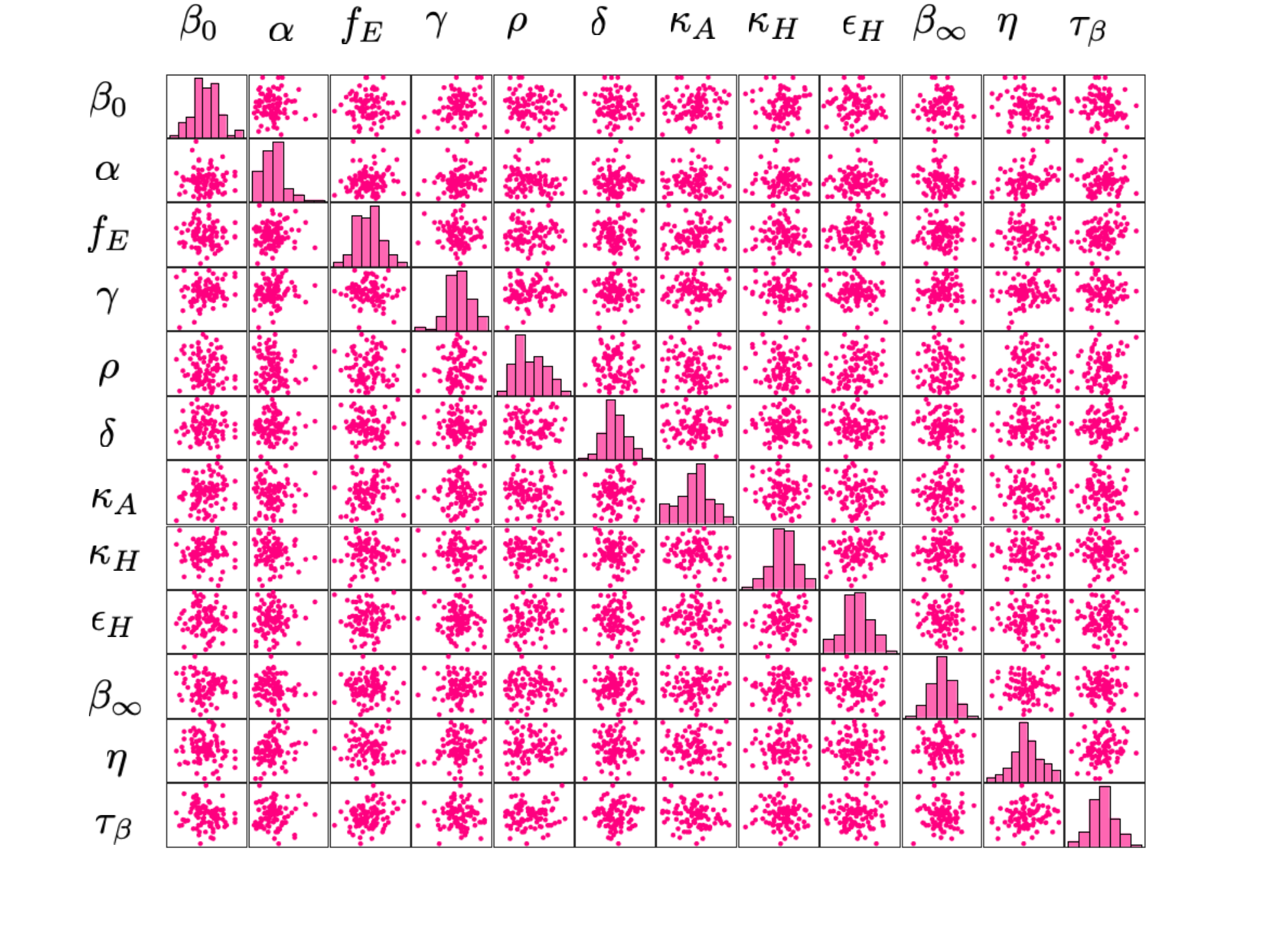}\\
    \includegraphics[scale=0.45]{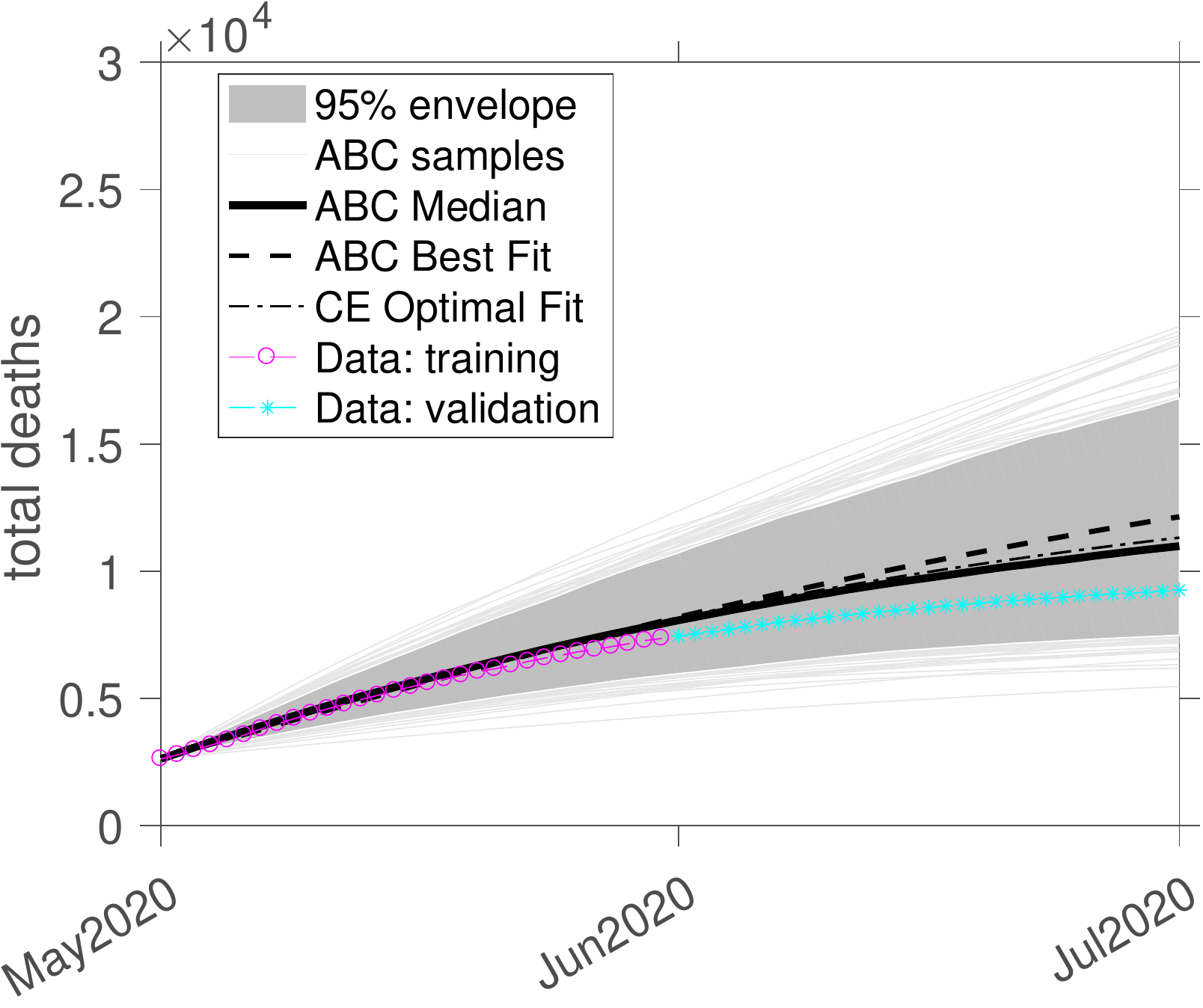}~~~~~~
    \includegraphics[scale=0.45]{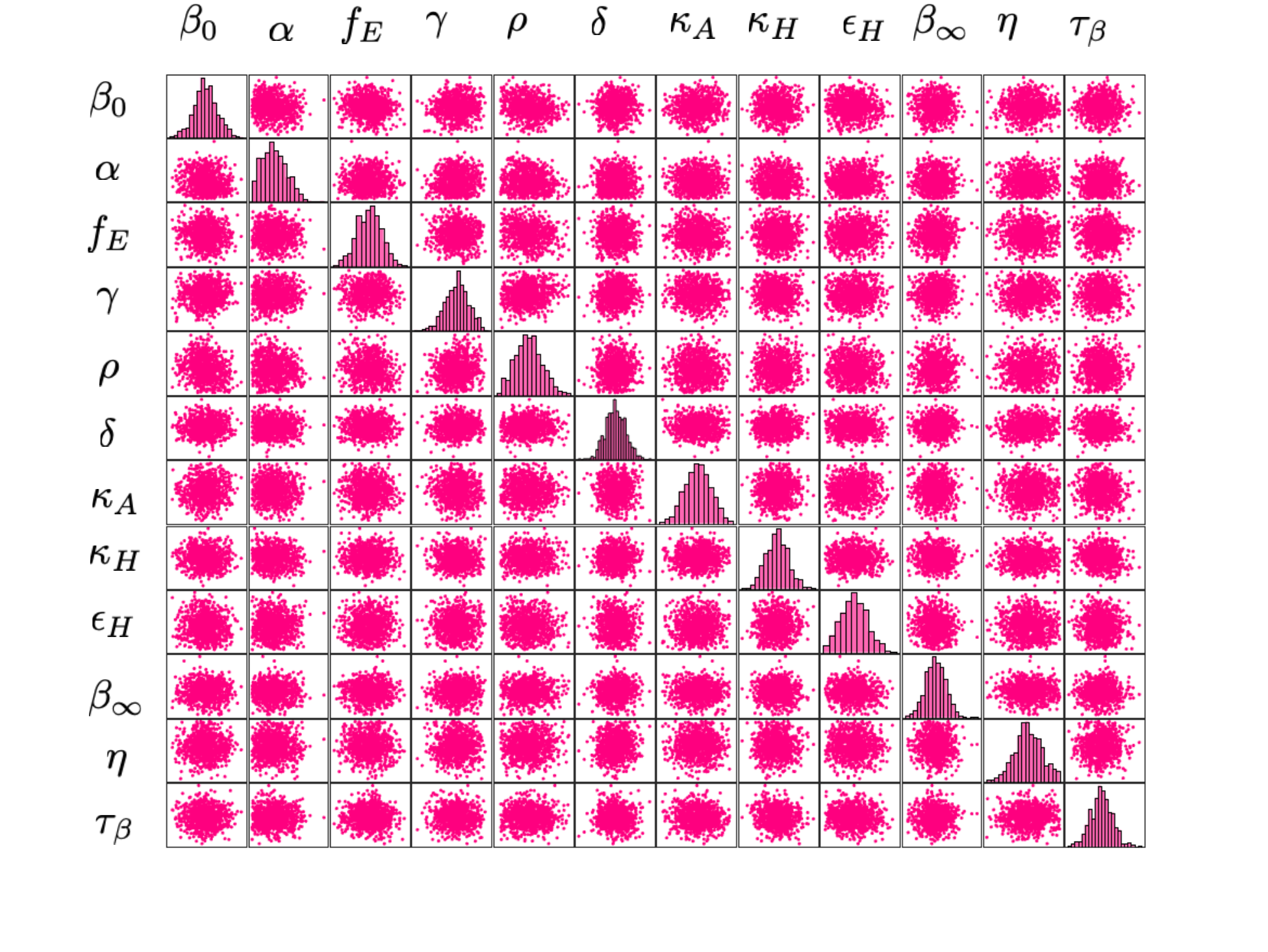}
    \caption{Effect of varying the number ABC simulation samples on the QoIs calculations and posterior inference. At the top $N_{abc} = 100$, while at the bottom $N_{abc} = 1000$.  Here the discrepancy function weight is $\omega = 0.75$ and $N_{ce} = 100$.}
    \label{fig_effect_N_abc}
\end{figure*}

The effect of varying the number ABC simulation samples $N_{abc}$ can be seen in Figure~\ref{fig_effect_N_abc}, where results are presented for $N_{abc} = 100$ (top) and $N_{abc} = 1000$ (bottom). Note that the number of accepted samples is directly proportional to the total number of simulated samples. This variation directly impacts the estimation of the posterior distribution, with a consequence in the obtained median and credibility band. Of course, these estimates are also affected by the tolerance $\texttt{tol}$, which improves or worsens the results as it is decreased or increased. As it is an obvious effect, numerical experiments in this sense are not shown. A tolerance of the order of 10\%, i.e., $\texttt{tol} = 0.1$, proved to be effective for the simulations in this paper.

The parameters $\texttt{atol}$, $\texttt{rtol}$, and $\texttt{maxiter}$ have an influence on how much faster the CE optimizer will stop, with consequent gain/loss of accuracy, followed by an increase/decrease in the computational cost. The reader is invited to do numerical experiments with these parameters to see their effect in practice. In the preliminary numerical studies conducted by the authors, we observed that there are no great gains in obtaining an estimate of the optimal parameters with great precision. Relative tolerance values of the order of 5\%, i.e., $\texttt{rtol} = 0.05$, provide a good compromise between accuracy and computational cost.

\begin{figure*}
    \centering
    \includegraphics[scale=0.45]{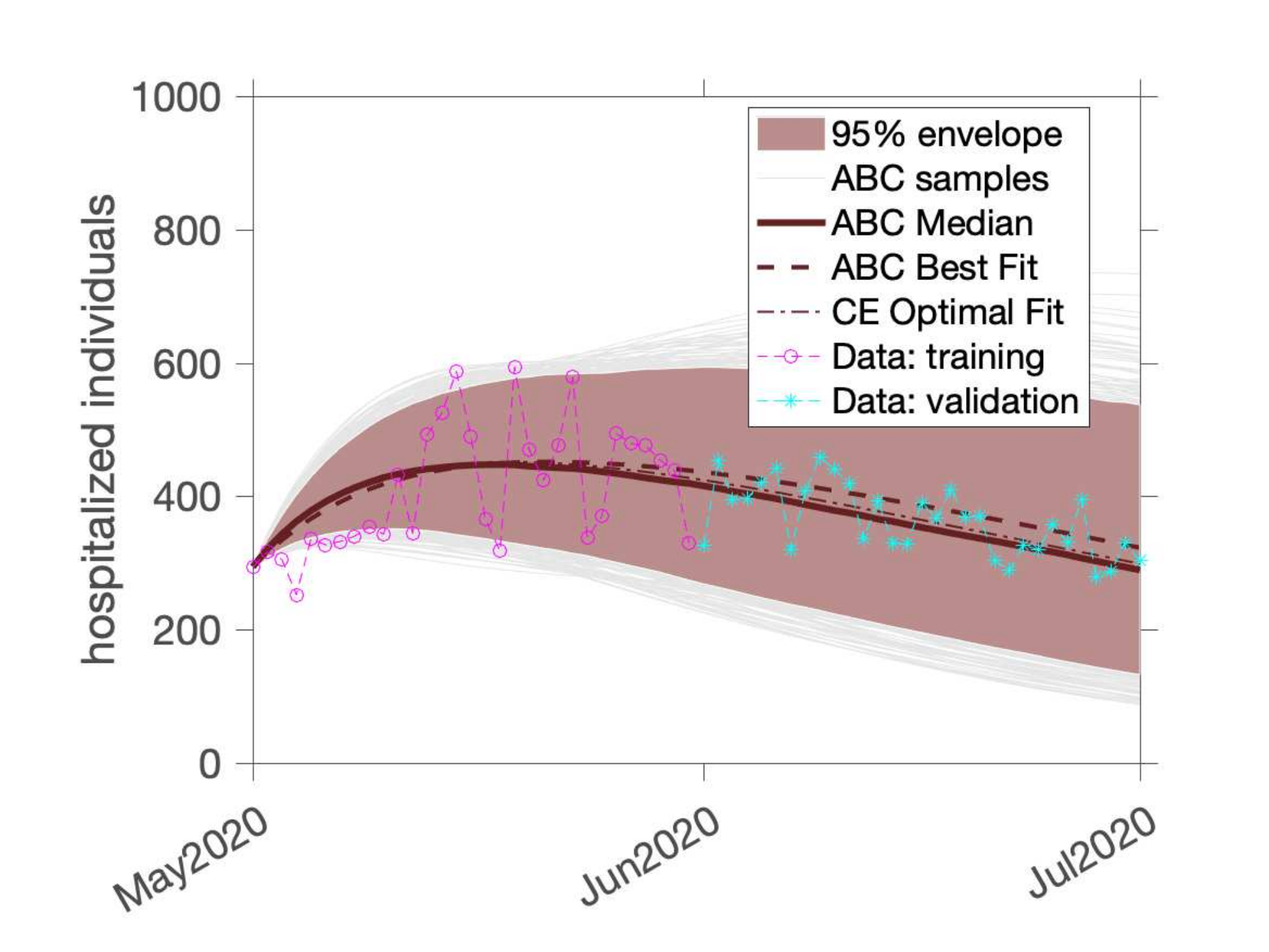}~
    \includegraphics[scale=0.45]{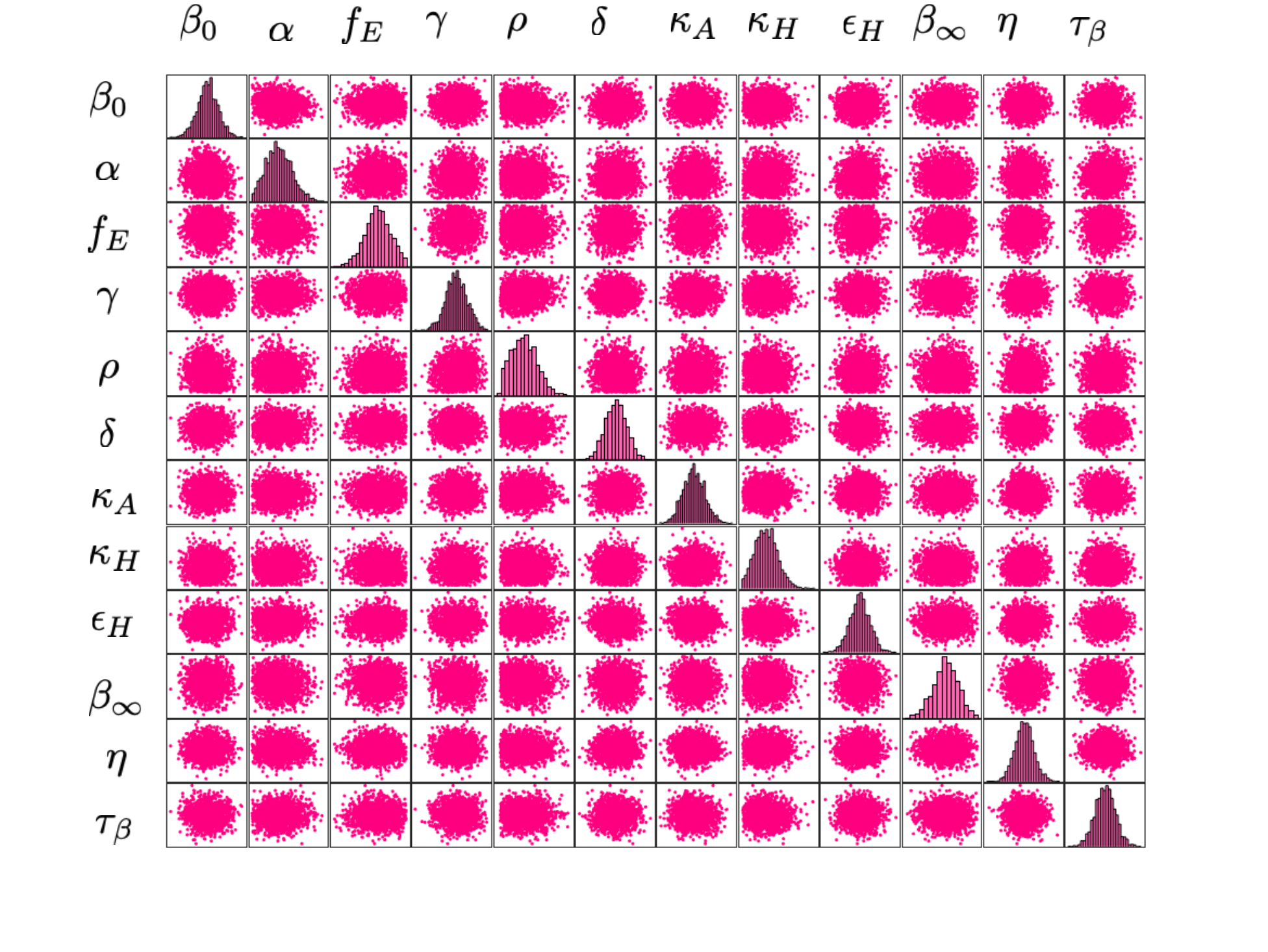}\\
    \includegraphics[scale=0.45]{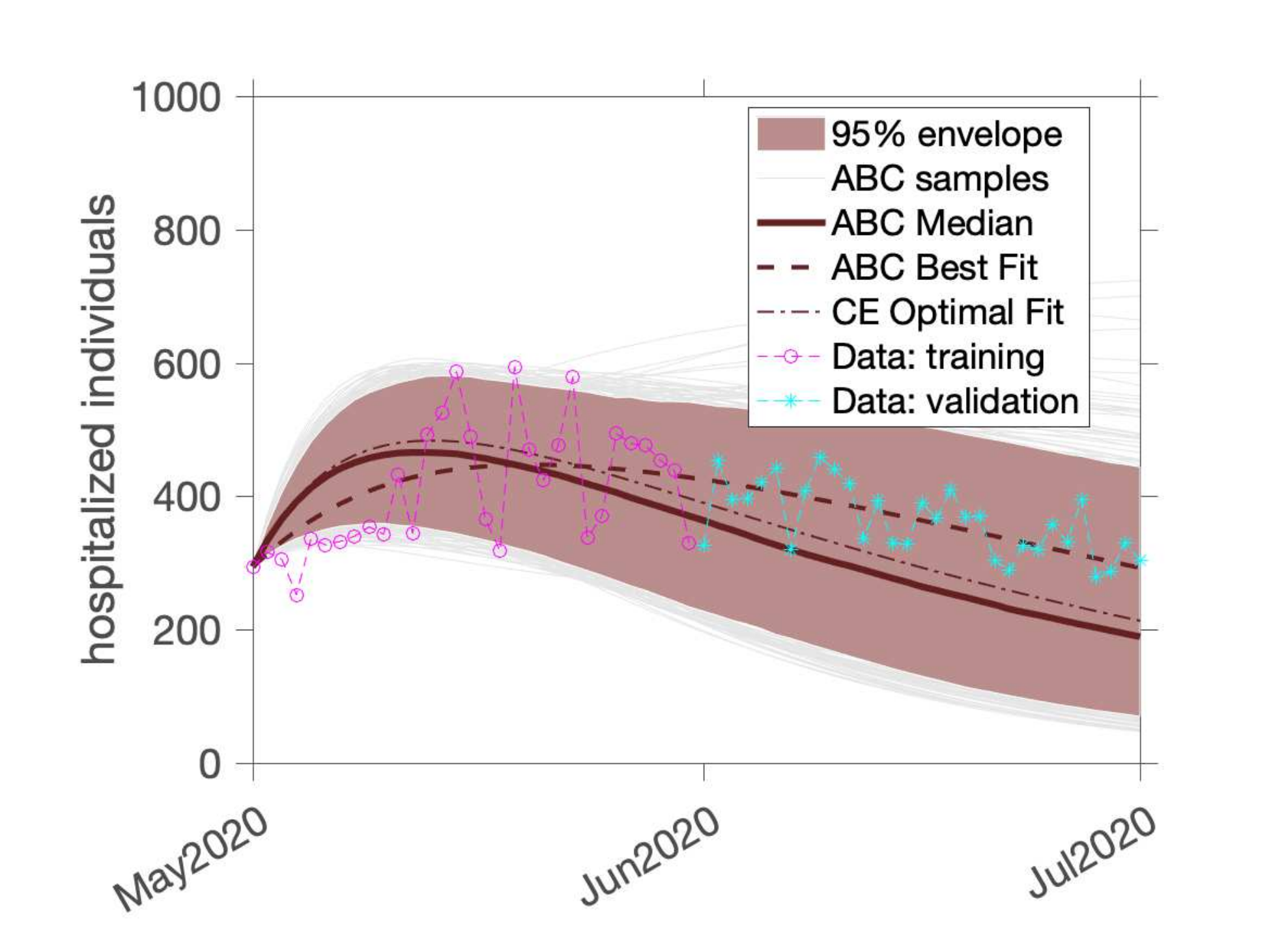}~
    \includegraphics[scale=0.45]{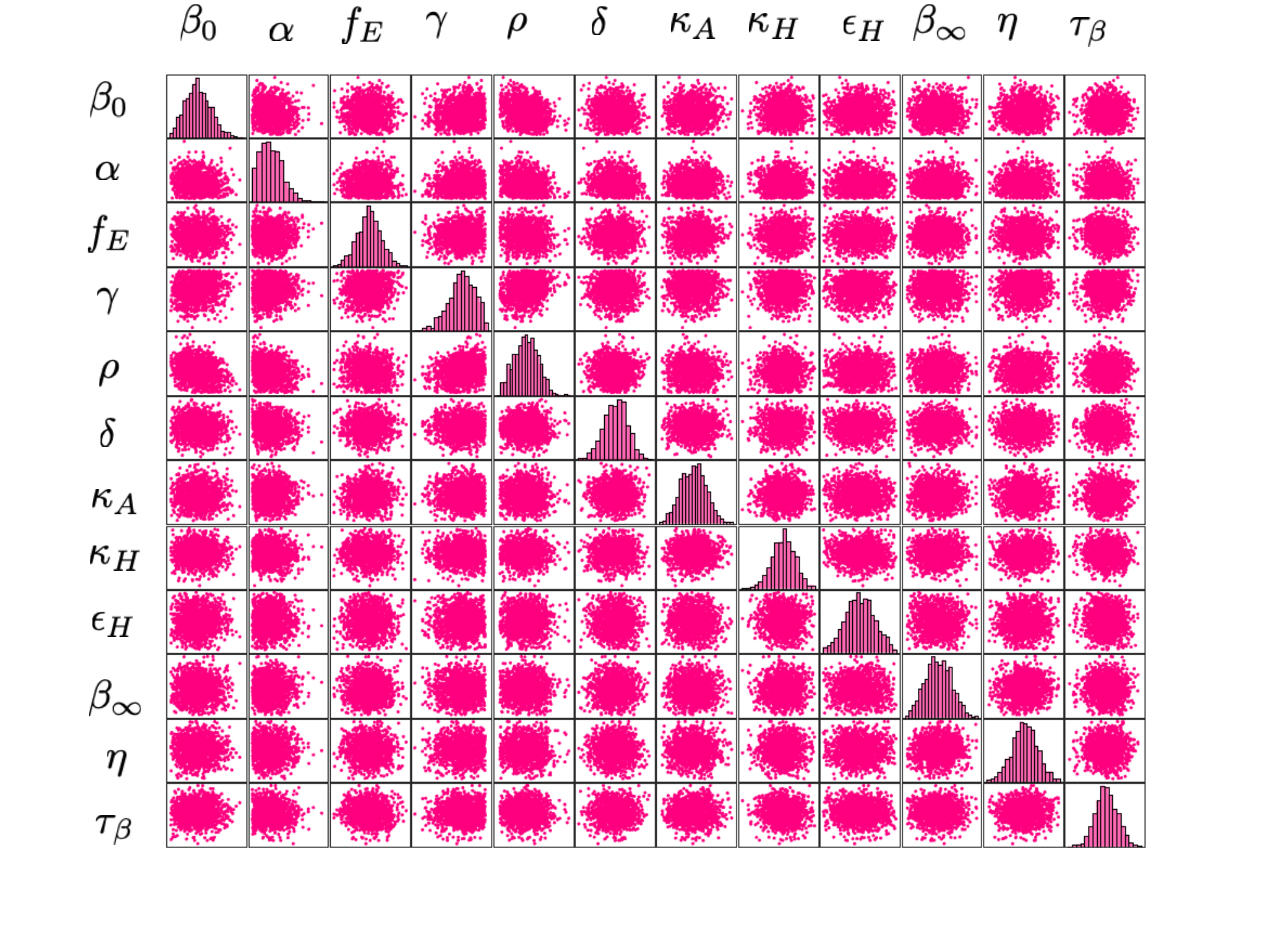}
    \caption{Effect of varying the number CE samples on the QoIs calculations and posterior inference. At the top $N_{ce} = 50$, while at the bottom $N_{ce} = 200$.  Here the discrepancy function weight is $\omega = 0.75$ and $N_{abc} = 2000$.}
    \label{fig_effect_N_ce}
\end{figure*}

However, it is interesting to observe the effect of $N_{ce}$ variation in practice, as shown in Figure~\ref{fig_effect_N_ce}, which considers different values for CE samples, $N_{ce} = 50$ (top) and $N_{ce} = 200$ (bottom). The number of CE samples influences the selection of the optimal set of parameters and the inference process performed by ABC, since the a priori distribution used by ABC is constructed with the help of statistics calculated by the CE. It is interesting to note that, unlike ABC, where a greater number of samples typically leads to a better inference result, this is not necessarily the case with CE optimization. In the example shown in Figure~\ref{fig_effect_N_ce}, the model calibrated with only 50 samples has better adherence to the data than the counterpart that uses $N_{ce} = 200$. A variation in the size of the elite set $N_{\mathcal{E}_{\ell}}$ can positively or negatively impact accuracy, depending on the case. Prior numerical experiments help identifying which case of the problem of interest.

%This is counter intuitive to me. Why should the results get worse if $N_{ce}$ increases? 

\begin{figure*}
    \centering
    \includegraphics[scale=0.45]{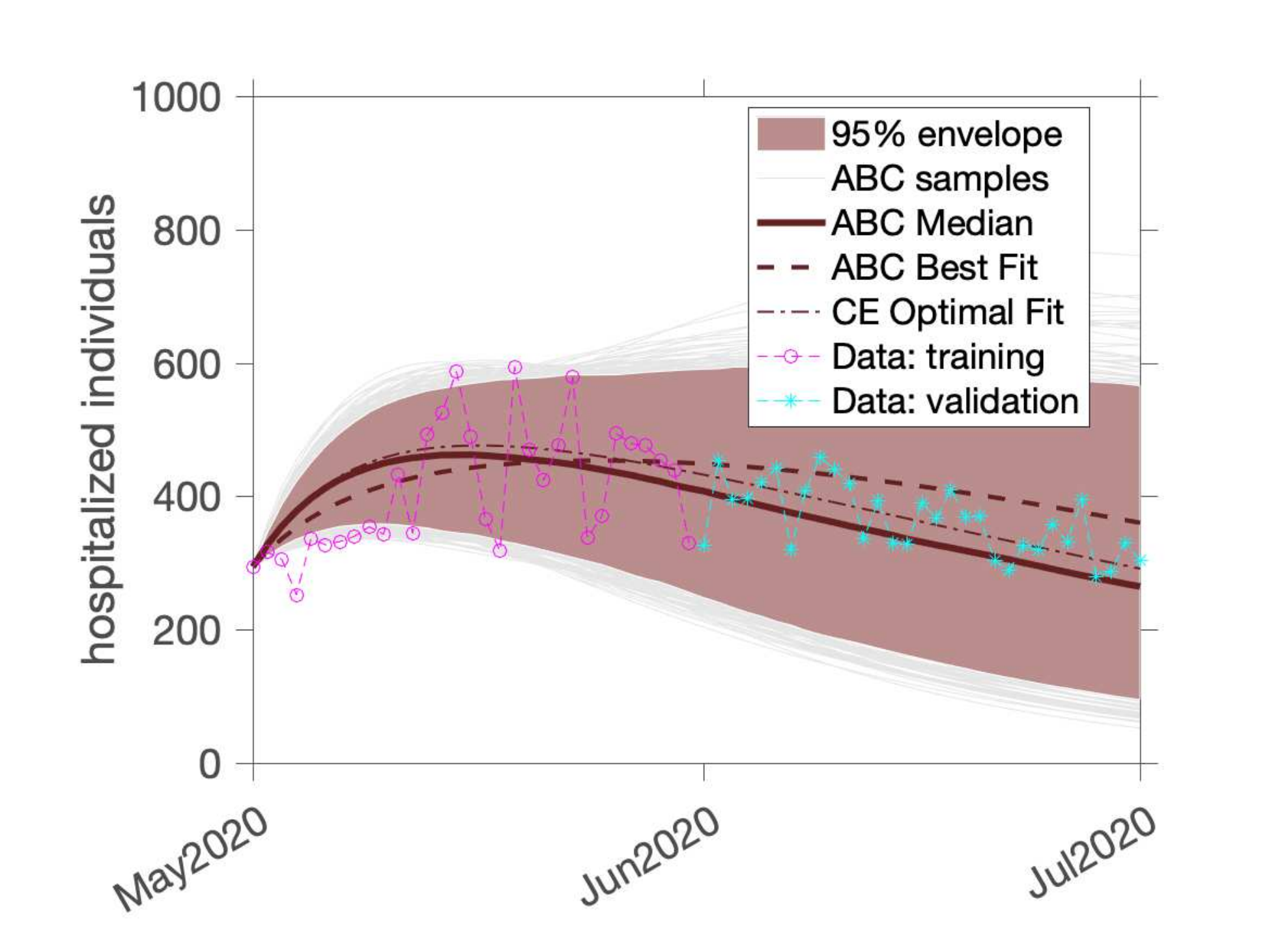}~
    \includegraphics[scale=0.45]{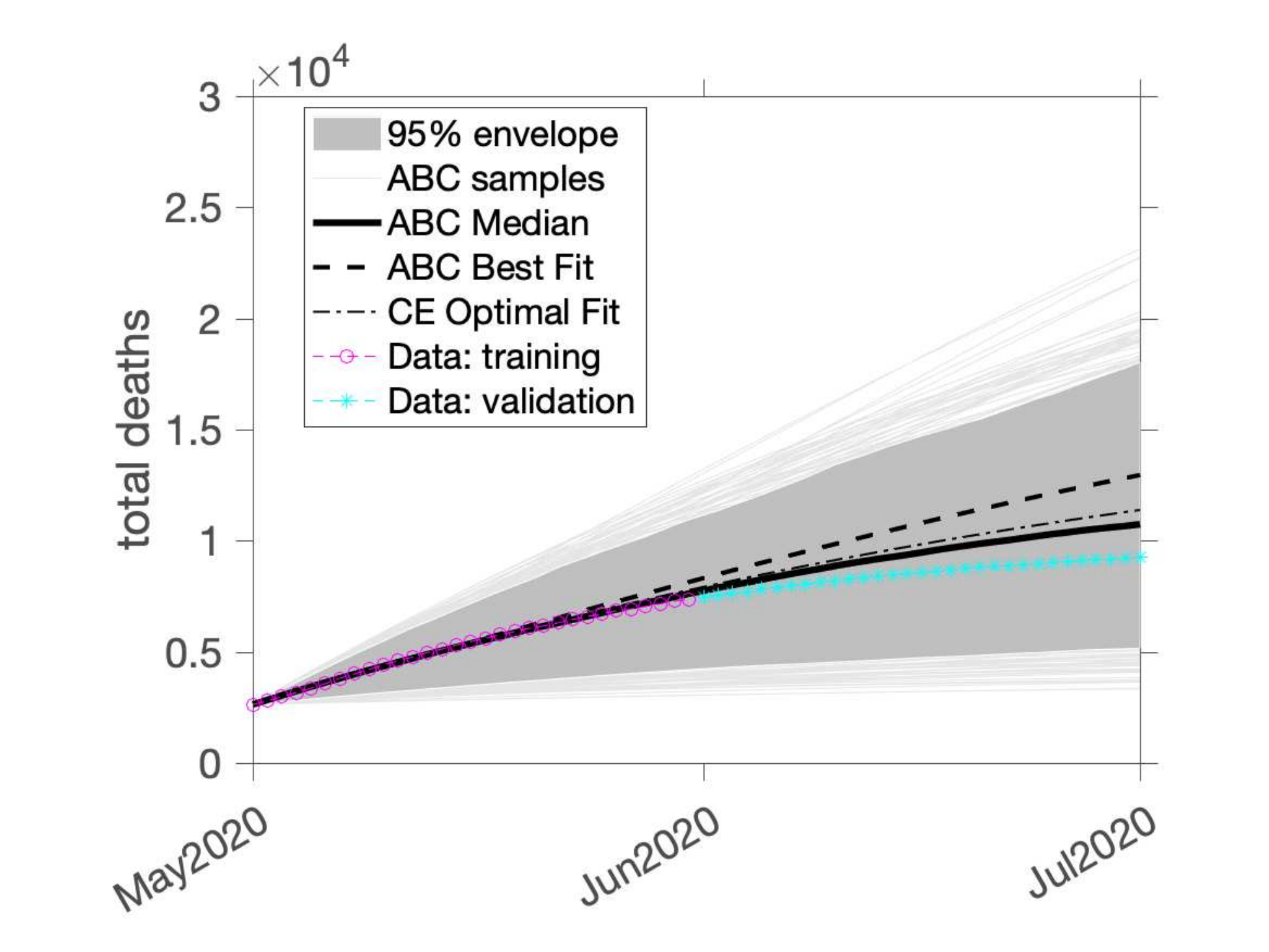}\\
    \includegraphics[scale=0.45]{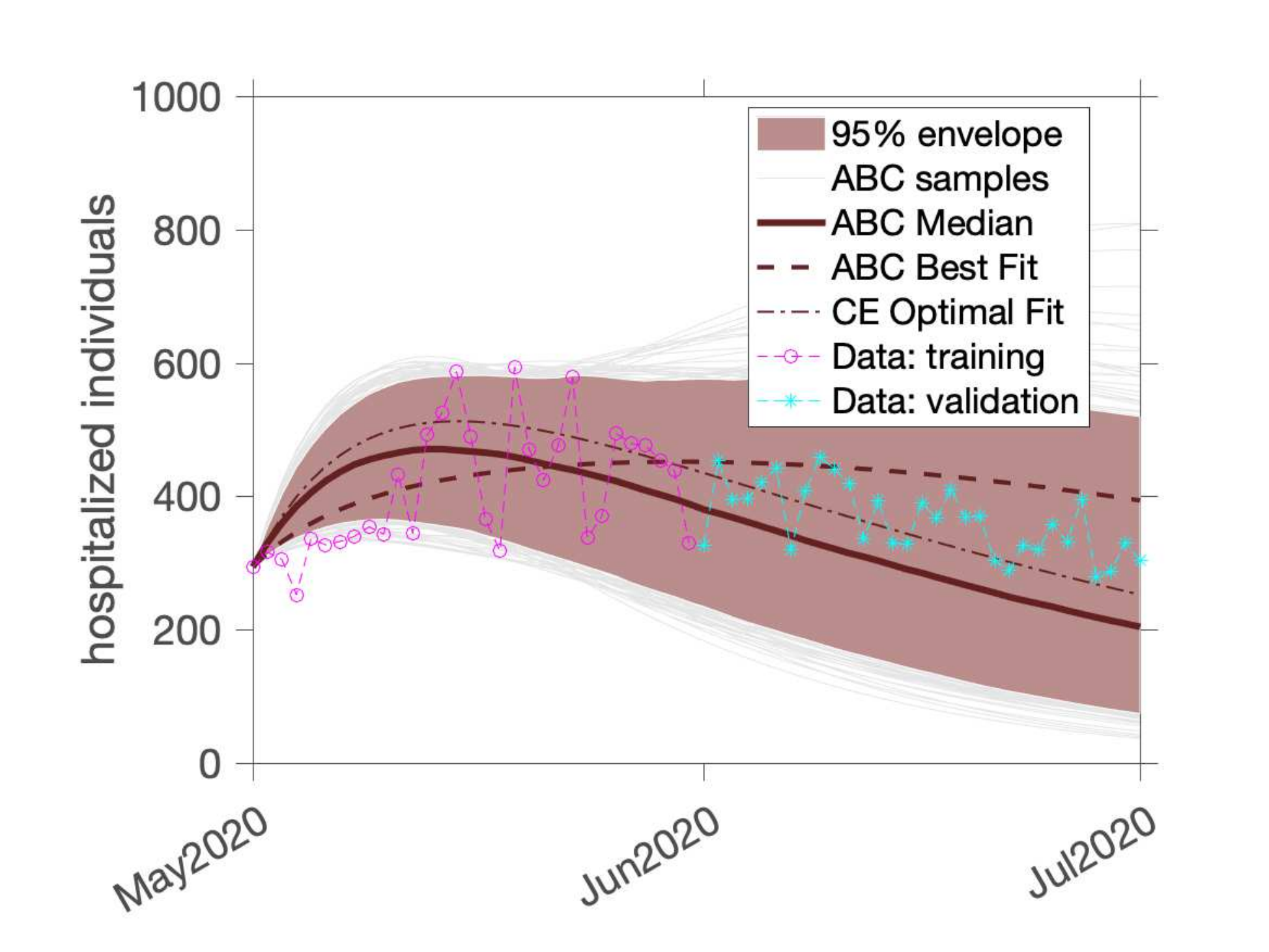}~
    \includegraphics[scale=0.45]{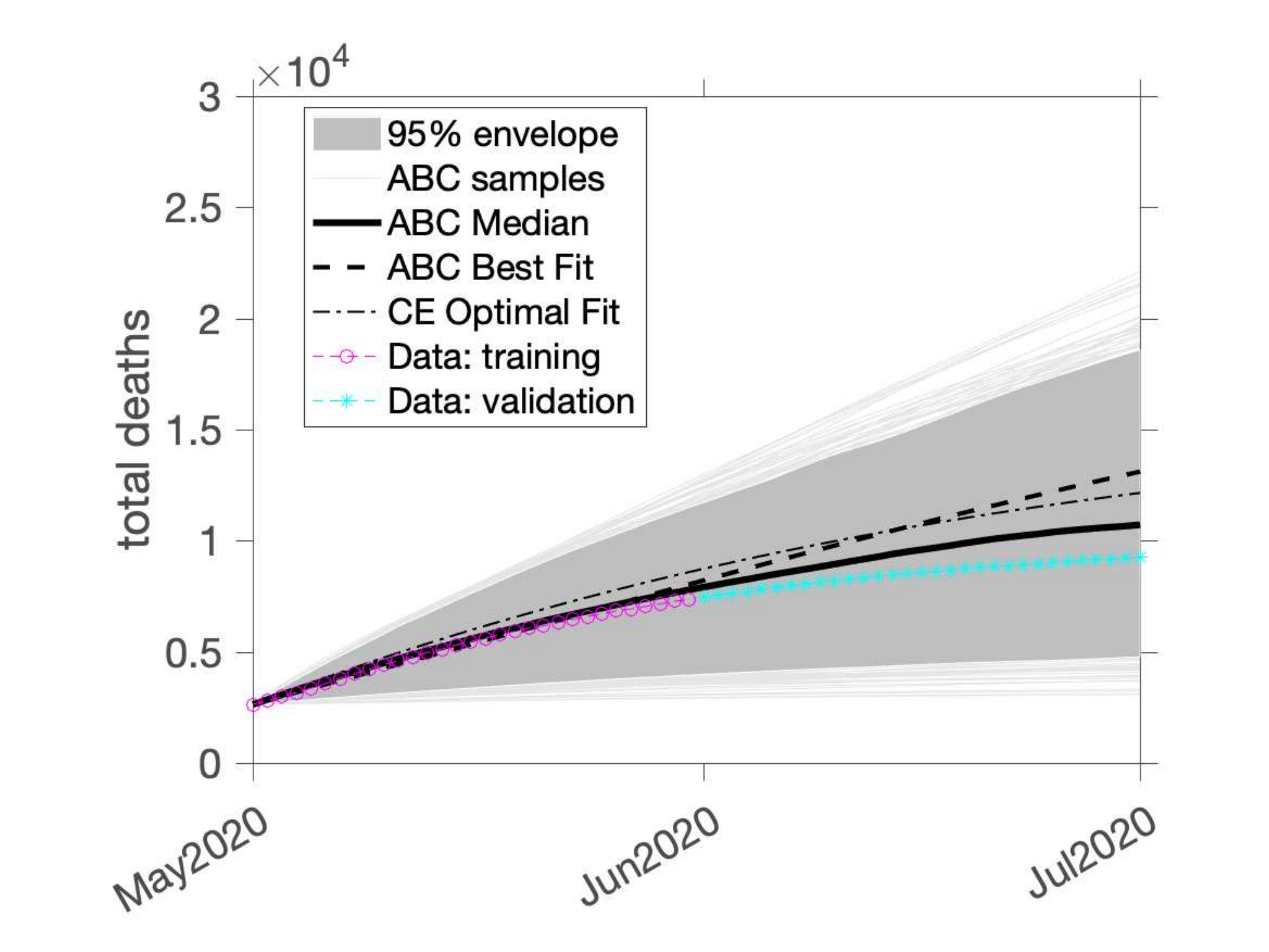}~
    \caption{Effect of the statistical seed on the QoIs calculations. Here the discrepancy function weight is $\omega = 0.75$, $N_{abc} = 2000$ and $N_{ce} = 200$.}
    \label{fig_effect_seed}
\end{figure*}

As it is a stochastic algorithm, obviously the results depend on the value of the statistical seed used. Figure~\ref{fig_effect_seed} shows two simulations with the same hyperparameters, but with slightly different inference results.

Finally, but not least, it is worth mentioning that the results are strongly influenced by choice of bounds $\bm{x}_{min}$ and $\bm{x}_{max}$. Indeed, in the authors' experience, these are the parameters that have the most significant impact on the quality of results (together with the epidemic surveillance data). A bad choice for the parameter bounds can lead to unreliable models for the actual behavior of the outbreak. Good choices for these parameters demand detailed knowledge about the biological aspects of the problem. Numerical experimentation can also be of great help in finding plausible values.

\subsection{Predictability limit for the epidemic dynamics using the CE-ABC and the SEIR(+AHD) model}

This section presents a study to delineate the predictability limit of the SEIR(+AHD) model as a tool to predict the dynamics of COVID-19 in the city of Rio de Janeiro in the year 2020. 

\begin{figure*}
    \centering
    \includegraphics[scale=0.3]{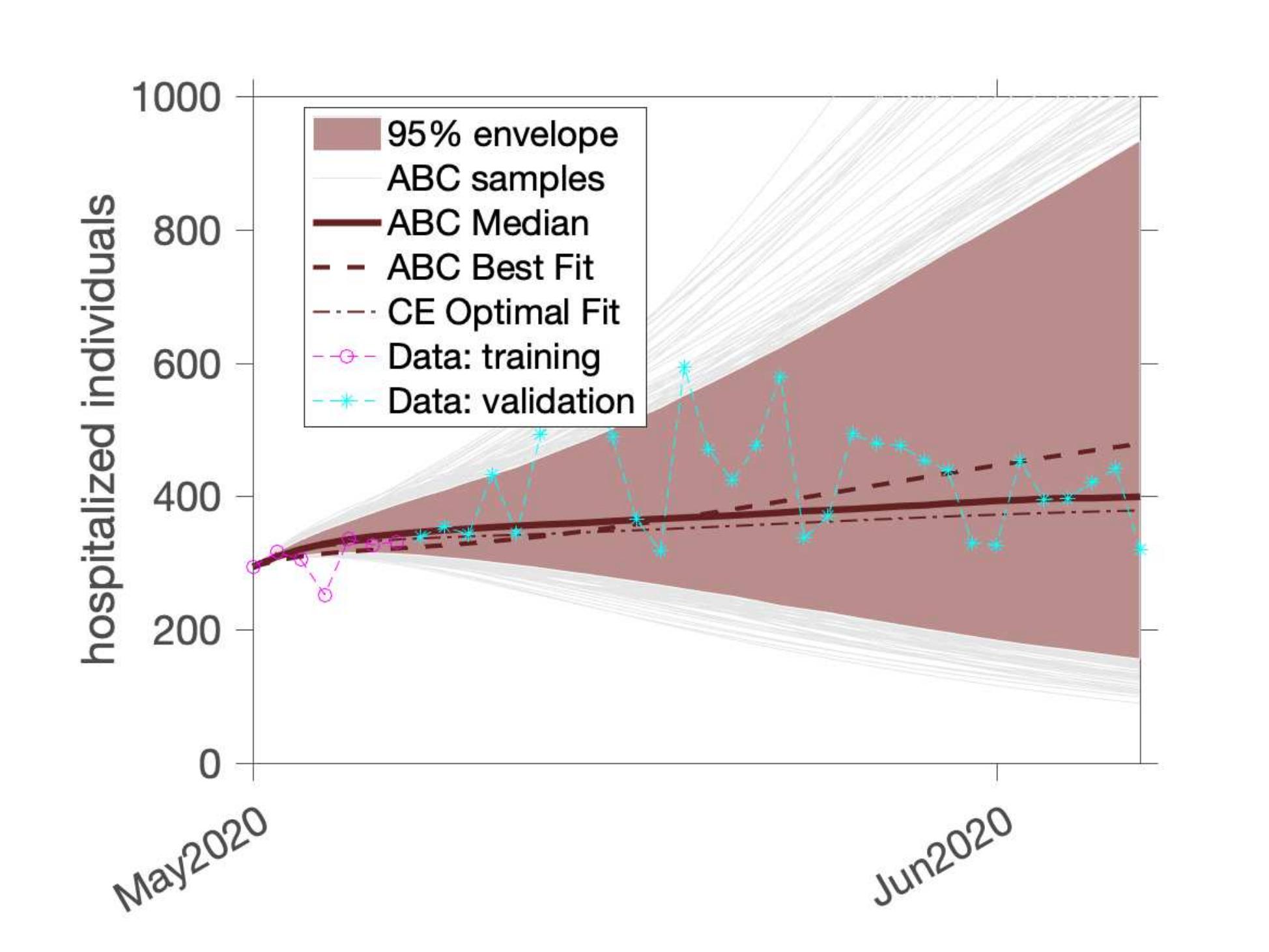}
    \includegraphics[scale=0.3]{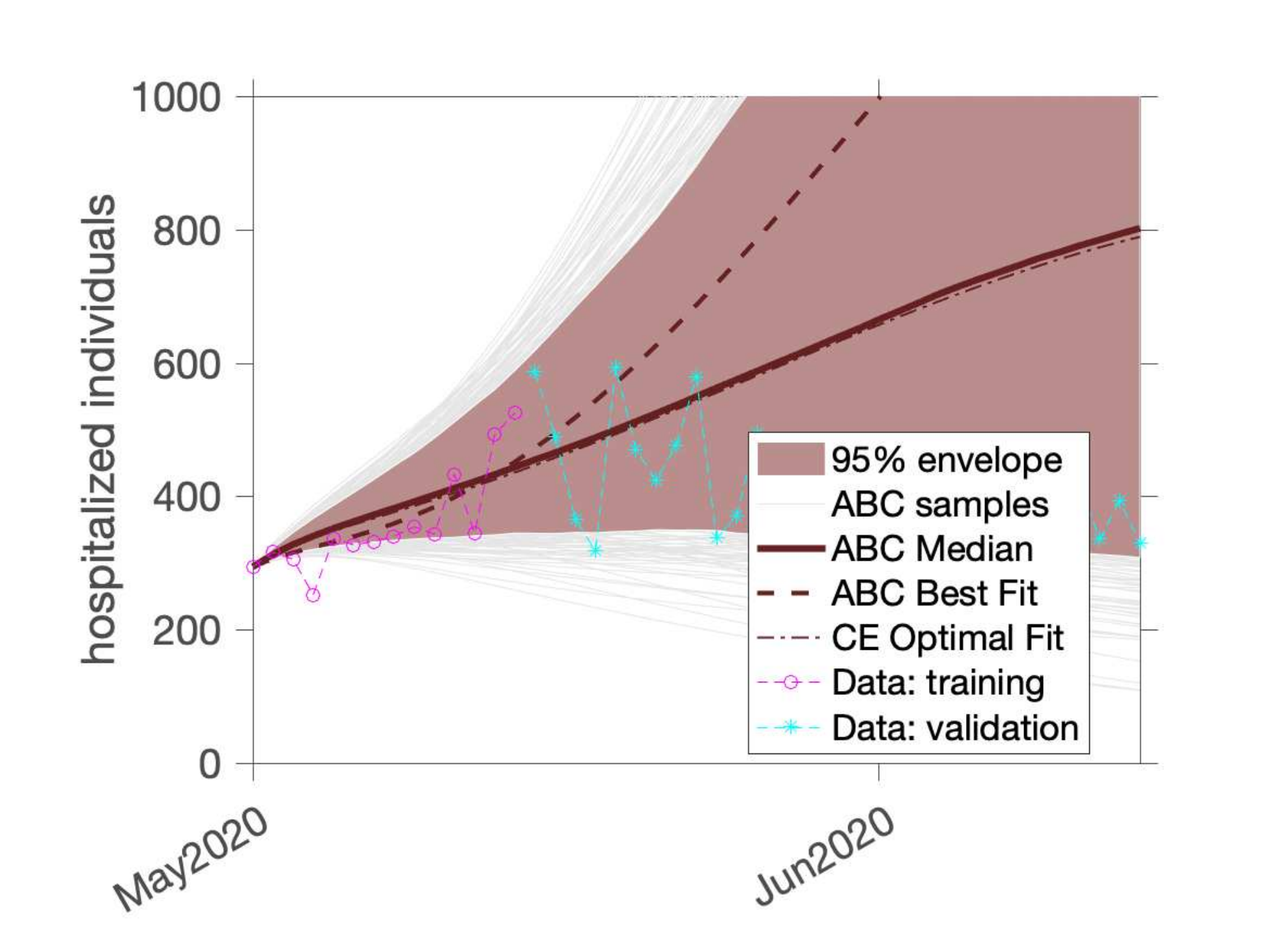}\\
    \includegraphics[scale=0.3]{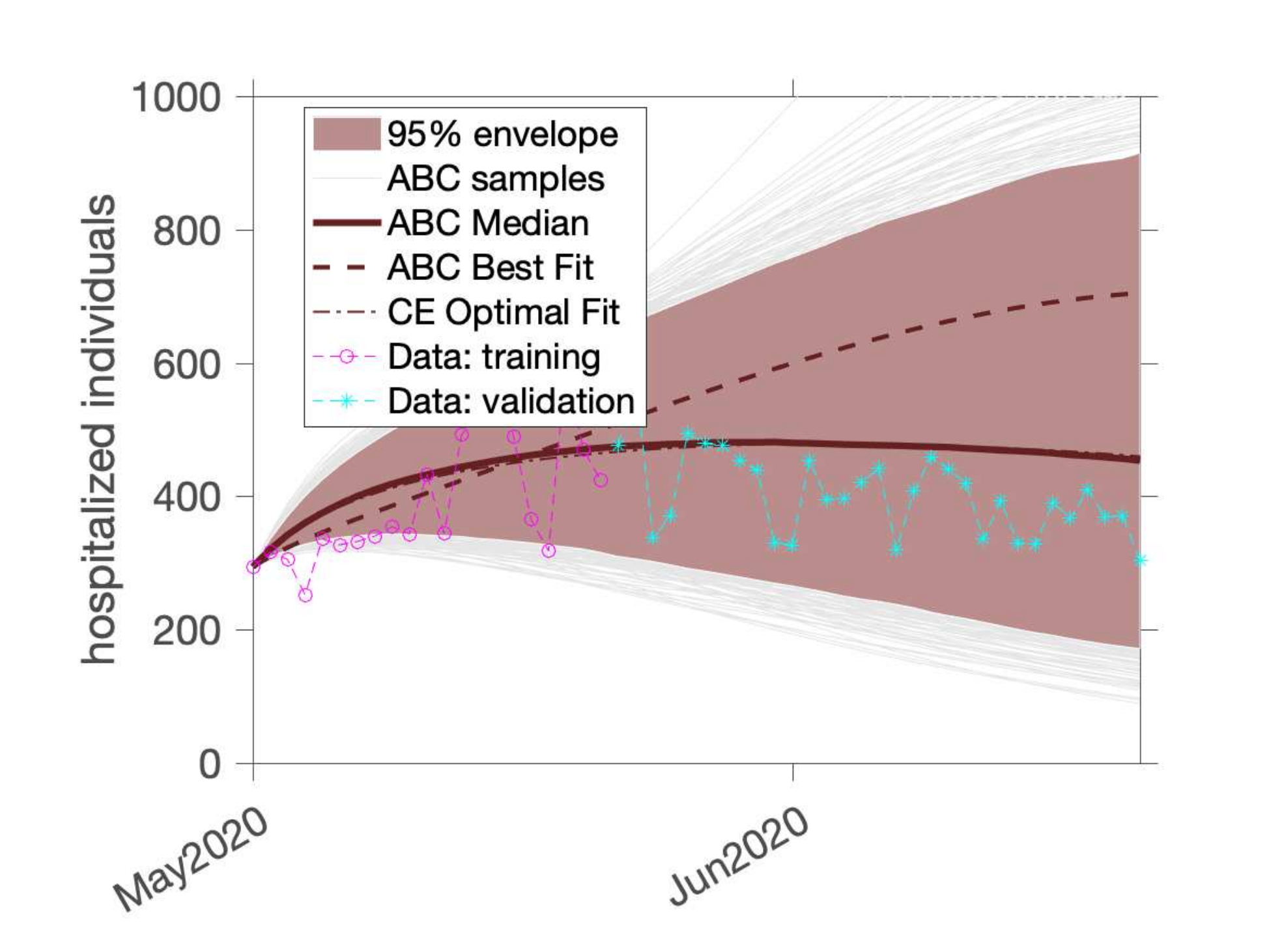}
    \includegraphics[scale=0.3]{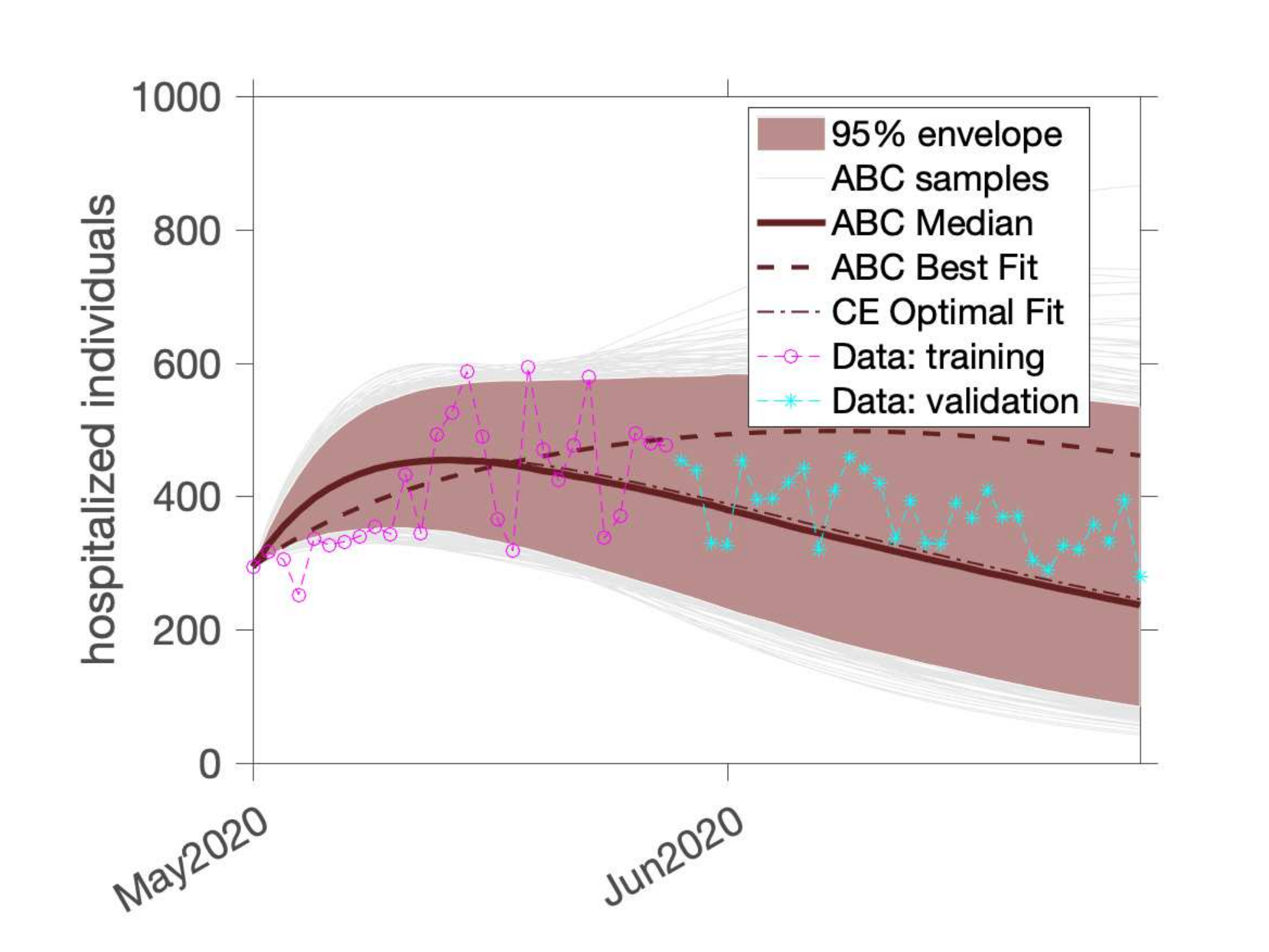}\\
    \includegraphics[scale=0.3]{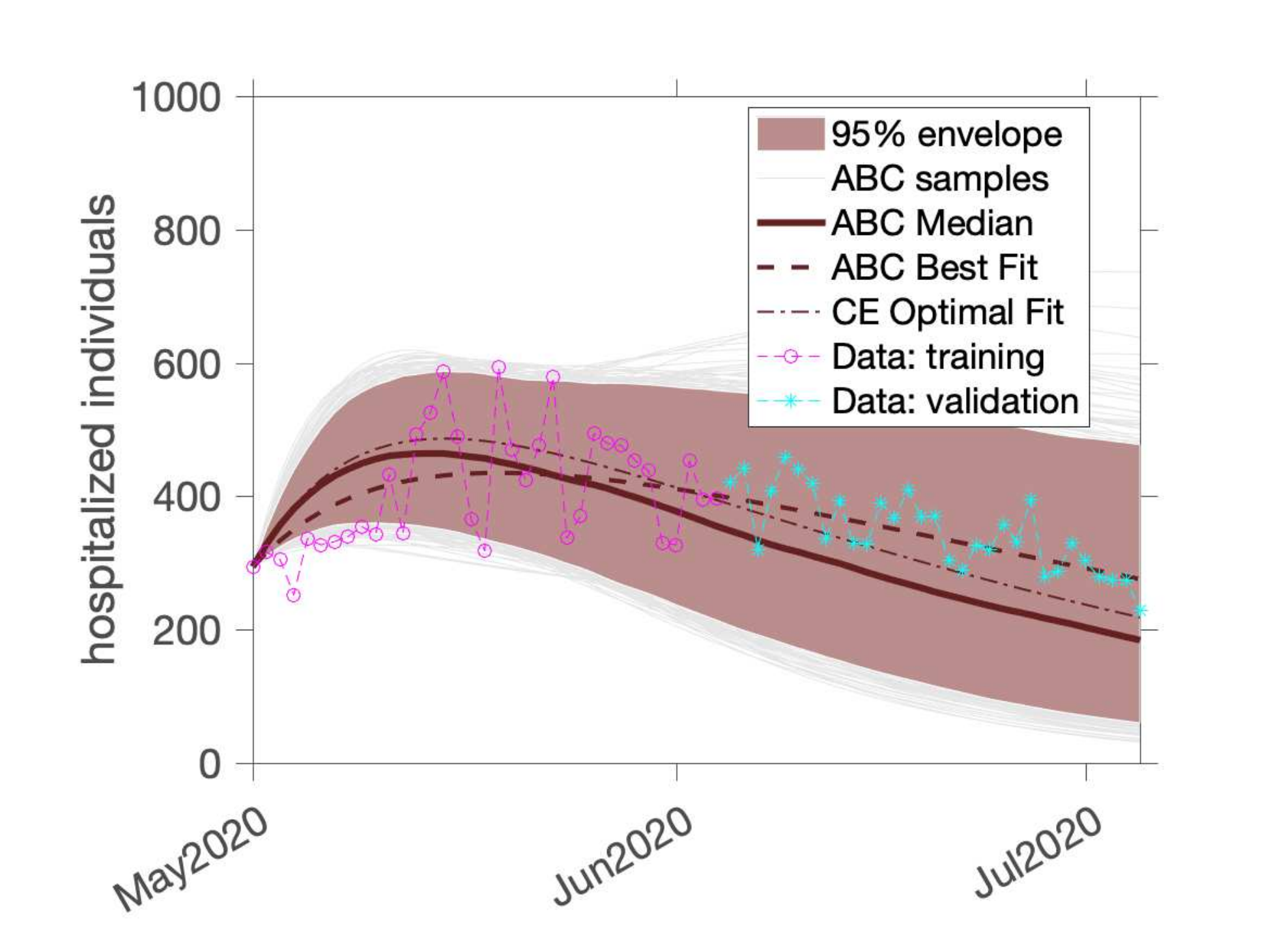}
    \includegraphics[scale=0.3]{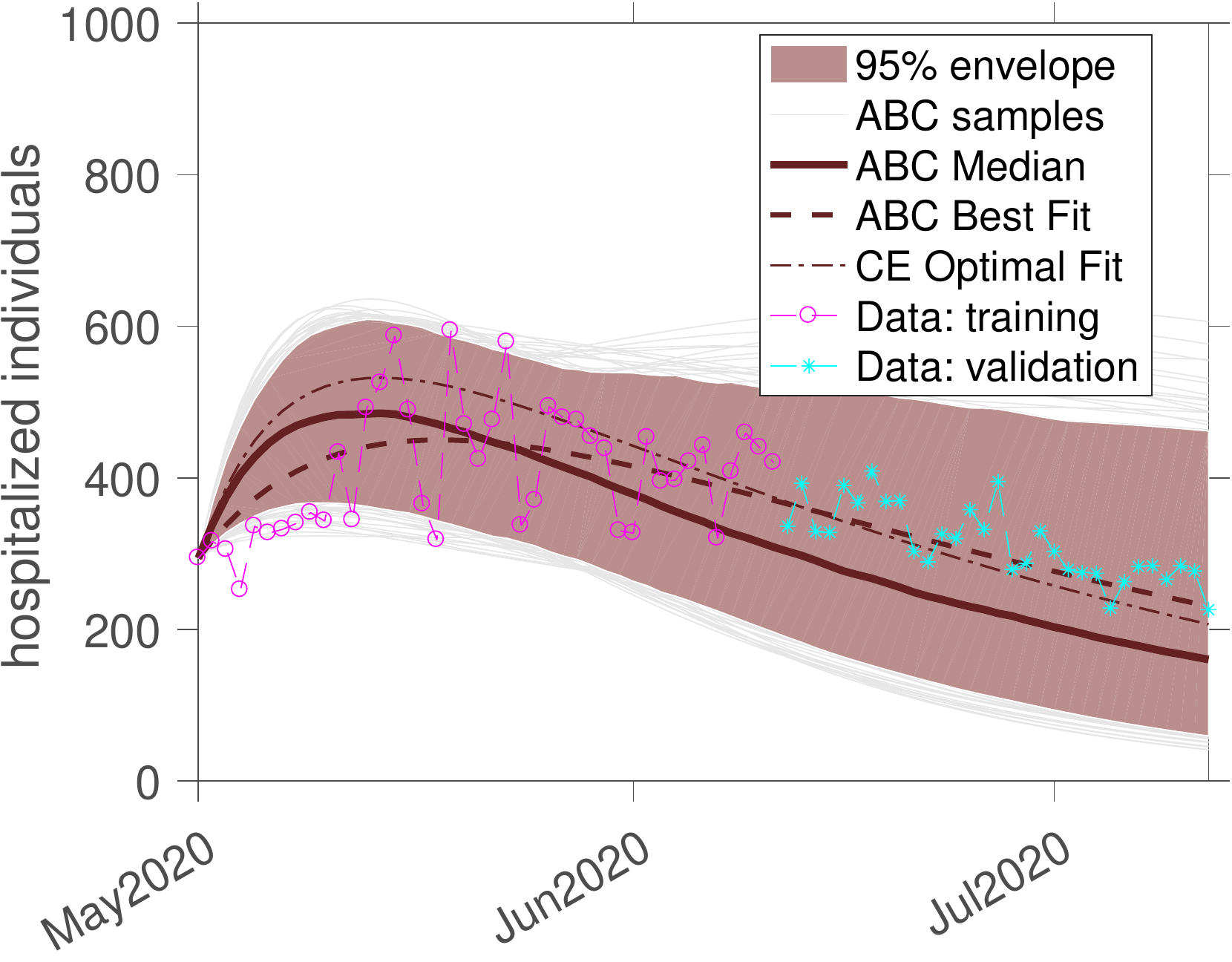}\\
    \includegraphics[scale=0.3]{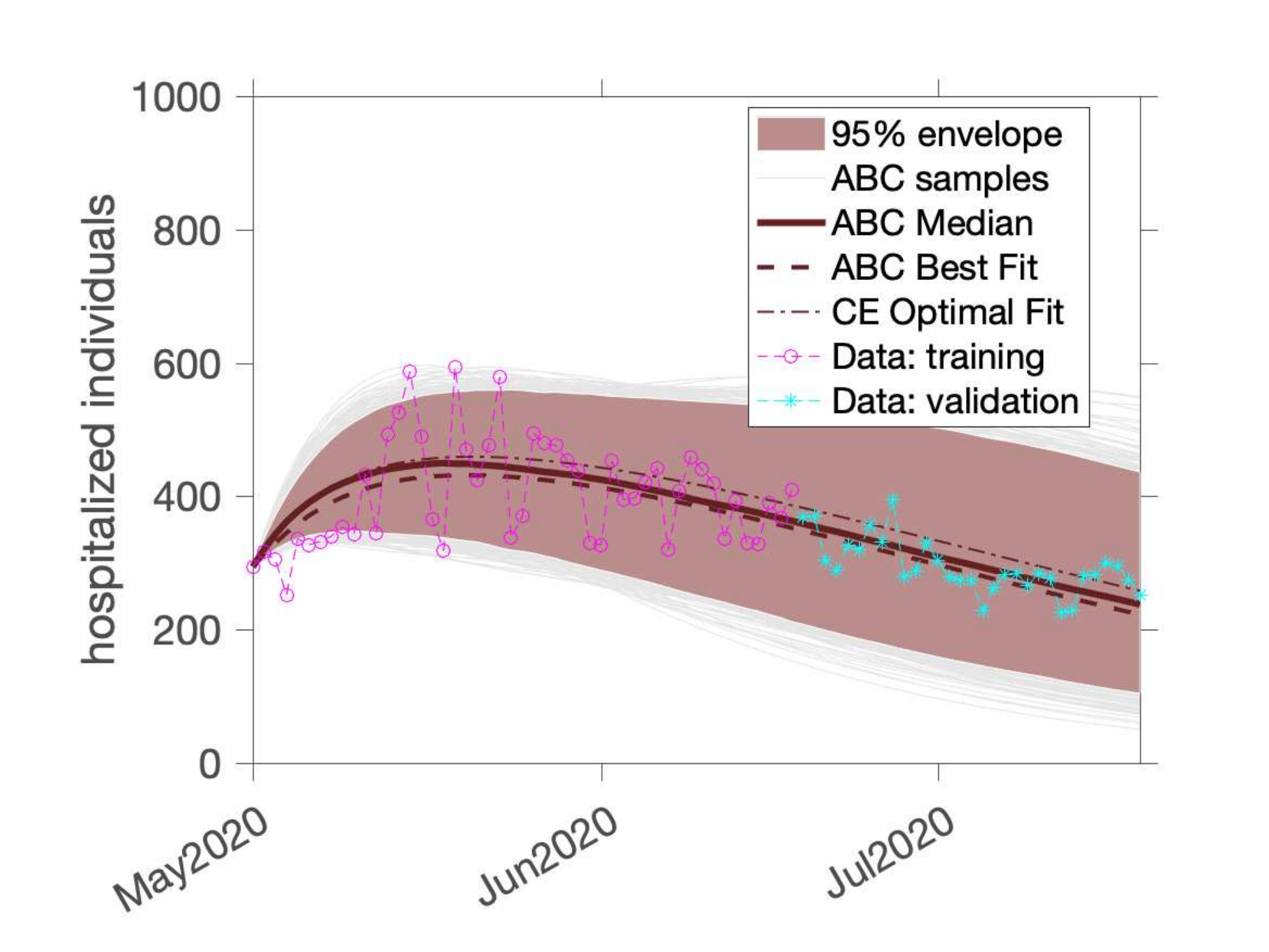}
    \includegraphics[scale=0.3]{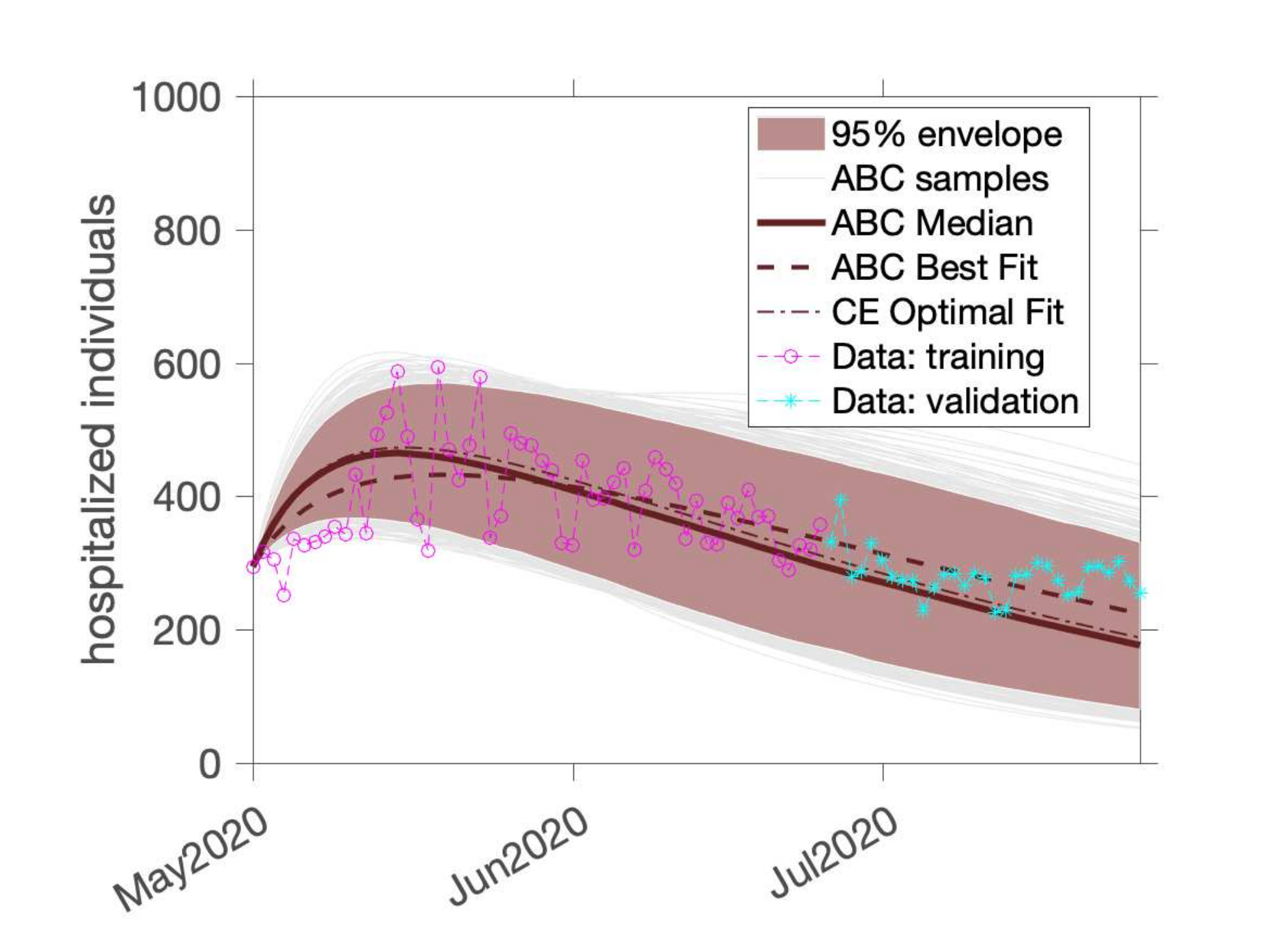}\\
    \includegraphics[scale=0.3]{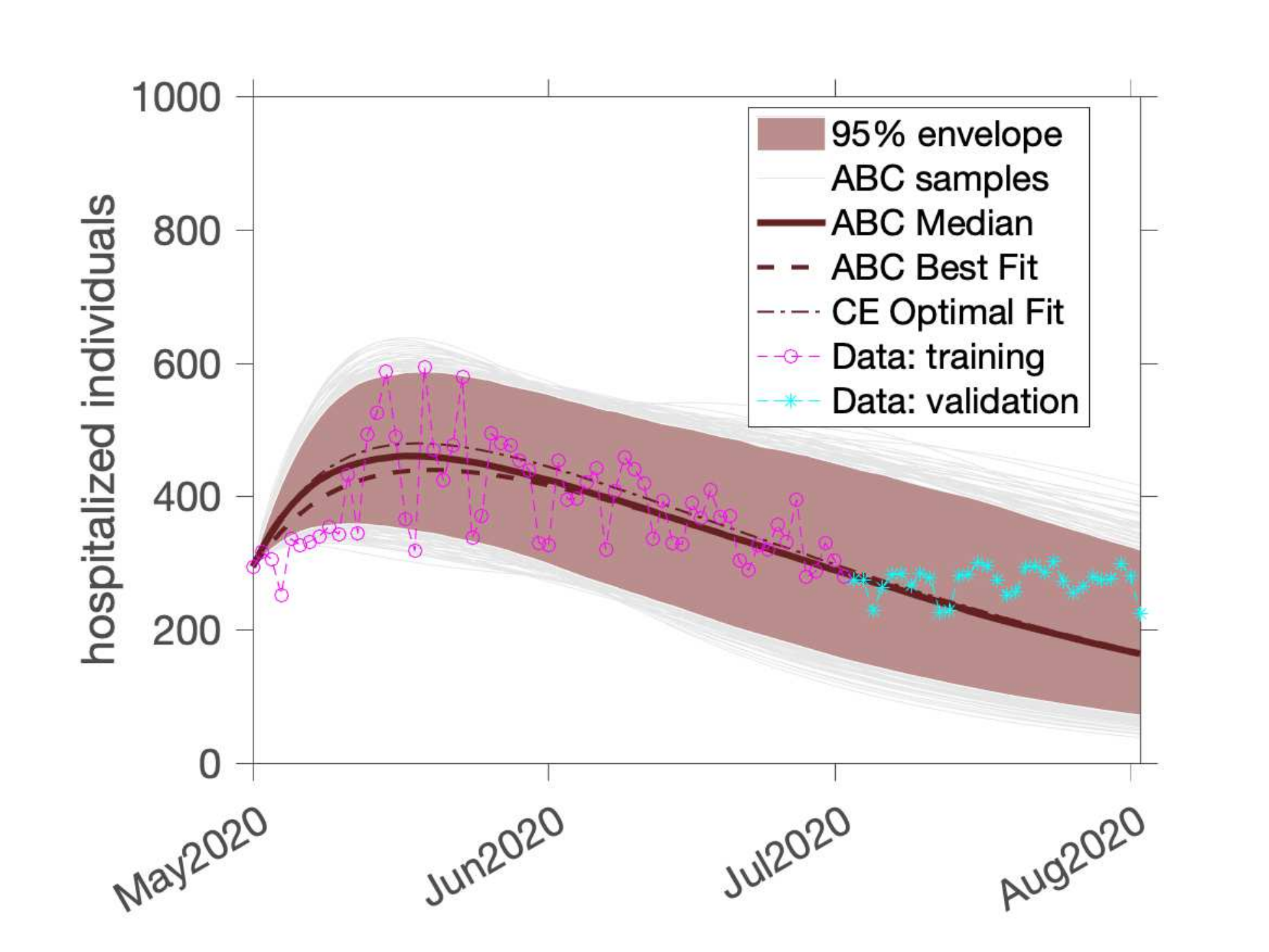}
    \includegraphics[scale=0.3]{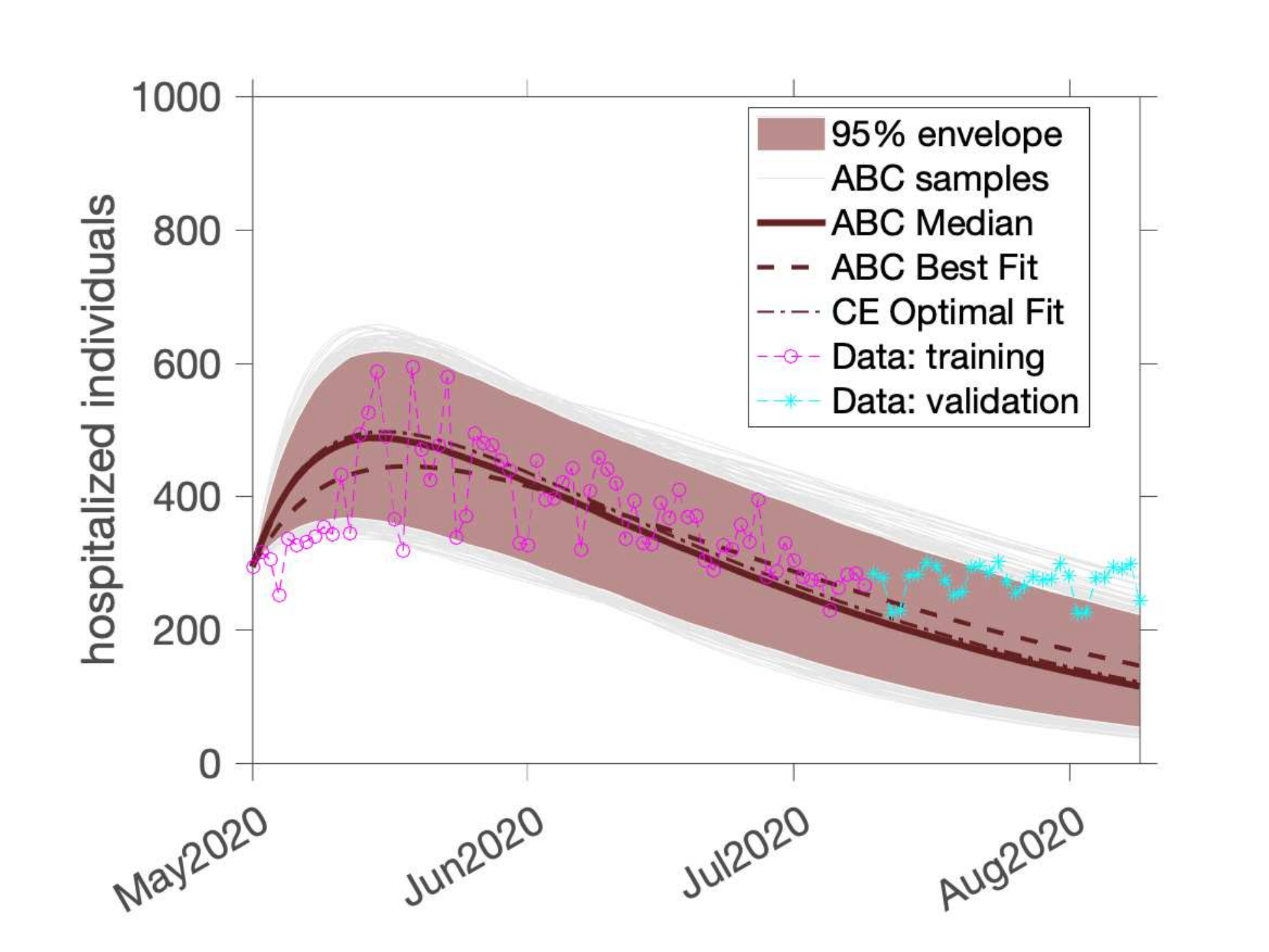}
    \caption{Evolution of the hospitalizations time series, calibrated with different datasets of Rio de Janeiro epidemic, extrapolating forecasts over a 30-day horizon. Training data are incremented every seven days, including information from the last seven days, starting with the period between May 1 and 7, 2020, and ending with May 1 and July 9, 2020.}
    \label{fig_pred_hor_H}
\end{figure*}

\begin{figure*}
    \centering
    \includegraphics[scale=0.3]{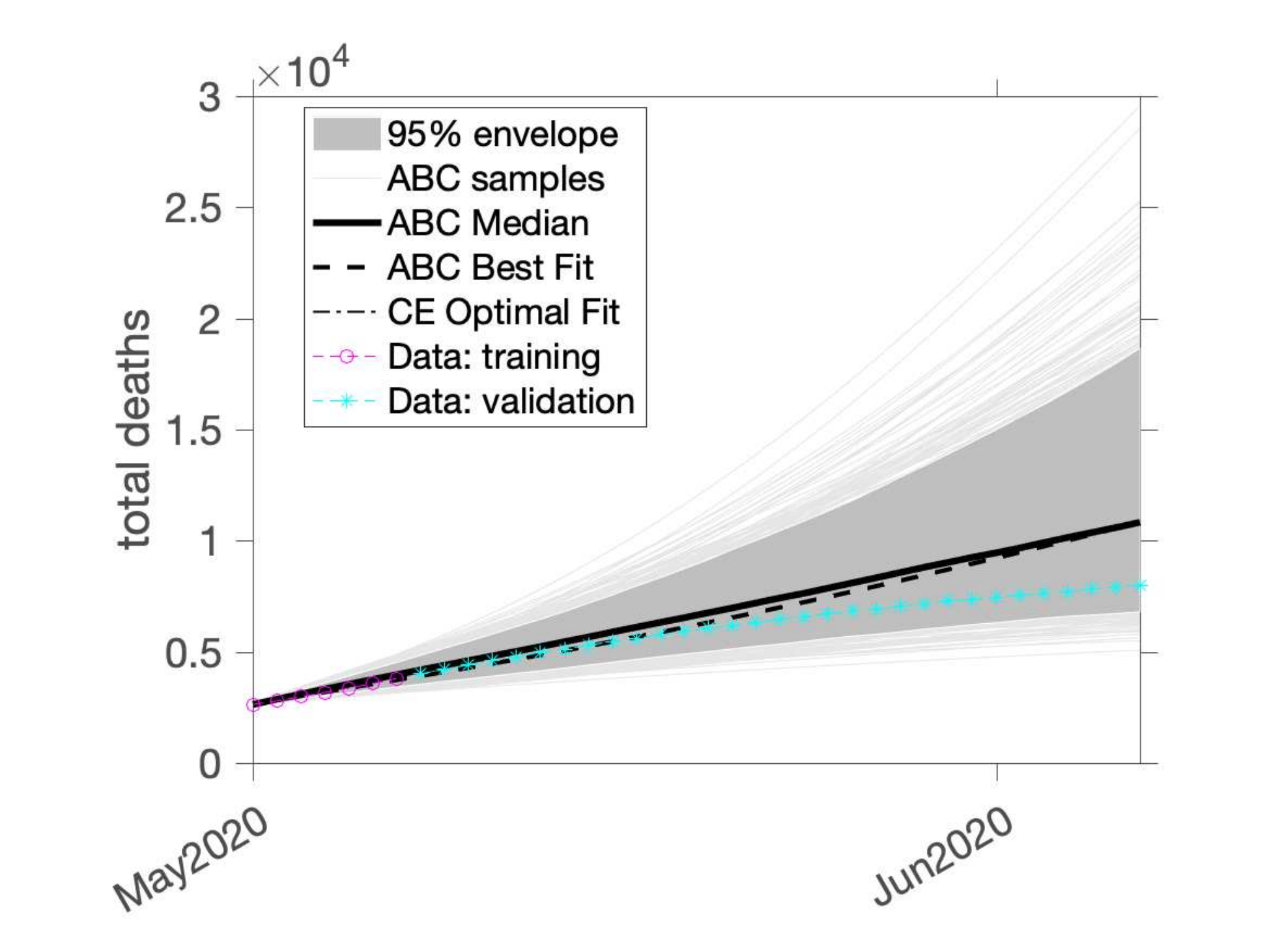}
    \includegraphics[scale=0.3]{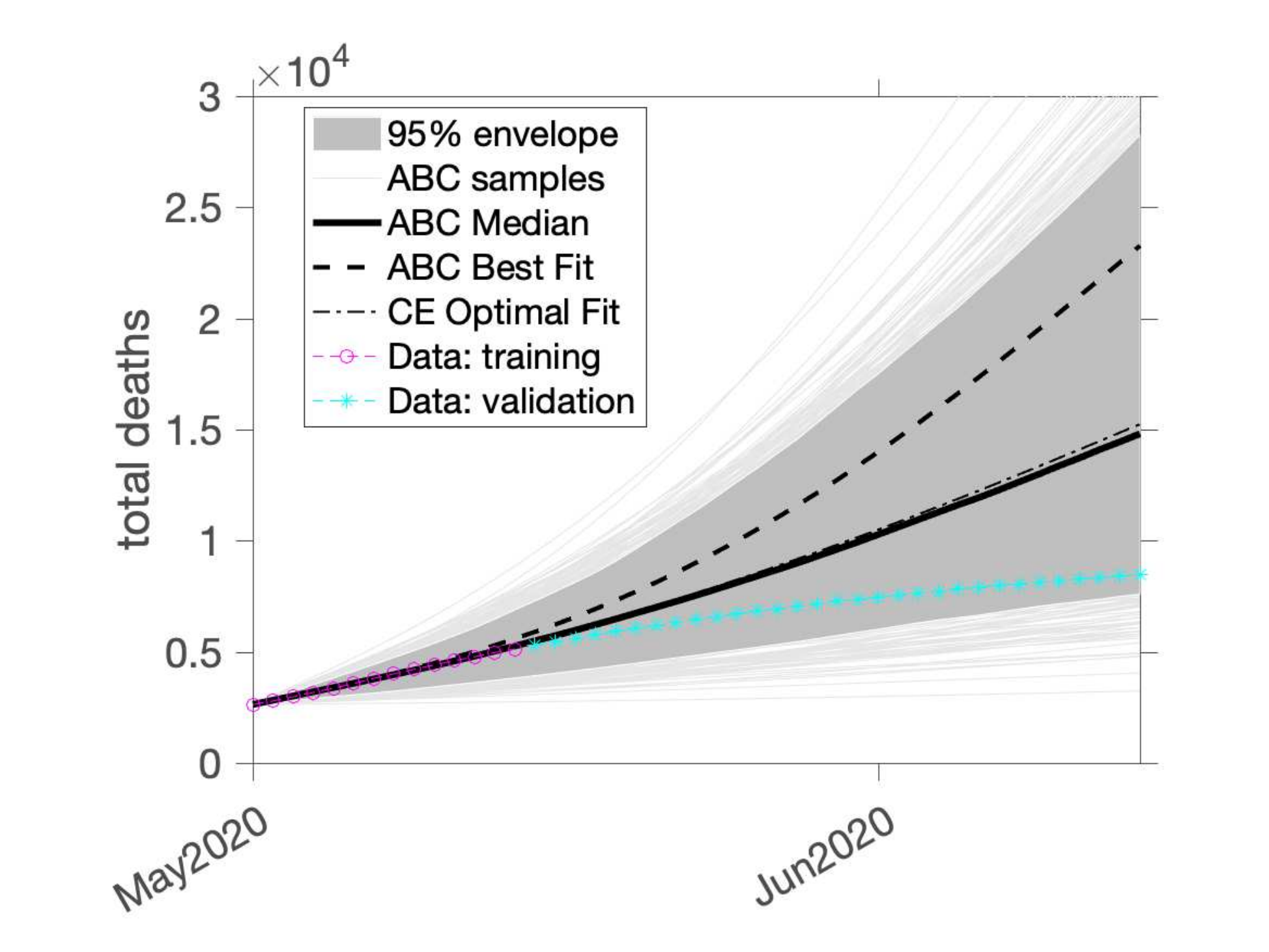}\\
    \includegraphics[scale=0.3]{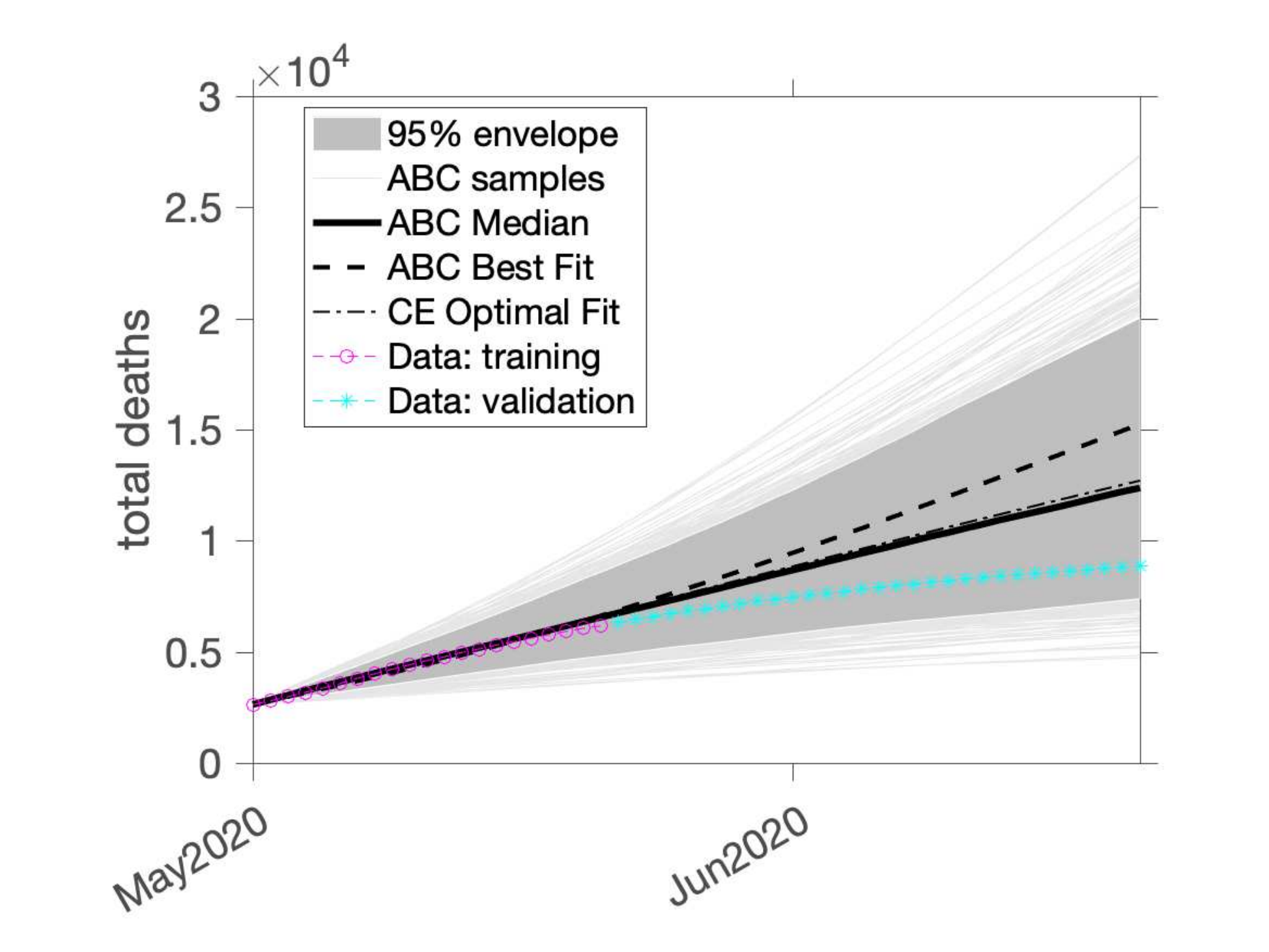}
    \includegraphics[scale=0.3]{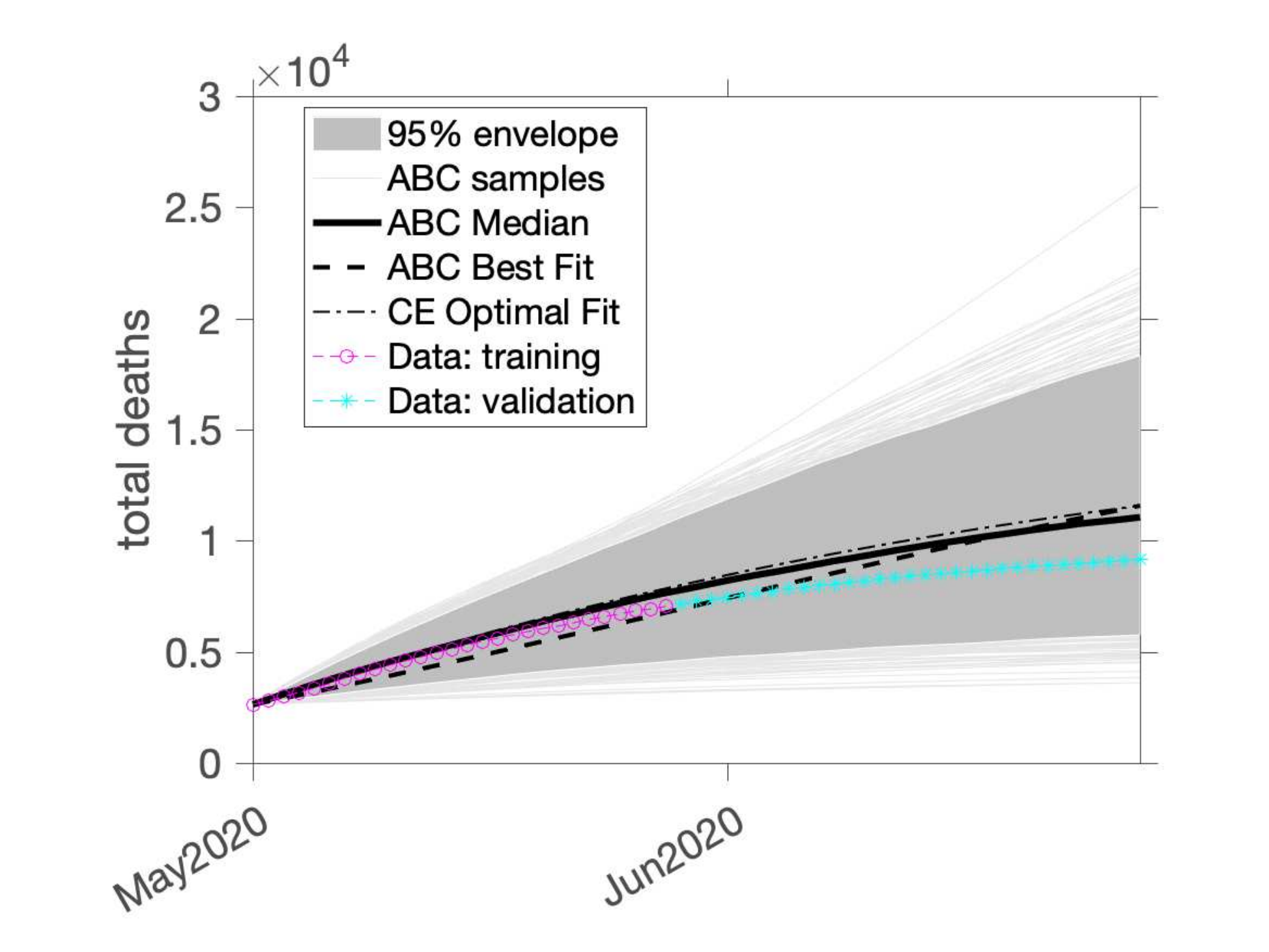}\\
    \includegraphics[scale=0.3]{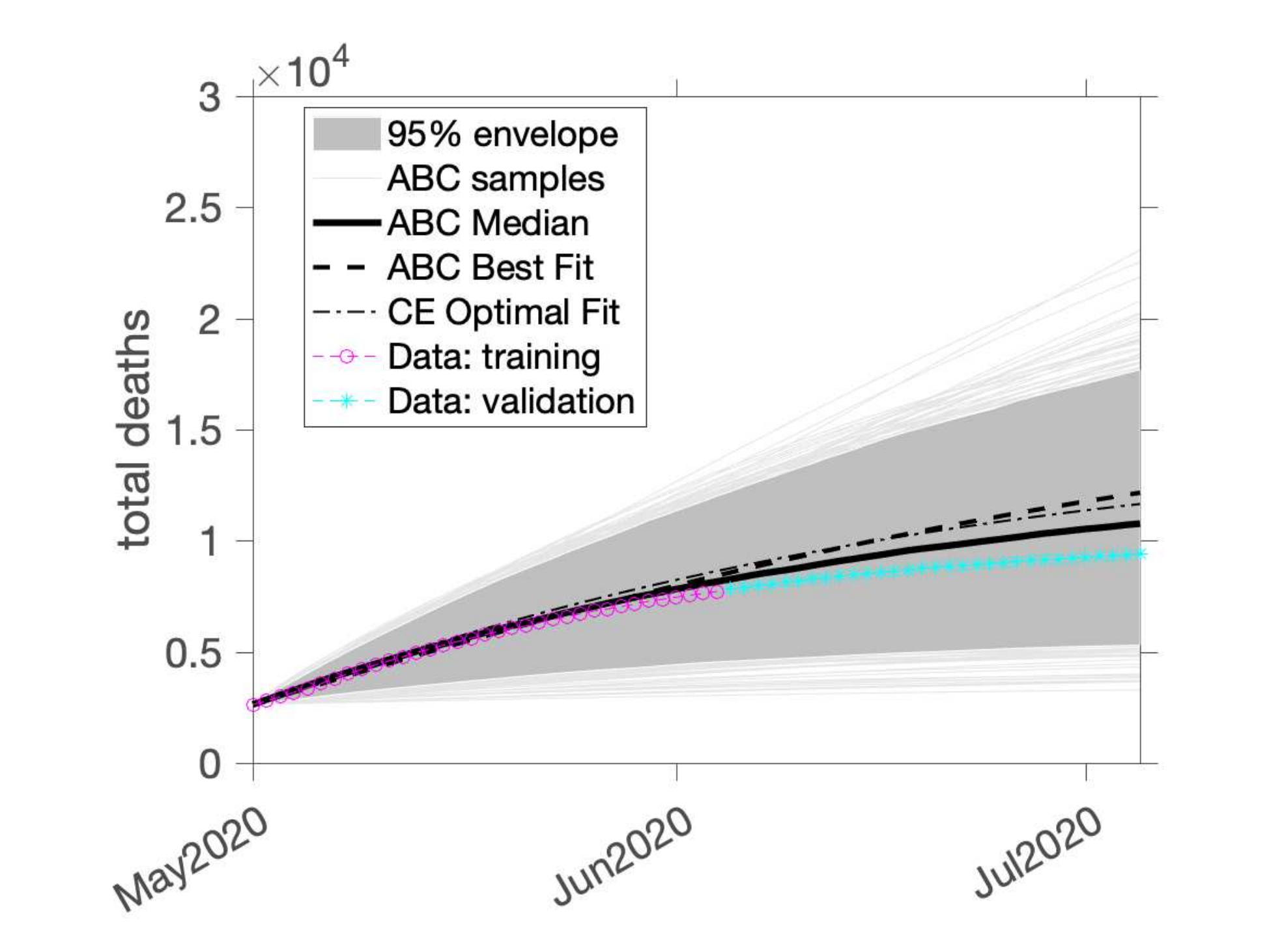}
    \includegraphics[scale=0.3]{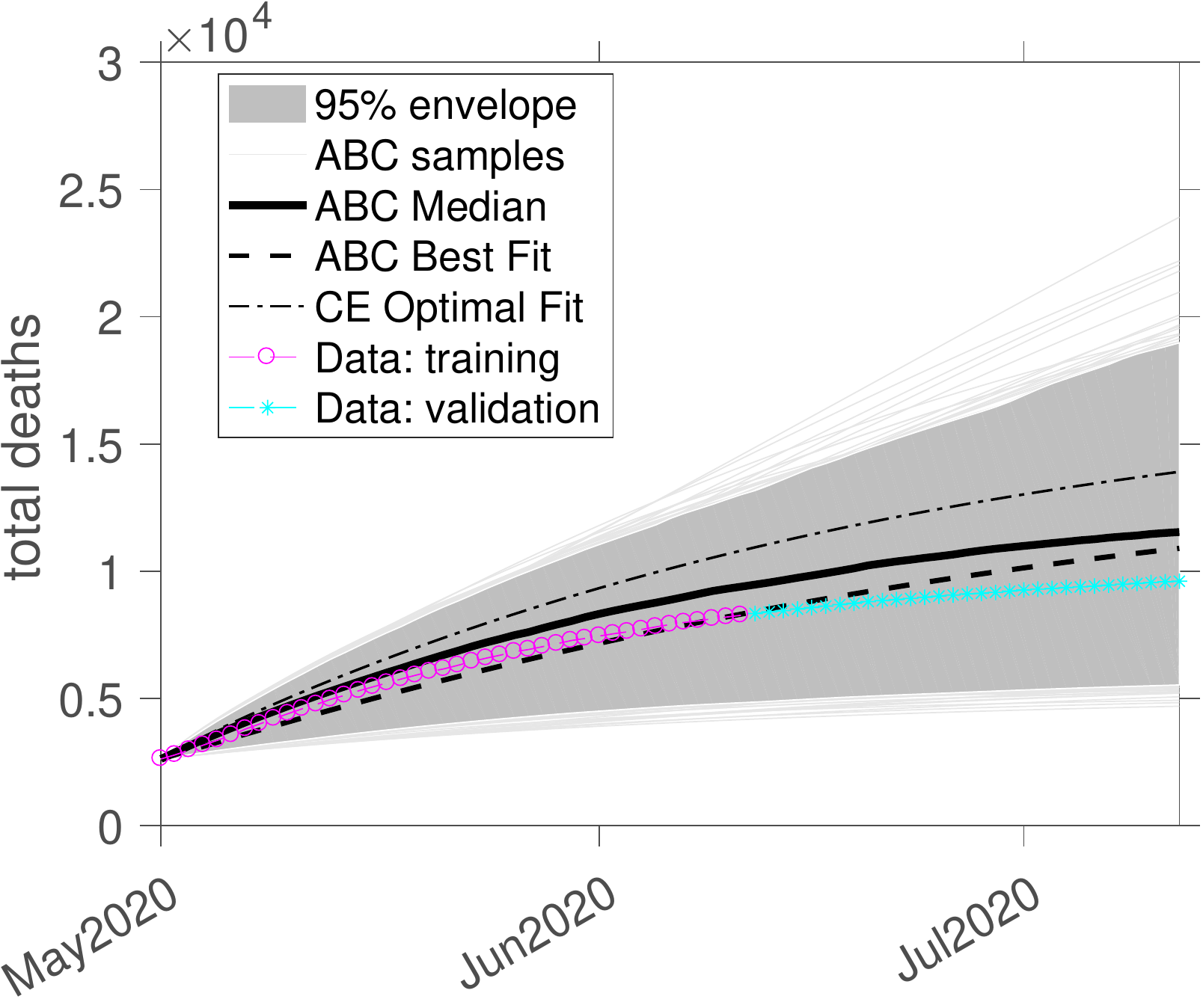}\\
    \includegraphics[scale=0.3]{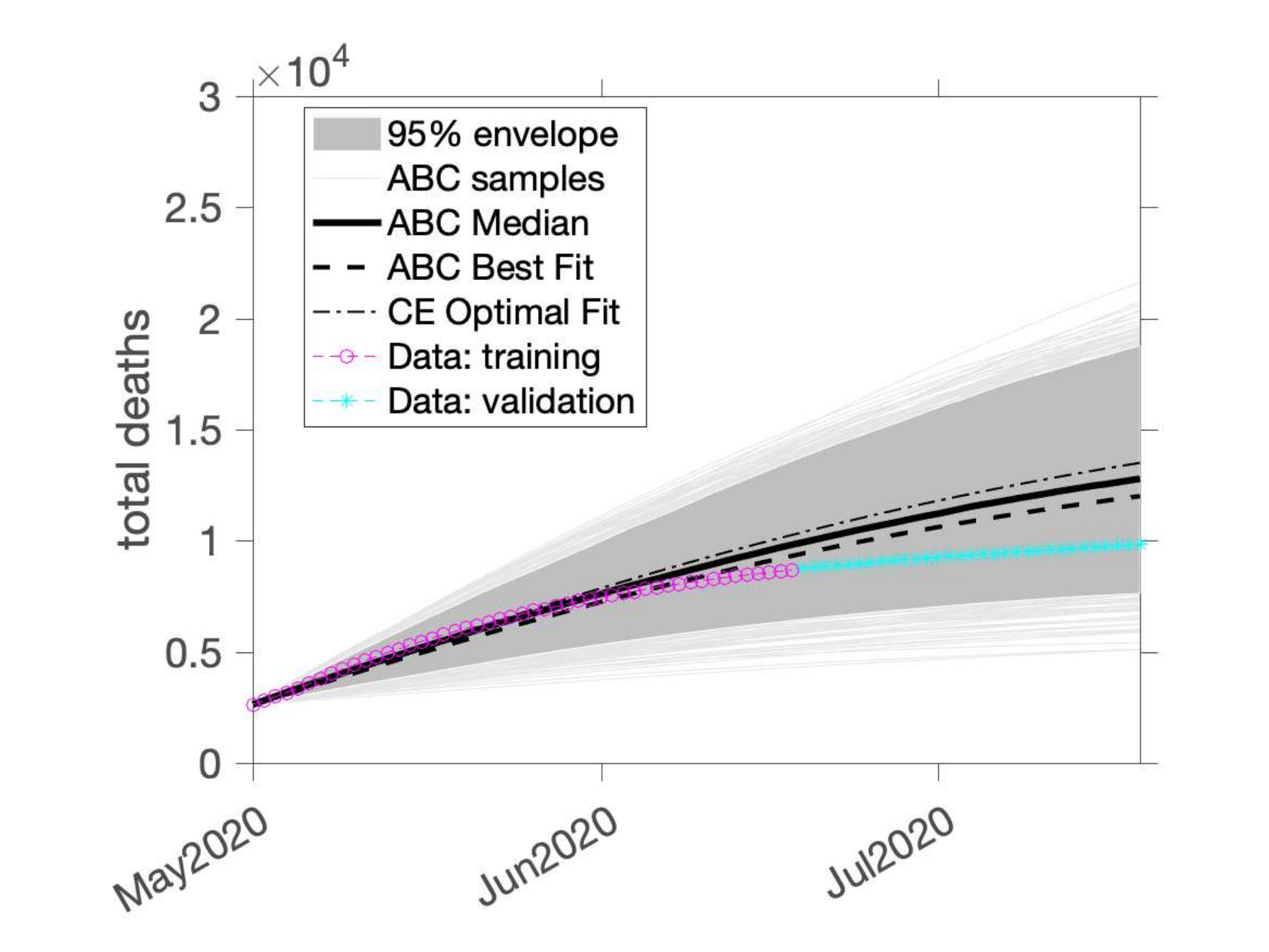}
    \includegraphics[scale=0.3]{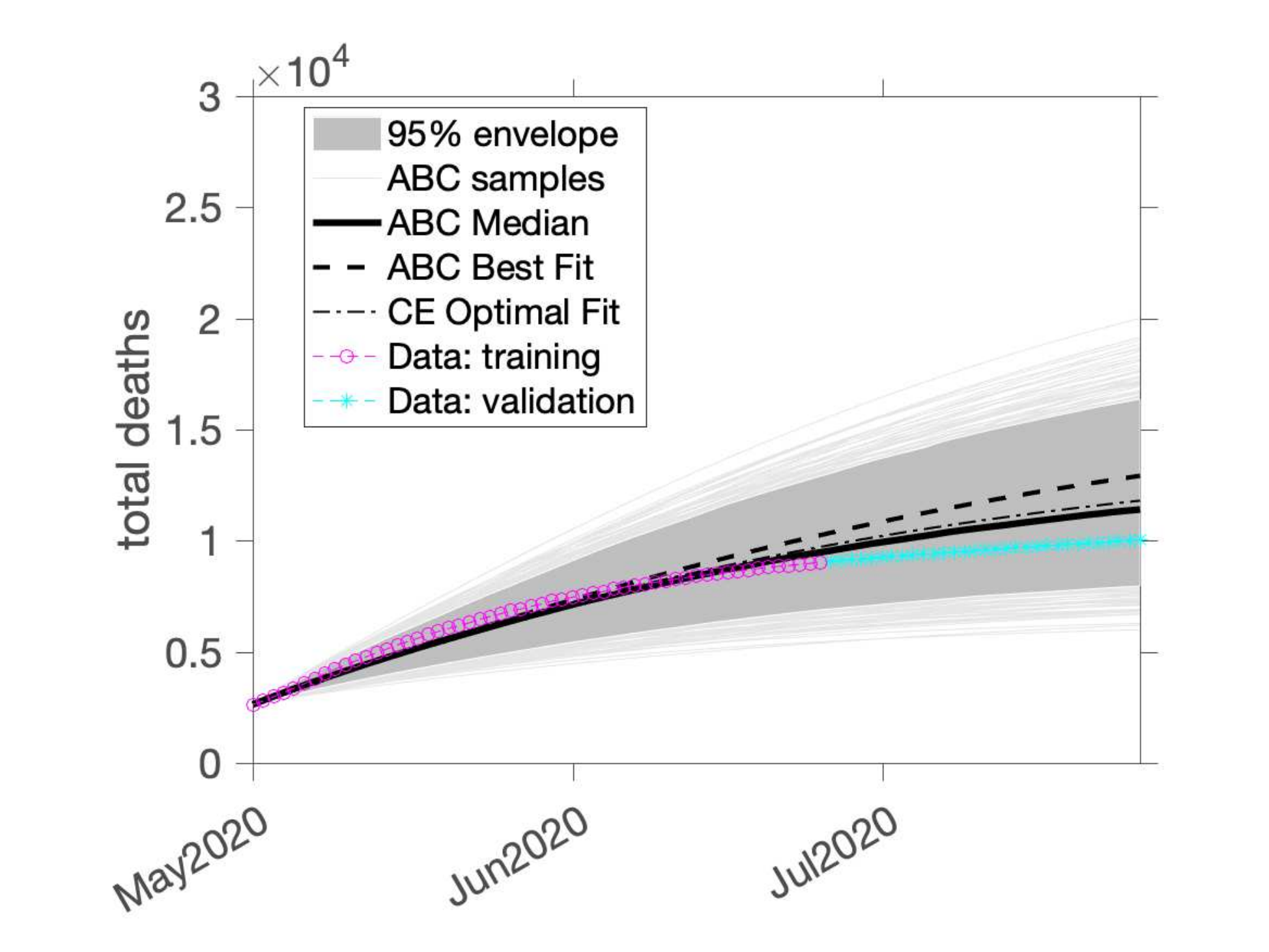}\\
    \includegraphics[scale=0.3]{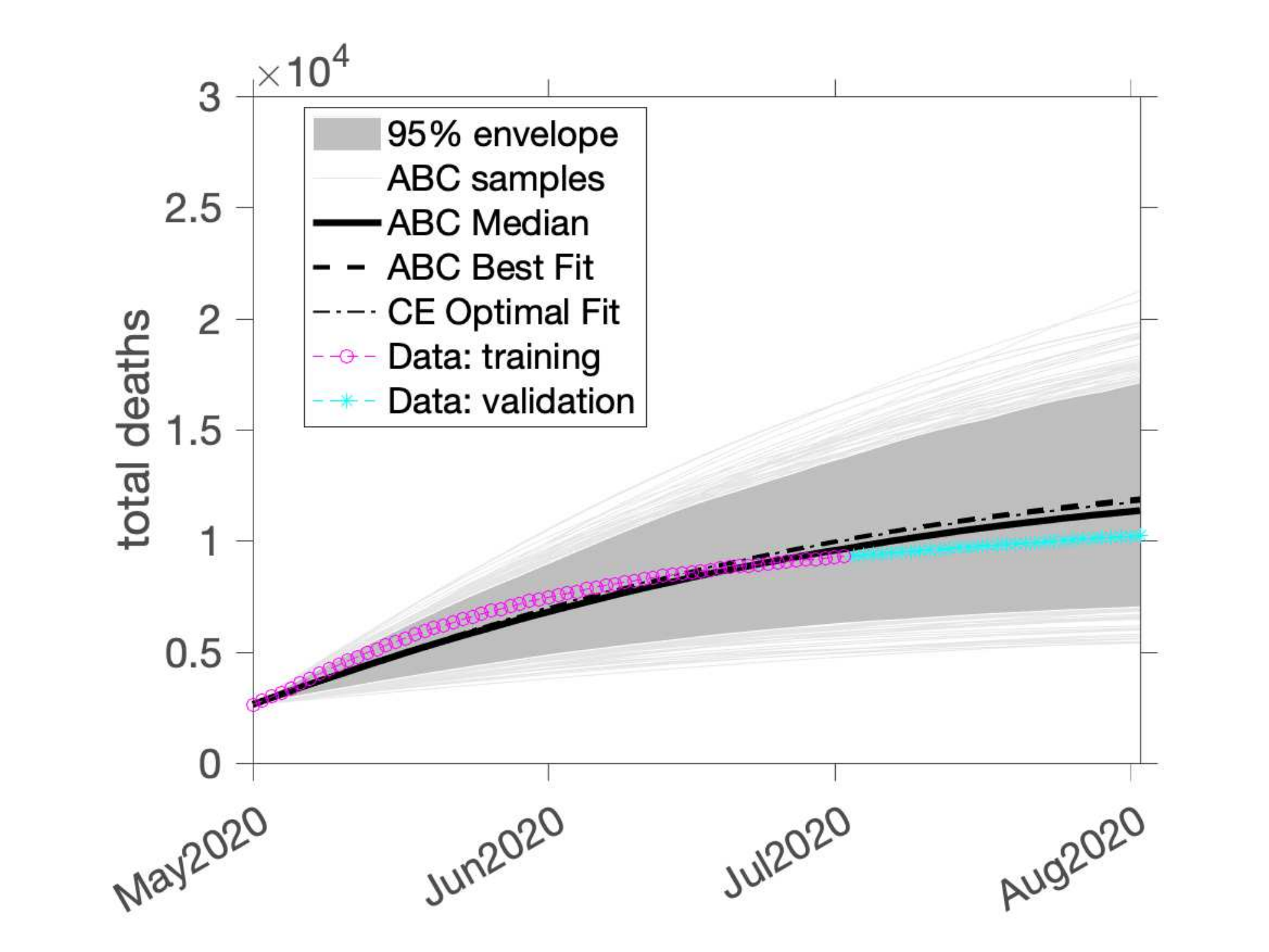}
    \includegraphics[scale=0.3]{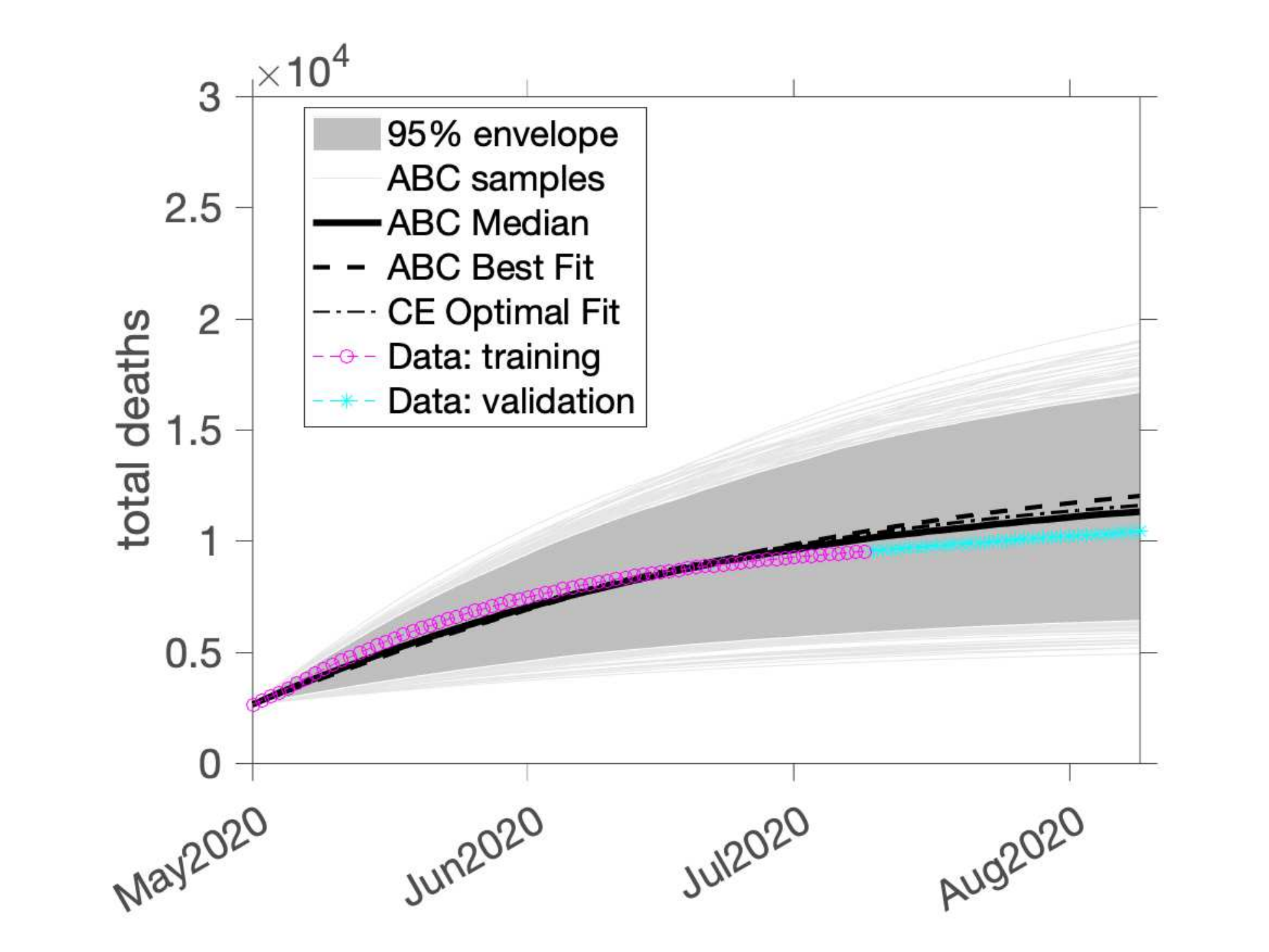}
    \caption{Evolution of the total deaths time series, calibrated with different datasets of Rio de Janeiro epidemic, extrapolating forecasts over a 30-day horizon. Training data are incremented every seven days, including information from the last seven days, starting with the period between May 1 and 7, 2020, and ending with May 1 and July 9, 2020.}
    \label{fig_pred_hor_D}
\end{figure*}

To this end, Figures~\ref{fig_pred_hor_H} and \ref{fig_pred_hor_D} show the evolution of the two QoIs, for various training data sets, extrapolating forecasts over a 30-day horizon. Training data are incremented every seven days, including information from the last seven days, starting with the period between May 1 and 7, 2020, and ending with May 1 and July 9, 2020.

For the first three calibrations (calibrations between the first and third week), the model presents a modest descriptive capacity of the data, with the trend of the short and mid-term forecasts being quite discrepant to that observed in the following weeks. However, the respective credibility intervals encompass the observations.

As the weeks go by, with more (and better quality) data feeding into the model (calibrations between the fourth and seventh week), the description of the training data improves substantially, as does the predictive ability. In such cases, within a week, the model's median predicts the numbers of hospitalizations and deaths with reasonable accuracy for epidemic estimates. However, it starts to lose accuracy from the second week of extrapolation gradually, although it continues to follow, more or less, the trend of the data for 30 days (and covers them via the credibility band). 

The descriptive capacity remains impeccable for the last three weeks of model calibration (weeks eight to ten), with good short-term (one week) prediction. However, it is possible to notice a significant divergence between predictions and observations after seven days of extrapolation. Such a loss of predictability is not directly related to the quantity or quality of the calibration data but rather to a structural change in the dynamic behavior of the outbreak. In July 2020, the second wave of contagion began in the city of Rio de Janeiro \cite{Gianfelice2022p031101}, drastically changing the trend of evolution of QoIs. As the infection rate $\beta(t)$ was modeled to only contemplate a single change in the infection plateau, the present model cannot accurately describe the new wave of infections. One possibility to make the model regain its predictive capacity would be the inclusion of a new infection plateau in Eq.(\ref{eq_beta}), as done in \cite{Vasconcelos2021}. In general, this strategy can be adopted to address not just one but several waves of infection. Due to space limitations, the authors did not include results in this sense in the manuscript, but it would be interesting to test this strategy in future work.

The above results show that the model's descriptive capacity and predictability limit are strongly influenced by the amount and quality of data used in the calibration process. 

In general, the more data, the greater the predictability horizon of the model. It presents an excellent or good capacity to extrapolate within a horizon of one or two weeks, with some capacity to predict the trend up to thirty days (depending on the outbreak's dynamic behavior). The quality of the outbreak's data also matters, as it is clear from the first three numerical experiments. In these cases,  hospitalization data do not show the typical fluctuation of this time series (probably due to underreporting, since at the beginning of May 2020, the surveillance system in Rio de Janeiro was still adapting to the pandemic), which affects the model's fitting.

Insufficient or poor quality data can compromise the model's fit, generating unrealistic predictions. However, structural changes in the dynamic behavior of the epidemic (e.g., the emergence of a new wave of contagion) have an even more pronounced effect in compromising the predictive capacity of the model. In the periods preceding such changes, even quality data cannot guarantee that the model will extrapolate the data well in the mid (or even in the short) term.

\subsection{Verification of efficiency for the CE-ABC framework for data-driven epidemic modeling}

To conclude the discussion of the results, this section presents a computational experiment to address the computational-statistical efficiency of the CE-ABC framework in comparison with a standard ABC approach (without using CE method to refine the prior).

\begin{figure*}
    \centering
    \includegraphics[scale=0.3]{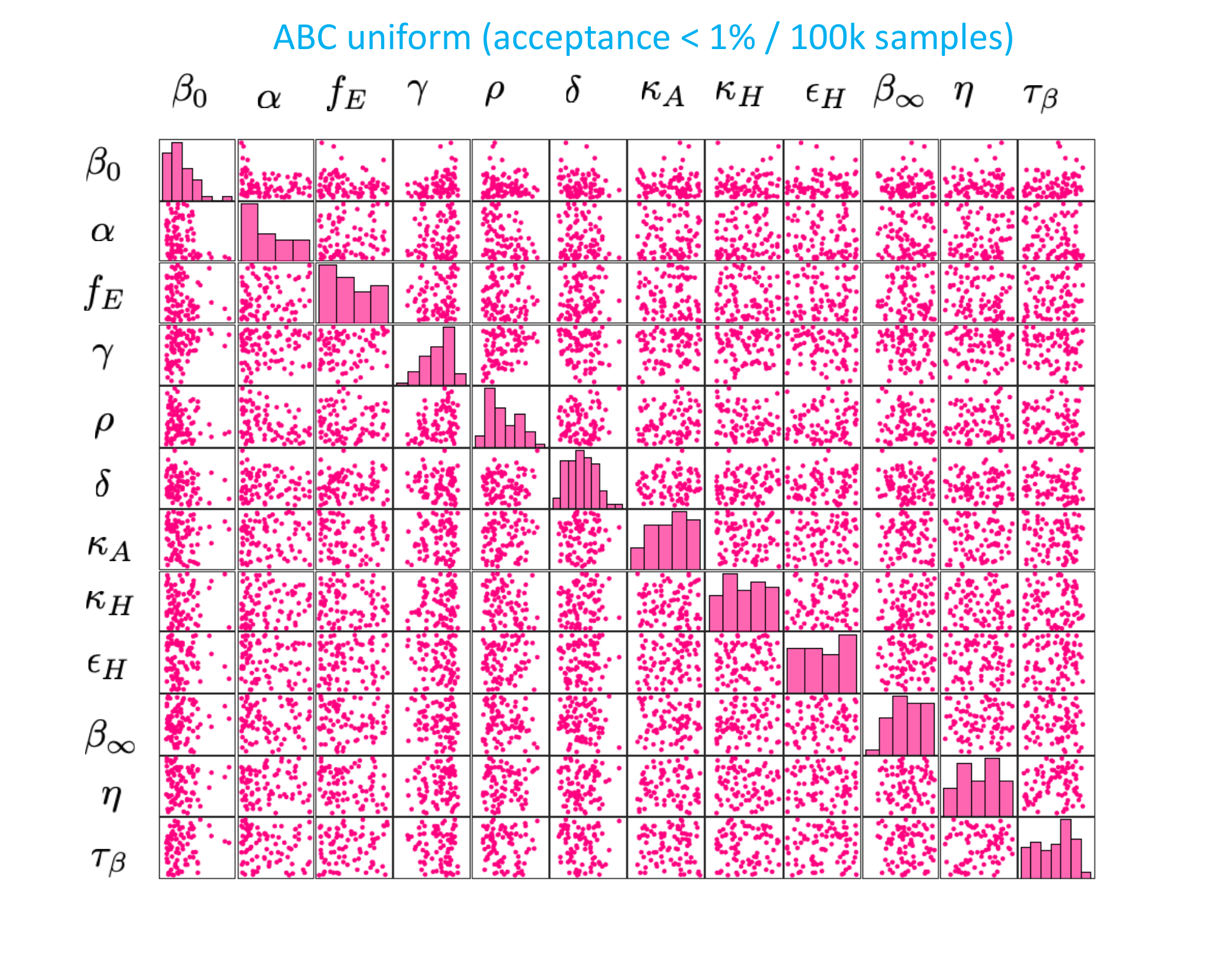}
    \includegraphics[scale=0.3]{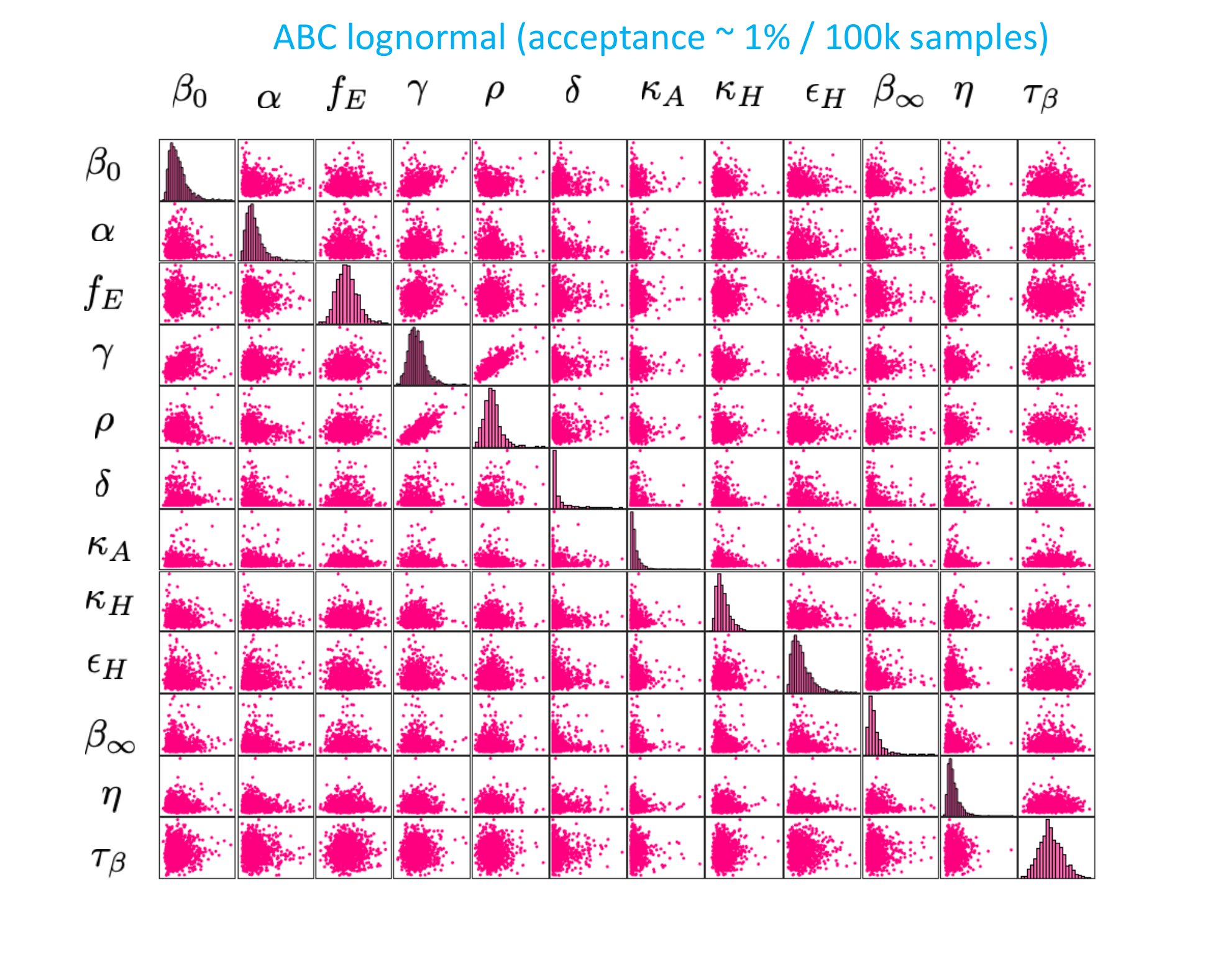}\\
    \includegraphics[scale=0.3]{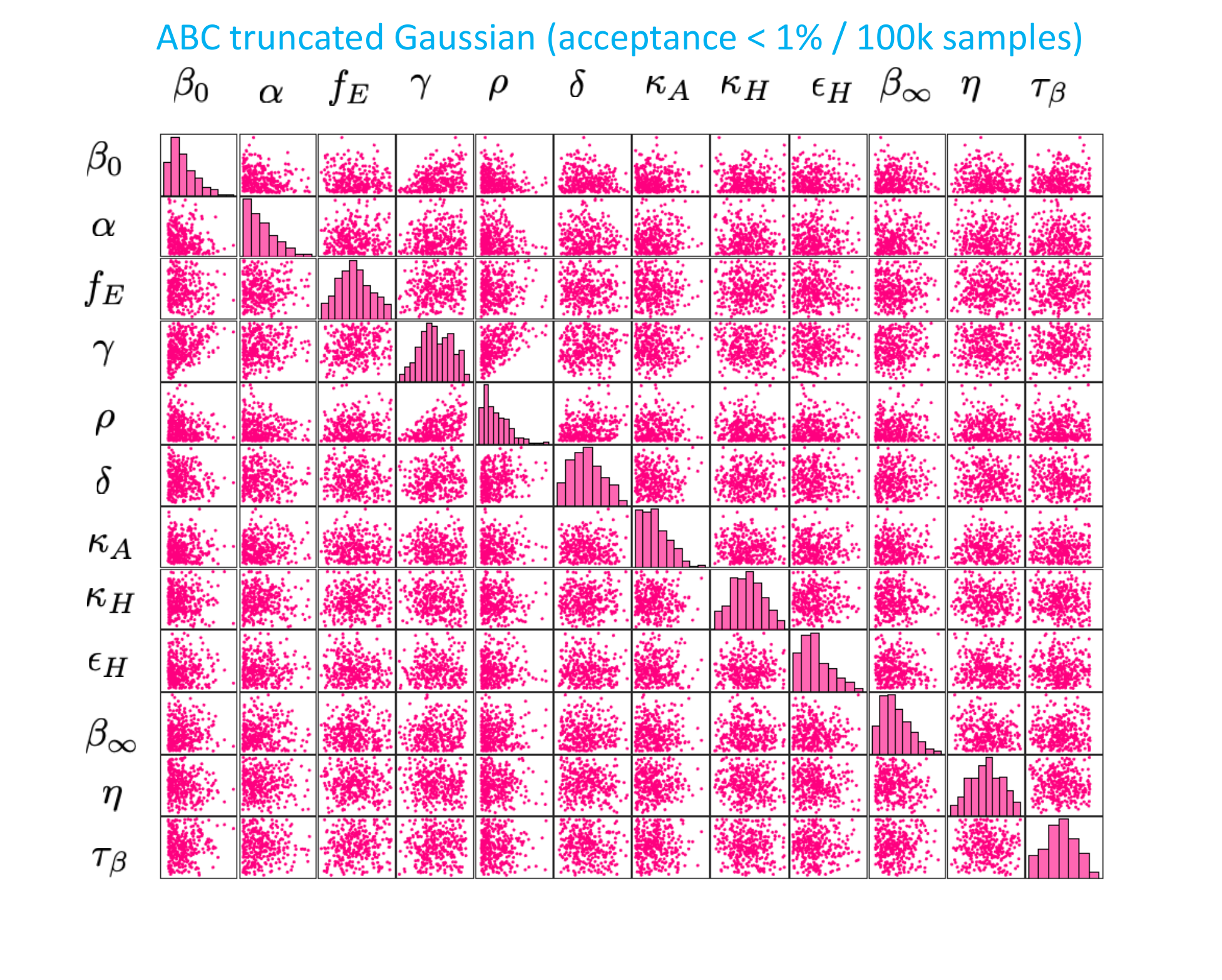}
    \includegraphics[scale=0.3]{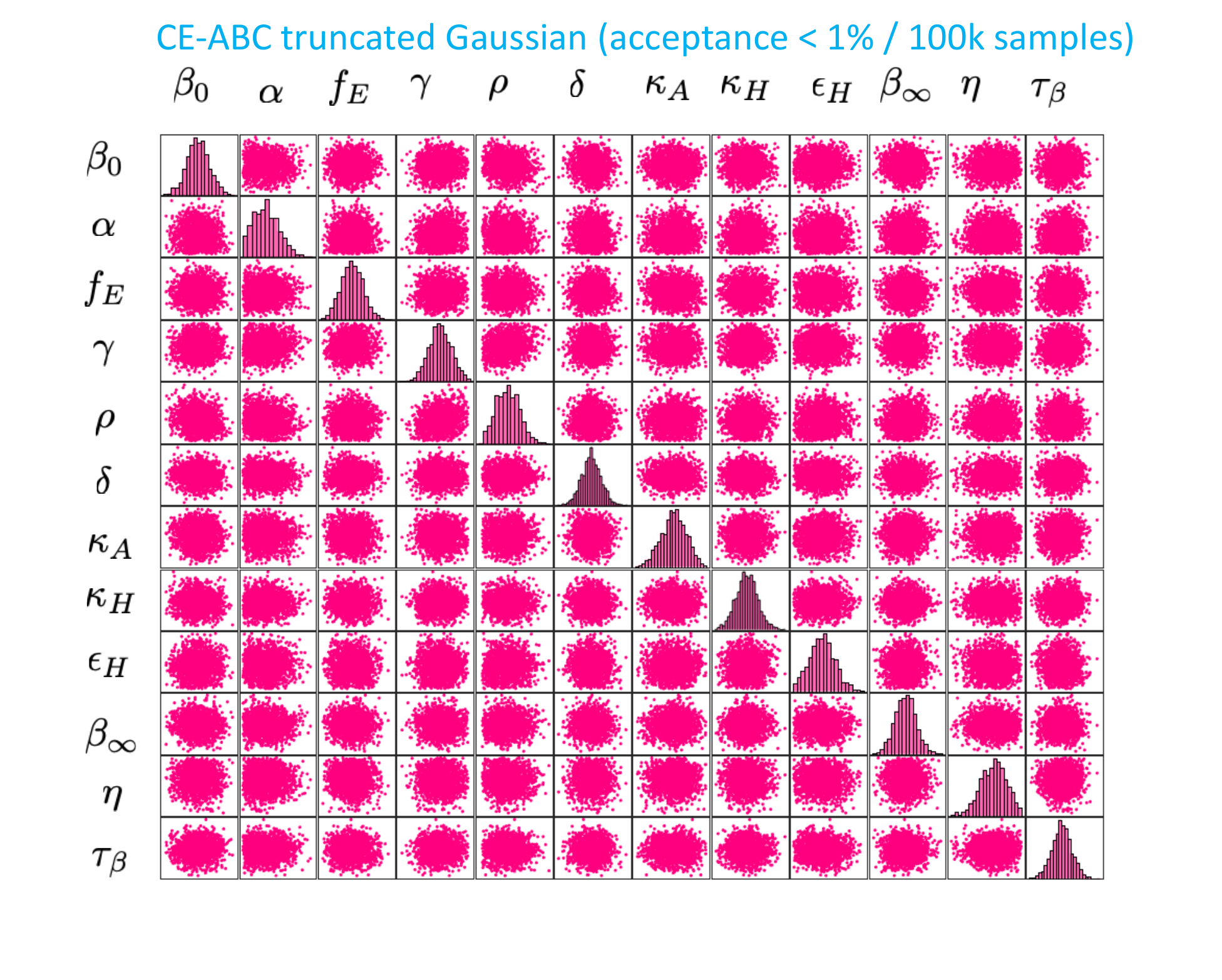}
    \caption{Histograms and scatter plots for the model parameters obtained with different statistical learning strategies. ABC with a uniform prior and 100k samples (top left); ABC with lognormal prior and 100k samples (top right); ABC with truncated Gaussian prior and 100k samples (bottom left); CE-ABC with truncated Gaussian prior and 2k samples (bottom right).}
    \label{fig_ce-abc_efficiency}
\end{figure*}

For this purpose, four test cases are compared in Figure~\ref{fig_ce-abc_efficiency}, which shows histograms and scatter plots for the epidemic model parameters obtained with different strategies of statistical learning:
\begin{enumerate}
\item ABC with a uniform prior;
\item ABC with lognormal prior;
\item ABC with truncated Gaussian prior;
\item CE-ABC with truncated Gaussian prior.
\end{enumerate}

In all these tests, the values adopted for the limits of the distributions (when of limited support) are shown in Table~\ref{tab_nominal_parameters}, where the mean values also come from. The respective standard deviations are defined as those corresponding to a uniform distribution on the finite support of this table. The other parameters are similar to the case discussed in section~\ref{calib_valid_model}, except the number of samples used in Bayesian inference, which is $N_{abc} = 100k$ for the three case that use standard ABC, and just $N_{abc} = 2k$ for the novel CE-ABC framework.

For the three cases where only ABC is used, the acceptance rate is at most a modest 1\%, while a substantial value of 87\% is obtained in the case that uses CE-ABC. These results show the efficiency of the new CE-ABC approach proposed here concerning the standard ABC method. It is also clear that the CE-ABC framework provided a gain in information about the distribution of parameter values much higher than the traditional method, with average values reasonably far from the boundaries. Both characteristics result from using the CE method to refine the prior, which becomes much more informative than the distribution used to sample the domain at the beginning of the optimization process that identifies the model parameters' nominal values.

% --------------------------------------------------------------------------

% --------------------------------------------------------------------------
\section{Concluding remarks}
\label{sec_concl}
% --------------------------------------------------------------------------

\subsection{Contributions}

This paper proposes a new framework for model calibration and uncertainty quantification that combines the cross-entropy (CE) method for optimization with approximate Bayesian computation (ABC) for statistical learning. In this approach CE is used to obtain an initial and informative estimation of the model parameters. Then, central tendency and dispersion information obtained from CE are used to construct a informative prior distribution for an inference process that uses ABC to refine the model calibration and propagate the underlying uncertainties via acceptance-rejection Monte Carlo sampling. This framework also employs a heuristic strategy for identification of the initial conditions by using plausible dynamic states that are compatible with observational data.

This combination of well-established algorithms gives rise to a framework for uncertainty quantification with several good features. CE and ABC are intuitive and straightforward algorithms. Their combination gives rise to a simplistic and robust framework, with few control parameters of clear interpretation, where no gradient computation is required. In the update step with ABC, the initial knowledge about the model parameters obtained by CE optimization is incorporated into the prior distribution and updated with the available data to produce an informative posterior distribution. Also, there is no need to assume an additive Gaussian error. The uncertainty propagation is performed when the parameters are identified, generating considerable computational savings. The methodology's major limitation is its sampling nature, so many simulations might be needed to achieve convergence. This characteristic is not a problem for applications using epidemic models based on differential equations, where each deterministic simulation has a low computational cost. But for other domains (e.g. computational mechanics), where models may take hours/days to run a single instance, the CE-ABC framework may not be competitive.

The proposed methodology was tested on an epidemic model with an SEIR-type structure that also considers asymptomatic individuals, hospitalizations, deaths, and time-dependent transmission rate. Actual data from COVID-19 outbreaks in Rio de Janeiro city were employed in the model calibration process. The results were consistent, and the methodology seems promising. They show that it is possible to perform good calibrations of the epidemic model with the CE-ABC formalism in scenarios that require a descriptive model (to explain past outbreaks) and those where the objective is to obtain a predictive model (to infer future behavior of epidemics). In scenarios where the epidemic model structure is a good abstraction of the contagion dynamics, a horizon of good quantitative predictability of up to two weeks can be achieved using CE-ABC for model calibration and uncertainty quantification, with a good capacity for qualitative description of the data trend for up to one month.

\subsection{Future directions}

There are some possibilities to continue this research. One can apply the proposed methodology to other dynamical systems, including other COVID-19 models with different data, or the possibility of contemplating reinfection. Another branch that can be explored is related to model (epistemic) uncertainties. For instance, it can be very appealing to combine our CE-ABC framework with methodologies that compensate for deficiencies in the structure of the mathematical model. For instance, the random matrix-based nonparametric probabilistic approach by C. Soize \cite{Soize2000}; the universal differential equations (UDE) for scientific machine learning by Rackauckas et al. \cite{Rackauckas2020}; or one of the physics-informed neural networks approached for epidemic modeling available on the literature \cite{Kharazmi2021p744,Raissi_Ramezani_Seshaiyer_2019,Shaier2021arxiv,Yazdani2020p1}. It would also be exciting and natural to insert the CE-ABC algorithm proposed here as a calibration/UQ module in the integrated framework for data-driven epidemic models developed by Zhang et al. \cite{Zhang2021p1}.

\subsection{Disclaimer}

A model is always wrong, but some of them are useful. This idea has a more pronounced meaning in computational epidemiology than in physics, as the first principles of epidemic dynamics are unknown. Although a mechanistic model such as the SEIR(+AHD) used here is a (typically very rough) approximation of epidemic dynamics, it allows exploring qualitative scenarios (short, medium, and long term) that can provide great insight into the evolution of the outbreak. Thus,  being an extremely valuable tool for epidemiologists \cite{Holmdahl2020p303}. Undoubtedly, such an approach is much more rational and conservative than being guided by the (well-intentioned or not) opinion of curious people with no training in the area or the general public (layman by definition).

However, a final observation is necessary before one uses an epidemic model to guide decision-making during a real-time outbreak. It concerns the interpretation of results. As epidemiology is a highly interdisciplinary area, it is practically impossible for a single professional to hold all the necessary skills to assess the results of an epidemic simulation unequivocally and, above all, understand the consequences of intervention measures that can be taken. Scientific, ethical, and humanistic aspects are equally important in this context and must be discussed by an interdisciplinary panel of professionals. In this sense, the authors of this paper strongly recommend that simulations of this nature, especially made with our model and framework, be evaluated and used with great caution when making decisions, preferably being scrutinized by a team of experts.
% --------------------------------------------------------------------------

% --------------------------------------------------------------
\section*{Dedication}

The authors dedicate this work to the memory of all the victims of the COVID-19 worldwide tragedy. In particular, the first author dedicates the paper to his friends Natasha Zadorosny and Victor Costa Silva, who left in their prime.
% --------------------------------------------------------------

% --------------------------------------------------------------
\section*{Acknowledgements}

The authors thank Prof. Luiz Max Fagundes de Carvalho (EMAp/FGV) and the anonymous reviewers for their critical reading of this text and for the excellent comments that helped improve the manuscript's final version.
% --------------------------------------------------------------

% --------------------------------------------------------------
\section*{Funding}

The first author received financial support from the Carlos Chagas Filho Research Foundation of Rio de Janeiro State (FAPERJ) under the following grants: 210.167/2019, 211.037/2019, and 201.294/2021. The second author would like to acknowledge the Engineering and Physical Sciences Research Council (EPSRC) via grant number EP/R006768/1. The last author would like to acknowledge the financial support from the Brazilian agencies \textit{Coordenação de Aperfeiçoamento de Pessoal de Nível Superior} (CAPES) under the grants Finance Code 001, PROEX 803/2018, and CAPES-PRINT 88887.569759/2020-00, and FAPERJ under the grant 201.183/2022.
% --------------------------------------------------------------

% --------------------------------------------------------------
\section*{Code availability}

To facilitate the reproduction of this paper's results the code used in the simulations is available on GitHub:\\
\textcolor{magenta}{\url{https://github.com/americocunhajr/CE-ABC}}
% --------------------------------------------------------------

% --------------------------------------------------------------
\section*{Declarations}
% --------------------------------------------------------------

% --------------------------------------------------------------
\section*{Conflict of Interest }
The authors declare they have no conflict of interest.
% --------------------------------------------------------------

% references
\bibliographystyle{spmpsci}      % mathematics and physical sciences

\bibliography{bibliography}
% --------------------------------------------------------------

\end{document}